\documentclass[12pt]{article}
\pdfoutput=1
\usepackage{mathtools, amsfonts, amsthm, latexsym, amssymb}

\usepackage{authblk}

\usepackage{graphicx, caption, subcaption}
\graphicspath{ {./Figures/} }

\usepackage[sc]{mathpazo}      
\usepackage[scaled=0.90]{helvet}
\usepackage[scaled=0.85]{beramono}
\usepackage[T1]{fontenc}
\usepackage{dsfont}
\usepackage[tracking=true]{microtype}

\SetTracking{encoding={*}, shape=sc}{50}

\usepackage[cm]{fullpage}
\numberwithin{equation}{section}

\usepackage{fancyhdr}
\fancypagestyle{empty}{
	\setlength{\headheight}{15.2pt}
	\fancyhf{}
	 
	\fancyhead[R]{UMD-PP-013-008} 
}

\usepackage{hyperref}
\hypersetup{linktocpage, colorlinks=true, linkcolor=blue, citecolor=blue, urlcolor=blue}

\newcommand{\CFT}{\textls[50]{CFT }}
\newcommand{\BTZ}{\textls[50]{BTZ }}

\renewcommand{\d}{\partial}
\DeclareMathOperator{\T}{T}
\DeclareMathOperator{\tr}{tr}
\DeclareMathOperator{\re}{Re}
\DeclareMathOperator{\im}{Im}
\DeclareMathOperator{\CPT}{CPT}
\renewcommand{\neg}{\negthickspace}

\newcommand{\alignStart}{ \begin{equation} \begin{aligned} }
\newcommand{\alignEnd}{ \end{aligned} \end{equation} }
\newcommand{\gatherStart}{ \begin{equation} \begin{gathered} }
\newcommand{\gatherEnd}{ \end{gathered} \end{equation} }

\newcommand{\TT}{\mathcal T}
\newcommand{\J}{\mathcal J}
\newcommand{\1}{\mathds 1}
\renewcommand{\L}{\mathcal L}
\renewcommand{\O}{\mathcal O}
\newcommand{\D}{\mathcal D}

\newcommand{\bra}[1]{\langle #1 \vert}
\newcommand{\ket}[1]{\vert #1 \rangle}

\newcommand{\matrixel}[3]{\langle #1 \vert #2 \vert #3 \rangle}
\newcommand{\vev}[1]{{\langle #1 \rangle}}

\title{\bfseries Holography of the \BTZ Black Hole, \\ Inside and Out}
\author{Anton de la Fuente {\small and} Raman Sundrum}
\affil{Maryland Center for Fundamental Physics \\
Department of Physics \\ 
University of Maryland \\
College Park, 20782 MD}
\date{}
\begin{document}

\maketitle

%%%%%%%%%%%%%%%%%%%%%%%%%%%%%%%%%%%%%%%%%%%%%%%%%%%%%%%%%%%%%%%%%%%%%%%%%%%%%%%%%%%%%%%%%%%%
\begin{abstract}
We propose a $1+1$ dimensional CFT dual structure for quantum gravity and matter on the extended $2+1$ dimensional BTZ black hole, realized as a quotient of the Poincar\'e patch of AdS$_3$. The quotient spacetime includes regions beyond the singularity, "whiskers",  containing timelike and lightlike closed curves, which at first sight seem unphysical. The spacetime includes the usual AdS-asymptotic boundaries outside the horizons
as well as boundary components inside the whiskers. We show that local boundary correlators with some endpoints in the whisker regions: (i) 
are a protected class of amplitudes, dominated by effective field theory even when the associated Witten diagrams appear to traverse the singularity, (ii) 
describe well-defined diffeomorphism-invariant quantum gravity amplitudes in BTZ, (iii) sharply probe some of the physics inside the horizon but outside the singularity, and (iv) are equivalent to correlators of specific \emph{non-local} CFT operators in the standard thermofield entangled state of two CFTs.  In this sense, the whisker regions can be considered as purely auxiliary spacetimes in which these useful non-local CFT correlators can be rendered as \emph{local} boundary correlators, and their diagnostic value more readily understood. Our results follow by first performing a novel reanalysis of the Rindler view of standard AdS/CFT duality on the Poincar\'e patch of AdS,  followed by  
 exploiting the simple quotient structure of BTZ which turns the Rindler horizon into the BTZ black hole horizon. 
While most of our checks are within gravitational effective field theory, we arrive at a fully non-perturbative CFT proposal to probe the UV-sensitive approach to the singularity. 
  \end{abstract}

\thispagestyle{empty}
\clearpage
\setcounter{page}{1}
\tableofcontents
\noindent\makebox[\linewidth]{\rule{\textwidth}{.5pt}} 

%%%%%%%%%%%%%%%%%%%%%%%%%%%%%%%%%%%%%%%%%%%%%%%%%%%%%%%%%%%%%%%%%%%%%%%%%%%%%%%%%%%%%%%%%%%%
%%%%%%%%%%%%%%%%%%%%%%%%%%%%%%%%%%%%%%%%%%%%%%%%%%%%%%%%%%%%%%%%%%%%%%%%%%%%%%%%%%%%%%%%%%%%
\section{Introduction}

Nearly a century after the discovery of the Schwarzschild metric, 
\begin{equation}
	ds^2 = \left(1 - \frac{r_S}{r} \right) dt^2 - \frac{dr^2}{1-\frac{r_S}{r}} - r^2 \left(d \theta^2 + \sin^2 \theta \, d\phi^2 \right), \label{schwarzschildMetric}
\end{equation}
black holes remain a source of mystery and fascination. In theoretical physics, they provide key insights for our most ambitious attempts to unify gravity, relativity and quantum mechanics. Viewed from the outside as robust endpoints of gravitational collapse, and decaying subsequently via Hawking radiation, black holes pose the information paradox. Falling inside, the roles of "time", $\tau$, and "space", $r$, apparently trade places,  the horizon now encompassing a universe within, with the future singularity  its "big crunch".  Understanding these
 dramatic phenomena seems tantalizingly close to our grasp, 
 just beyond the horizon, a region comprised of familiar, smooth patches of spacetime.  And yet, the local simplicity of the horizon belies its global subtlety, which still lacks an explicit inside/outside description within a fundamental framework for quantum gravity (as exemplified by the recent "firewall" paradox \cite{amps,amps2,amps3}\footnote{See also \cite{samB} for a prediction similar to firewalls from different assumptions.}) regarding evaporating black holes. Nevertheless, powerful ideas and results in holography \cite{holo1} \cite{holo2}, complementarity \cite{stretched}, string theory and AdS/CFT duality 
 \cite{juancft} \cite{gubsercft} \cite{wittencft} (reviewed in \cite{cftrev} \cite{juantasi} \cite{myrev}), 
 have combined with gravitational effective field theory
 (EFT) to give us a much clearer picture of the central issues (reviewed in \cite{genrev} \cite{mathur}). 
 
 In such a situation, it is natural to look for an "Ising model", a special case that enjoys so many technical advantages that we can hope to solve it  exactly, and whose solution would test and crystalize tentative grand principles, and brings new ones to the fore. For this purpose, the $2+1$-dimensional BTZ black hole \cite{btz, btz2} is, in many ways, an ideal candidate.
The BTZ  geometry solves Einstein's Equations with negative cosmological constant in $2+1$ dimensions, and is given in Schwarzschild coordinates by,
\begin{align}
	ds^2_\text{BTZ} &= \frac{r^2 - r_S^2}{R_\text{AdS}^2} d\tau^2 -\frac{R_\text{AdS}^2}{r^2 - r_S^2} dr^2 - r^2 d\phi^2 &&(-\pi \leq \phi \leq \pi,\; r > 0), \label{BTZmetric}
\end{align}	
not that dissimilar from \eqref{schwarzschildMetric}. 
The geometry asymptotes for large $r$ to that of global anti-de Sitter spacetime, AdS$_\text{3 global}$, with radius of curvature $R_\text{AdS}$ and AdS boundary at $r = \infty$. 
The horizon is at the Schwarzschild radius, $r = r_S$. 
 It is the simplest of the "large" AdS Schwarzschild black holes, eternal in that they 
do not decay via Hawking radiation, but rather are 
  in equilibrium with it \cite{adsbh}. It retains many of the key interesting features of black holes in general. In what follows it will be more convenient to rescale coordinates, 
\begin{align}
	\frac{R_\text{AdS}}{r_S} r \to r && \frac{r_S}{R_\text{AdS}} \tau \to \tau &&  \sigma \equiv r_S  \phi,
\end{align}
and to switch to 
$R_\text{AdS} \equiv 1$ units,  so the metric becomes
\begin{align}
ds^2_\text{BTZ}	&= (r^2-1) d\tau^2 - \frac{dr^2}{r^2-1} - r^2 d\sigma^2   && \left( -\pi r_S  \leq \sigma \leq \pi r_S, \; r > 0 \right). \label{AdSschwarzschildMetric}
\end{align}
The horizon is now at $r =1$.
 
 Although pure $2+1$-dimensional general relativity does not contain propagating gravitons, it does have gravitational fluctuations and backreactions, and coupled to propagating matter the EFT is non-renormalizable as in higher dimensions (in fact, it may be a compactification of higher dimensions, and contain propagating Kaluza-Klein gravitons), requiring UV completion. 
 It also shares with higher-dimensional eternal AdS Schwarzschild black holes, the central consequence of AdS/CFT duality: 
 as an object inside AdS$_\text{global}$ the black hole inherits a holographic dual in terms of a "hot" conformal field theory (CFT) (for BTZ, a $1+1$ CFT on a spatial circle), the CFT temperature being dual to the BTZ Hawking temperature. 
More precisely \cite{juanbh} (see also the  earlier steps and insights of  \cite{israel} \cite{horowitzmarolf} \cite{vijay}), the duality is framed in terms of the Kruskal extension of BTZ, 
\begin{align}
	ds^2 &= \frac{4dudv}{(1+uv)^2} - \left( \frac{1-uv}{1+uv} \right)^2 d\sigma^2 	&&(|uv| < 1). \label{kruskalMetric}
\end{align}
The horizon, "singularity" and AdS boundaries are now as follows:
\alignStart
	\text{boundary:} \quad & uv = -1 \\
	\text{horizon:} \quad & \text{$u = 0$ or $v = 0$}  \\
	\text{singularity:} \quad & uv = 1. 
\alignEnd
\begin{figure}
	\centering
	\includegraphics[width=0.25\textwidth]{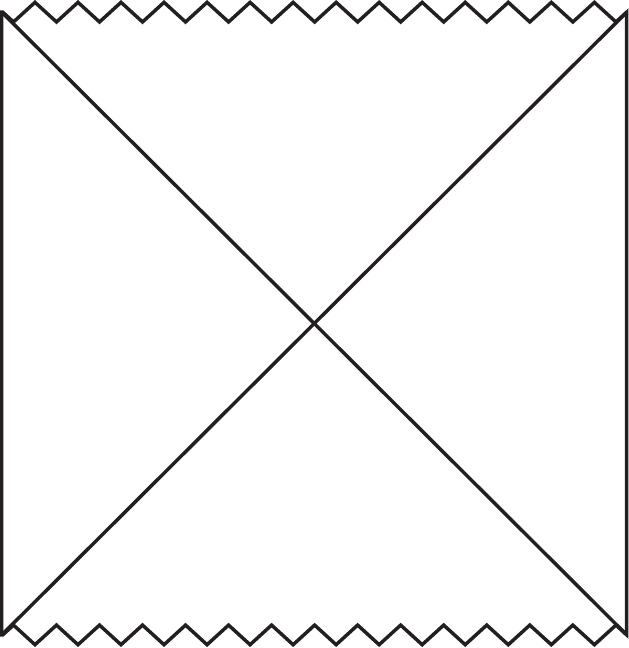}
	\caption{The Penrose diagram of the extended BTZ black hole spacetime. The vertical lines represent the boundaries of two asymptotically AdS regions.}
	\label{penrose}
\end{figure}
The Penrose diagram of this spacetime is shown in Fig.\@~\ref{penrose}.   BTZ is seen to interpolate between two distinct asymptotically-AdS$_\text{global}$ boundary regions. 
The holographic dual is then given by two CFTs, dynamically decoupled, but in a state of  "thermofield" \cite{thermo1,thermo2,thermo3,thermo4,thermo5,thermo6,thermo7} entanglement,
\begin{equation}
	\ket{\Psi}_\text{BTZ}  \equiv \sum_n e^{-\pi E_n} \ket{\bar n} \otimes \ket{n}. \label{BTZthermofieldState}
\end{equation}
The entangled state is dual to the Hartle-Hawking choice of vacuum \cite{hh} for the BTZ black hole.

There remains the puzzle of detailing just how this CFT description incorporates processes inside the BTZ horizon. We know that in asymptotic AdS spacetimes,  the set of local boundary correlators gives a beautiful diffeomorphism-invariant quantum gravity description of scattering which generalizes the S-matrix construction of asymptotic Minkowski spacetimes, and, in the sense described in \cite{wittenqgds}, is even richer in structure. Furthermore, these boundary correlators have a non-perturbative and UV-complete description in terms of correlators of local CFT operators "living" on the AdS boundary, $\partial$AdS. But in AdS-Schwarzchild spacetimes like BTZ it is not apparent what CFT questions give a diffeomorphism-invariant and non-perturbative description of scattering inside the horizon: one can send in wavepackets from outside the horizon aimed to scatter within, but the products of any scattering must causally end up at the singularity rather than returning to the exterior AdS boundaries. While one can connect Witten diagrams from interaction points in the interior of the (future) horizon to the boundaries shown in Fig.\@~\ref{penrose}, these connections cannot sharply capture the fate of such interactions since they are at best spacelike.

This does not mean that the interior of the horizon is out of bounds to the CFT description. 
In a sense, what is required is 
a set of "out states" consisting of approximately decoupled bulk particles located on a spacelike hypersurface before the (future) singularity, with which one can compute the overlap with the state resulting from the scattering process. Even in (the simpler) AdS spacetime, particles inside the bulk are described by \emph{non-local} disturbances of the CFT, so one can anticipate that any holographic description of scattering inside the horizon will necessarily involve correlators of \emph{non-local} CFT operators. But specifically which non-local CFT operators correspond to the simplest basis of "out states", so that their correlators (with other CFT operators) provide a sharp diagnostic of scattering inside the horizon?  In this paper, we identify such non-local CFT operators and demonstrate that they correspond to the intuitive notion of scattering inside the horizon. Our proposal is precisely and non-perturbatively framed. We test it by applying it to scattering inside the horizon but far from the singularity where, at short distances $\ll R_\text{AdS}$, the behavior is very much like scattering outside the horizon or in flat spacetime, and so we know what to expect. We then show how to apply our proposal to probe  the more mysterious regime near the singularity, where EFT breaks down and even perturbative string theory may be blind to important non-perturbative effects (see for example, \cite{juanbh}). 
Since the interior of the horizon is a cosmological spacetime, finding
 the non-local CFT operators can be thought of as giving the holographic description of a quantum cosmology with singularity, a signficant step beyond the more familiar holography of static AdS.

\section{Overview and Organization} 

\subsection{Diffeomorphism invariance in Non-perturbative Formulation}

The issue of diffeomorphism invariance, and the challenge it poses for a description of the interior of the horizon,  may seem unfamiliar to those who routinely use local field operators to sharply describe processes in the real world (which of course includes quantum gravity in some form).  
This would naively suggest that in the BTZ context we should use local bulk operators acting on the Hartle-Hawking state to create "in/out" states inside the horizon, and then translate these operators to (non-local) operators of the CFT.  However, fundamentally   \emph{all} local fields (composite or elementary)  violate the diffeomorphism gauge symmetry of quantum gravity (their spacetime argument at least is not generally coordinate-invariant), just as the local electron and gauge fields violate gauge invariance in QED. 
Of course, we are used to using gauge non-invariant local operators within a gauge-fixed formalism, but these are, in essence, non-local constructions in  the gauge-invariant data.
For example in electromagnetism, the gauge-invariant data (in Minkowski spacetime) are provided by specifying some field strength, $F_{\mu \nu}(x)$, subject to the Bianchi identity, $\epsilon^{\mu \nu \rho \sigma} \partial_{\nu} F_{\mu \nu}(x) = 0$. This uniquely determines a "local" gauge potential, $A_{\mu}(x):  \partial_{\mu} A_{\nu} -  \partial_{\nu} A_{\mu}  = F_{\mu \nu}$, once we stipulate some gauge-fixing condition (and behavior at infinity), such as
\begin{equation}
\partial^{\mu} A_{\mu} =c(x). 
\end{equation}
$A_{\mu}(x)$ is thereby a \emph{non-local} functional of $F_{\mu \nu}(y)$. In this way, gauge-fixing is seen as a method for giving  non-local gauge-invariant operators a superficially local (and useful) form. In gravity, the gauge-fixing approach is useful for perturbatively small fluctuations of the metric, but not when there are violent fluctuations of the metric (or when the notion of spacetime geometry itself breaks down). And yet it is precisely large fluctuations of the metric that we are interested in when we are concerned with non-perturbative effects (in $G_{\rm Newton}$) saving us from  information loss (see discussion in \cite{juanbh}), or in the approach to the singularity. 
Therefore, in the non-perturbative framing of our proposal we avoid the intermediate step of gauge-fixed local bulk fields, instead exploiting the greater simplicity of BTZ over other black holes to directly identify the (diffeomorphism-invariant) CFT observables. 

Nevertheless, it is useful to see how our approach reduces to gauge-fixed EFT of bulk fields, when that is valid, and this also provides an arena for testing the proposal. To this end, we will show that correlators of local field operators inside the horizon can be re-expressed as correlators of non-local EFT observables outside the horizon (in principle accessible to an outside observer). Even though this "dictionary" is between gravitational EFT descriptions, the "translating" operation is non-perturbative in form. It resonates with the ideas of complementarity \cite{stretched}, where the interior of the horizon is not independent of the exterior, but rather a very different probe of it. 

\subsection{Strategy for BTZ}

BTZ is particularly well-suited to address the above issues for two reasons. First, the
enhanced conformal symmetry of $1+1$-dimensional CFTs over higher dimensions provides us with a better understanding 
of their properties. The second reason is that BTZ can be realized as a quotient of AdS spacetime itself, by identifying 
points related by a discrete AdS isometry \cite{btz, btz2}. At the technical level, BTZ Green functions can be easily obtained from the highly symmetric AdS Green functions using the method of images \cite{ichinose} \cite{vakkuri}. 
Most importantly, the BTZ horizon emerges as the quotient of a "mere" Rindler horizon, as would be seen by a class of accelerating observers in  AdS \cite{juanandy}. (See \cite{radscft} for a related discussion, and \cite{raamsdonk} for a higher-dimensional discussion.) 
  The Rindler view of AdS,  the BTZ "black string", is given by \eqref{BTZmetric} and \eqref{AdSschwarzschildMetric} again, but now with non-compact $\sigma \in ( - \infty, +\infty)$,  $r_S \equiv \infty$. 
Our approach is based on a novel reanalysis of Rindler AdS/CFT \cite{juanbh} \cite{raamsdonk}, in a manner that can then be straightforwardly quotiented to the BTZ case of interest. 

\begin{figure}
        \centering
        \begin{subfigure}[b]{0.4\textwidth}
                \centering
                \includegraphics[width=\textwidth]{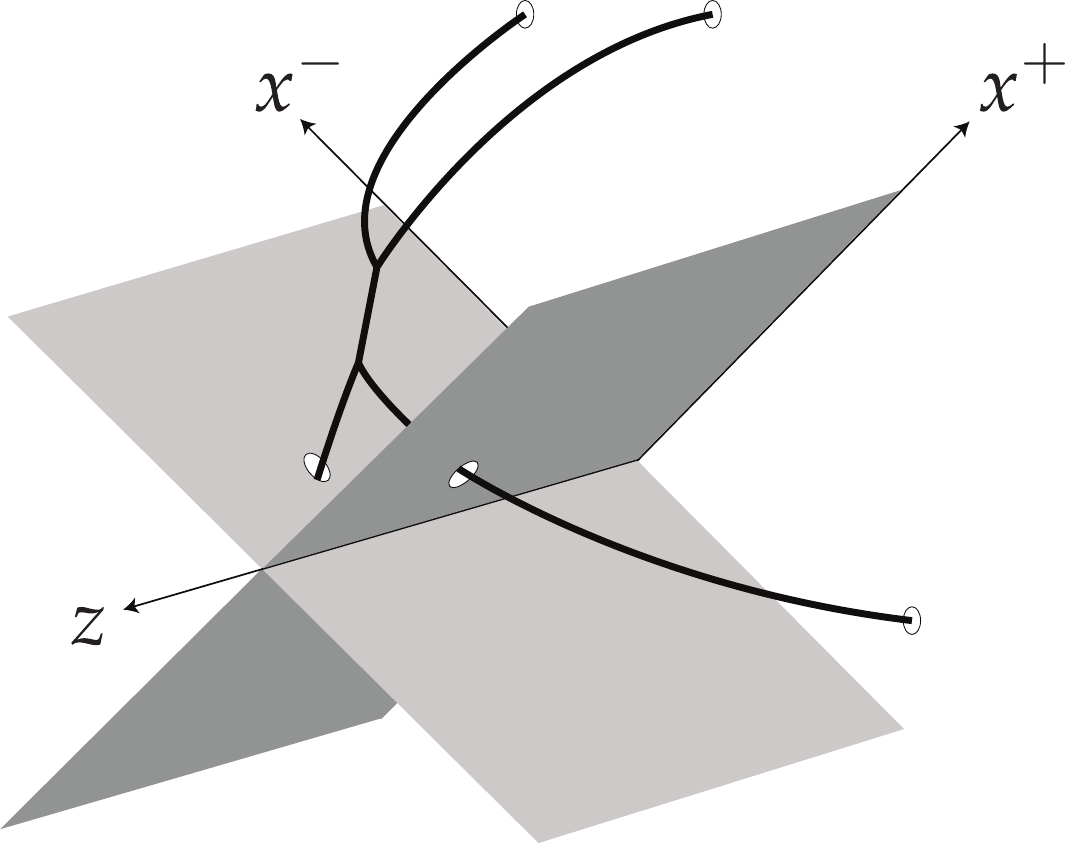}
                \caption{The two boundary operators at the top are timelike separated from the scattering event.}
                \label{sharpProbe}
        \end{subfigure}
        \qquad
        \begin{subfigure}[b]{0.4\textwidth}
                \centering
                \includegraphics[width=\textwidth]{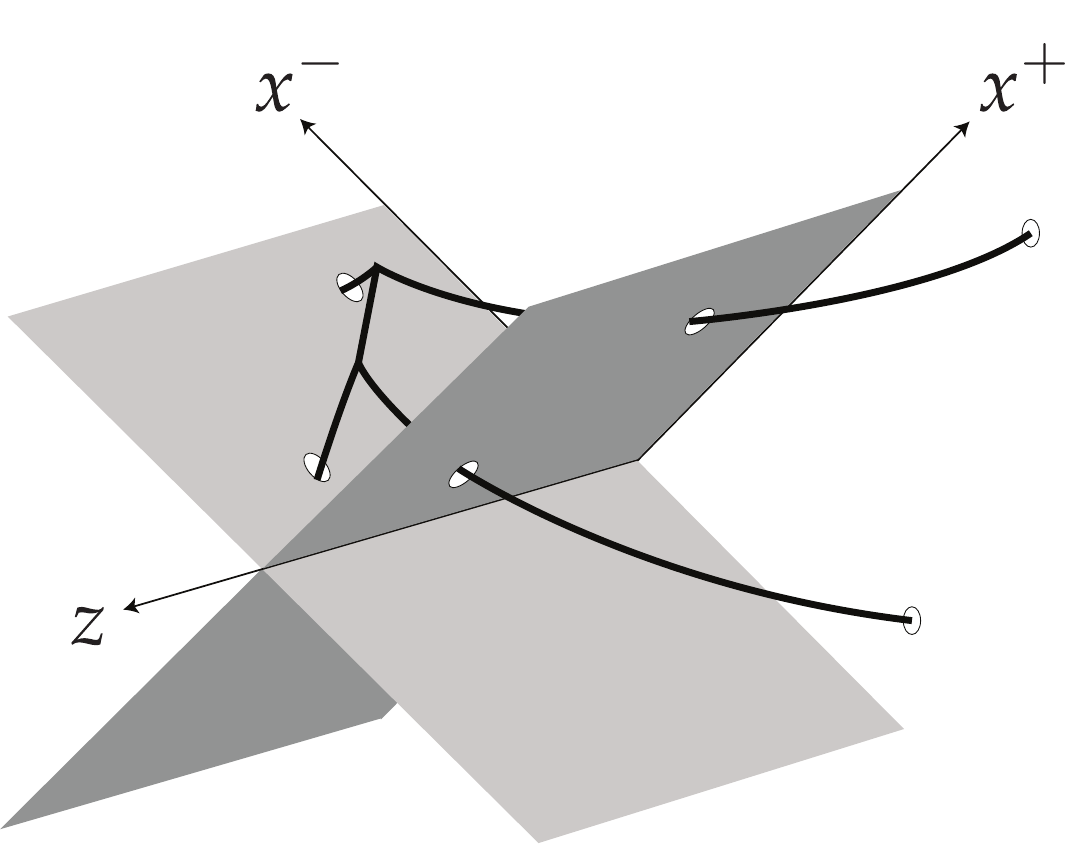}
                \caption{Boundary operators on the left and right Rindler wedges are spacelike separated from the scattering event.}
                \label{dullProbe}
        \end{subfigure}
        \caption{Boundary operators in the future region are needed to sharply probe scattering behind the horizon.}
\end{figure}

The central issue from the Rindler view can be seen in Fig.\@~\ref{sharpProbe}, depicting the Poincar\'e patch of AdS, where the intersecting planes 
are the Rindler horizons, light rays travel at 45 degrees to the vertical time axis, the boundary is at $z=0$, and 
\begin{equation}
x^{\pm} \equiv t \pm x
\end{equation}
are boundary ($1+1$ Minkowski) lightcone coordinates.
 Two particles are seen to enter the future horizon, scatter inside, and then the resultant particle lines  "measured" by local boundary correlators ending in the boundary region inside the horizon. Such correlators  can sharply diagnose the results of the scattering because the endpoints are causally connected to the scattering point (or region). In Minkowski CFT these endpoints correspond to local operators in the "Milne" wedge inside the Rindler horizon.
However, the dual Rindler CFT picture corresponds to two CFTs "living" only in the left and right regions outside the horizon (entangled with each other in the thermofield state), so that correlators of \emph{local} CFT operators correspond to boundary correlators only ending in the boundary regions outside the horizon. As seen in Fig.\@~\ref{dullProbe}, such local Rindler CFT correlators correspond to boundary correlators with endpoints at best \emph{spacelike} separated from the scattering point, not useful for a sharp diagnosis of the scattering (as we already saw from the Penrose diagram of Fig.\@~\ref{penrose}). 

However, the desired local operators of the Minkowski CFT (as opposed to the Rindler CFTs)  inside the horizon have the form,
\begin{equation}
	{\cal O}(t, x) \equiv e^{i H_\text{Mink} t} {\cal O}(0, x) e^{-i H_\text{Mink} t},  \qquad |t| > |x|, \label{HMink}
\end{equation}
where the operator at $t=0$ is now within the Rindler region and equivalent to a local Rindler CFT operator.
 The Minkowski CFT Hamiltonian $H_\text{Mink}$  is also some operator on the tensor product of the Hilbert spaces of the two Rindler CFTs ($=$ Hilbert space of the Minkowski CFT, as is apparent at $t=0$), so  
  ${\cal O}(t, x)$  must also be some operator of the Rindler CFTs. But 
because $H_\text{Mink} \neq H_\text{Rindler}$, ${\cal O}(t, x)$ is not simply a \emph{local} Heisenberg operator of the Rindler CFTs, but rather \emph{non-local} from the Rindler perspective. We conclude that non-local correlators of the Rindler CFTs are able to sharply capture scattering inside the Rindler horizon, the same way that local correlators of the Minkowski CFT ending inside the horizon do. The problem in taking the BTZ quotient of this nice story is that the quotient of $H_\text{Mink}$ does not exist: the associated $t$-translation isometry is broken by quotienting. 

An important  result of ours is to reproduce the correlators of \eqref{HMink}, which sharply capture scattering inside the Rindler horizon, with a new set of non-local Rindler CFT operators, 
\begin{equation}
	\O_\text{non-local} \equiv  e^{\frac{\pi}{2} (H_\text{Rindler} - P_\text{Rindler})}   \O_\text{local}  e^{-\frac{\pi}{2} (H_\text{Rindler} - P_\text{Rindler})}, \label{rindlerNonlocal}
\end{equation}
constructed from local Rindler CFT operators  ${\cal O}_\text{local}$ and the Rindler Hamiltonian and momentum, $H_\text{Rindler}, P_\text{Rindler}$. 
Note that we are not equating these new non-local operators with those of \eqref{HMink}; they will have different matrix elements within generic states. 
We only show that they have the same matrix elements in a fixed, special state, namely the thermofield state of the  two Rindler CFTs, 
namely $\ket \Psi$ for $r_S = \infty$.
 This suffices to capture scattering inside the Rindler horizon. But unlike \eqref{HMink}, the new operators are straightforwardly "quotiented" to the CFT dual of BTZ. 
Indeed, this quotient is simply the compactification of the spatial Rindler direction, so that $P_\text{Rindler}$ becomes the conserved \emph{angular} momentum of the thermofield CFTs on a spatial circle, and $H_\text{Rindler}$ becomes their Hamiltonian. We will show that the resulting non-local operators in the thermofield CFTs have correlators which provide some sharp probes of scattering inside the BTZ horizon.

\begin{figure}
	\centering
	\includegraphics[width=0.25\textwidth]{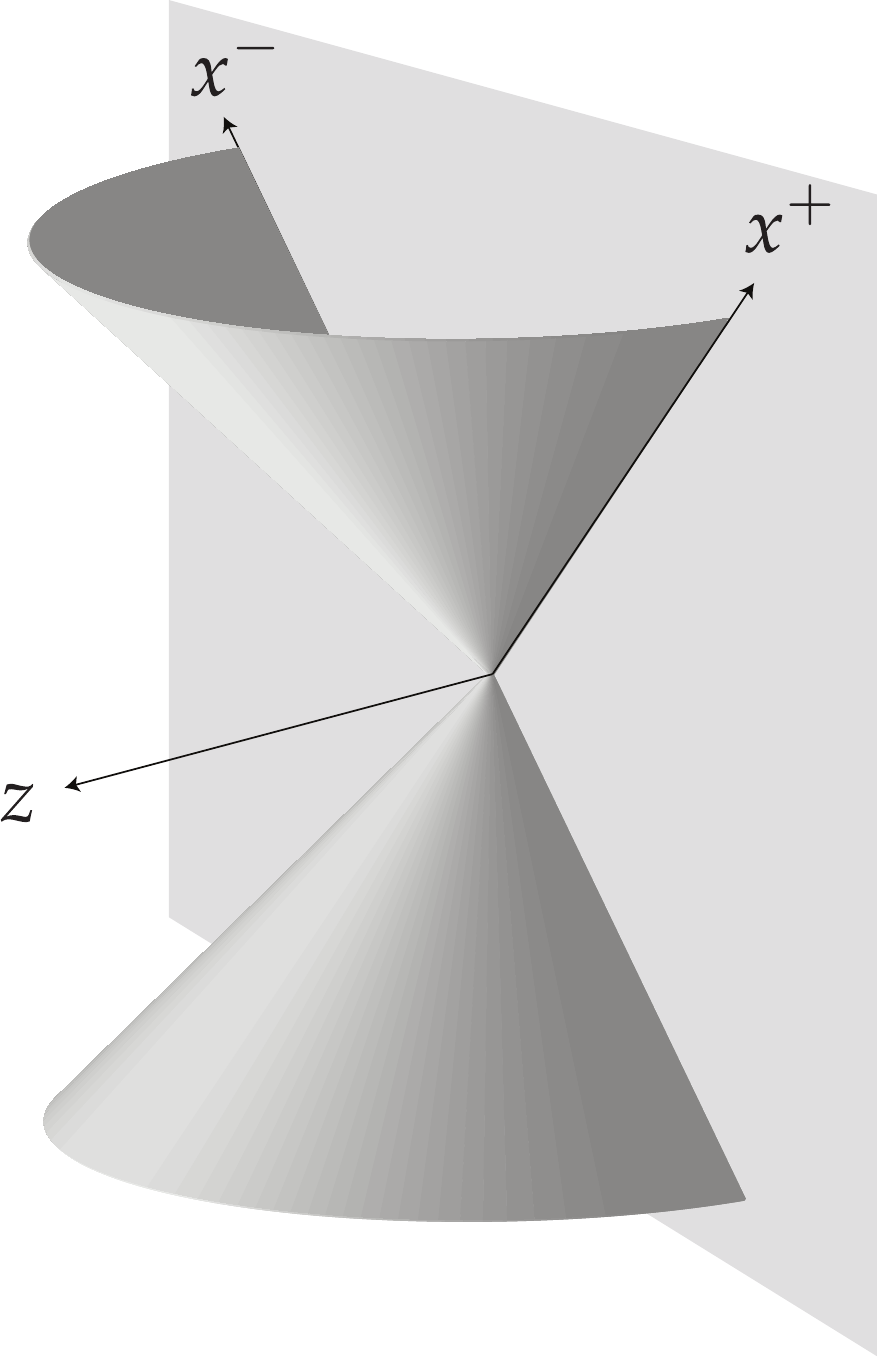}
	\caption{These lightcones becomes the singularity of the BTZ black hole after the quotient.}
	\label{lightlikeCircles}
\end{figure}

This conclusion is certainly subtle and delicate, as illustrated in Fig.\@~\ref{lightlikeCircles}. After quotienting AdS to BTZ, the lightcones in Fig.\@~\ref{lightlikeCircles} become the future and past singularities. So it would appear that the quotient construction of correlators to "see" the scattering inside the horizon will correspond to the analog of Fig.\@~\ref{sharpProbe} in BTZ, a diagram that necessarily traverses the singularity. This raises the question of whether the quotienting procedure outlined above is straightforward and trustworthy. Indeed we claim it is, but to double-check this requires studying the singularity more closely, and Feynman diagrammatics in its vicinity. 

\subsection{Through the Singularity: the "Whisker" Regions} 

Purely at the level of the spacetime geometry (before any dynamics is considered),  
the quotient construction
 gives BTZ  a perfectly smooth passage (with finite curvature) through the 
 "singularity", 
the quotient of the lightcones of Fig.\@~\ref{lightlikeCircles} ($r = 0$ or  $uv =1$ in Schwarzchild and Kruskal coordinates respectively).
However, after the quotient the regions inside these lightcones contain
closed timelike curves, dubbed "whiskers" in \cite{whiskers}. 
 (In the Rindler limit, $r_S  \rightarrow \infty$, these closed curves
become infinitely long and the whiskers revert to just ordinary parts of AdS.)
The smoothness of the quotient geometry is also deceptive, and the singularity well deserves its name once one makes any attempt to physically probe it.  After quotienting the lightcones of Fig.\@~\ref{lightlikeCircles}, they  are comprised of closed lightlike curves
   where 
 even small  (quantum) fluctuations \cite{steif} \cite{ortiz} can backreact  divergently with divergent curvatures, and general considerations imply the breakdown of ordinary (effective) field theory \cite{wald}. See \cite{visser} for a concise review of these general considerations.
 Similar singularities have also been studied in the context of string theory. Attempts to scatter \emph{through} the singularity in string theory failed to obtain well-defined amplitudes (see \cite{berkooz}  for a  concise review and original references). Ref.\@~\cite{eva} found that stringy effects involving the twisted sector smoothed out the large backreactions, but so as to isolate the spacetime regions outside the singularity from the whisker (and other) regions beyond the singularity. In any case, much of the literatures suggests that the whisker regions are both wildly unphysical and inaccessible because of the singularity. This seems at odds with our claim that diagrams ending in the whisker regions are the dual of the non-local CFT correlators described above, and that these capture scattering inside the horizon.
 
  However, we will show that local boundary correlators with some endpoints in the whisker regions are in fact well-defined, and a protected sub-class are dominated within EFT, parametrically insensitive to what happens very close  to the singularity, even when the associated (Witten) diagrams traverse the singularity. This protected sub-class is specified by first noting that the maximal extension of the BTZ black hole spacetime is given by \cite{btz2}
\begin{equation}
	\BTZ = \text{AdS}_\text{global}/\Gamma,
\end{equation}
where $\Gamma$ is a quotient discrete isometry group of AdS. An intermediate extension of the black hole spacetime is then given by replacing $\text{AdS}_\text{global}$ with just the Poincar\'e patch, $\text{AdS}_\text{Poincar\'e}$. This still includes the entire Kruskal extension of BTZ as well as two whisker regions, 
\begin{equation}
\BTZ_\text{Kruskal} \subset \text{AdS}_\text{Poincar\'e}/\Gamma.
\end{equation}
The protected class of boundary correlators  is precisely the set confined to $\text{AdS}_\text{Poincar\'e}/\Gamma$, rather than all of 
$\text{AdS}_\text{global}/\Gamma$. For this reason, we confine ourselves in this paper to $\text{AdS}_\text{Poincar\'e}/\Gamma$, and simply identify it in what follows as the "BTZ spacetime". We will return in  future work to a treatment of the boundary correlators of the maximally extended BTZ spacetime given by $\text{AdS}_\text{global}/\Gamma$ \cite{btzglobal}.   
  
 Technically, in the $\text{AdS}_\text{Poincar\'e}/\Gamma$ realization of BTZ, naive divergences appear when Witten diagram interaction vertices approach the singularity, but are rendered finite by (a) using and tracking the correct "$i \epsilon$" 
 prescription in BTZ propagators, following from AdS propagators by the method of images, and (b) including the whiskers in the integration region for interaction vertices.  Roughly, 
\begin{equation}
\int_{r _1<0}^{r_2>0} dr  \frac{\ln^p r}{r^q}  \rightarrow \int_{r_1 <0}^{r_2>0} dr  \frac{\ln^p(r + i \epsilon)}{(r + i \epsilon)^q}  < \infty, \label{iEpsilonCancellations}
\end{equation}
where $r$ is the Schwarzchild radial coordinate for $r >0$ and a related coordinate inside the whisker region for $r<0$.  Clearly, the finiteness of such expressions as $\epsilon \rightarrow 0$ requires integrating into the whisker region, $r<0$.
   (Similar cancellations were noted in \cite{shenkerkraus}). 
   
   More strongly, we will show that many of the BTZ local boundary correlators are 
    well-approximated by the analogous diagrams in (unquotiented) AdS$_\text{Poincar\'e}$ itself, where the interpretation in terms of scattering behind the (Rindler) horizon is unambiguous. This is  the basis of our claim that we have found a class of correlators sensitive to scattering behind the BTZ horizon.

% Before quotienting, correlators ending in the Poincar\'e patch of AdS$_\text{Poincar\'e}$ can be computed using two prescriptions: one can use the AdS$_\text{Poincar\'e}$ 
 % $i \epsilon$ prescription and integrate interaction vertices just within AdS$_\text{Poincar\'e}$, or one can think of AdS$_\text{Poincar\'e}$ as sitting inside    AdS$_\text{global}$ and use the   AdS$_\text{global}$  $i \epsilon$  prescription and integrate interactions vertices over all of 
 % AdS$_\text{global}$. These two prescriptions yield the same AdS$_\text{Poincar\'e}$ correlators. However,  after quotienting to BTZ the two prescriptions no longer agree because the limit of large image number does not commute with the limit $\epsilon \rightarrow 0$. 
  % Our work shows that it is the former prescription that is well-defined, while the latter misbehaves.    
  % And yet it is the (equivalent of the) latter prescription that appears in the literature. For this reason and for its technical simplicity and closeness to our Rindler analysis, we restrict ourselves to realizing BTZ as a quotient of AdS$_\text{Poincar\'e}$   \cite{carlip} \cite{juanandy} \cite{danielsson}. 
%This is sufficient to find whisker boundary correlators 
%sensitive to scattering inside the horizon (but outside the 
%singularity).

\subsection{Space $\leftrightarrow$ Time inside the Horizon}

Despite these good features,  correlators in regions with timelike closed curves seem at odds with a physical interpretation and 
connection to the standard thermofield CFT dual. Relatedly, it is puzzling \emph{why} we are lucky enough that the associated Witten diagrams should be insensitive to what is happening close to the singularity.
We show that these correlators can be put into a more canonical form by performing a
 well-defined "space $\leftrightarrow$ time" transformation which takes local operators inside the horizon into non-local operators outside the horizon (and thereby make them accessible to external observers).  This transformation is particularly plausible in the dual 
 $1+1$ CFT where the causal (lightcone) structure is symmetric between space and time, and indeed we show that the transformation can be viewed as a kind of "improper" conformal transformation. It is this transformation that ultimately leads to the non-local operators arising from local ones, seen in \eqref{rindlerNonlocal}. Such a symmetry seems much less manifest from the AdS perspective where there is no such isometry, but we prove that it indeed exists as an unexpected symmetry of boundary correlators, by a careful Witten-diagrammatic analysis. 
 
 In more detail, the transformation is also accompanied by complex phases that are necessary for ensuring relativistic causality constraints in correlators, naively threatened because "spacelike $\leftrightarrow$ timelike".)
We thereby interpret our results as having found (i) non-local CFT operators that simply describe scattering inside the BTZ horizon (but outside the singularity), and (ii) an auxiliary but bizarre spacetime extension of the BTZ black hole, "whiskers", in which these non-local CFT operators are rendered  as \emph{local} operators, and in which some of their properties become more transparent. Whether or not one thereby considers the whiskers to be  "physical" regions is left to the reader. 

\subsection{Comparing Whiskers and Euclidean space as auxiliary spacetimes}

The notion of an auxiliary spacetime grafted onto the physical spacetime, where one uses path integrals and operators in the former to implant certain types of wavefunctionals in the latter, is already familiar when the auxiliary spacetime is Euclidean. For example, 
such constructions are used to create the Hartle-Hawking wavefunctional \cite{hh} or its perturbations in the physical spacetime, and can have a non-perturbative CFT dual \cite{juanbh}.   Indeed, they too can be used to create quite general bulk states in the interior of BTZ,  \emph{in principle} including the kind of  "out states" for scattering that we seek. However, the simple Euclidean constructions yield physical states at the point of time symmetry, $u+v=0$ (or $\tau = 0$). We would need to evolve these states to late time and take superpositions in order to find "out states" that consist of several approximately free bulk particles. The problem then is that identifying such superpositions is equivalent to solving the scattering dynamics itself!  By contrast, the virtue of our Lorentzian auxiliary spacetime "whisker" is that it allows us to create simple out states with simply defined operators. In this way, we can pose explicit (non-local) CFT correlators which capture the fate of scattering inside the horizon. A well-programmed "CFT computer" would then output the answers to such questions without first requiring equally difficult computations as input. 

\subsection{Whisker correlators as generalizing "in-in" correlators}

It is not simply fortuitous that Witten diagrams are insensitive to the singularity, even with
 some endpoints on the boundary of the whisker regions. Rather, we will show that the approach to the singularity in the bulk EFT is given by  
\begin{equation}
	... U^{\dagger} e^{- \frac{\pi}{2} (H_{\tau} - P_{\sigma})} U... ~,
\end{equation}
where $U$ is a time evolution approaching the (future, say) singularity, and $H_{\tau}, P_{\sigma}$ are the isometry generators corresponding to  $\tau$ and $\sigma$ translations in Schwarzschild coordinates. (Of course, $\tau$ represents a spacelike direction near the singularity, and therefore $H_{\tau}$ is really a "momentum" here, despite the notation.) 
The $U^{\dagger}$ factor arises from the whisker region. The exponential weight is a  non-trivial consequence of 
our "space $\leftrightarrow$ time" transformation, where the timelike circles become standard spacelike circles. 
One can think of the whisker-related factor, $... U^{\dagger} e^{- \frac{\pi}{2} (H_{\tau} - P_{\sigma})}$, as setting up a useful "out" state inside the horizon of the physical region. 

If there are no sources (endpoints of correlators) in the vicinity of the singularity,  the time evolution $U$ commutes with the isometry generators, $H_{\tau}, P_{\sigma}$ and hence cancels against $U^{\dagger}$. This cancellation, which also can be seen non-perturbatively in the CFT description, is the deep reason behind the insensitivity of boundary correlators to the details of UV physics. It matches the cancellations in Witten diagrams (before massaging by space $\leftrightarrow$ time) in the manner of \eqref{iEpsilonCancellations}. Such $U^{\dagger} U$ cancellation in the far future is reminiscent of what happens for correlators in the "in-in" formalism \cite{schwinger1, schwinger2} (see \cite{weinberg} for a modern discussion and review).  Indeed, we will show using the space $\leftrightarrow$ time transformation that local boundary correlators traversing the singularity are equivalent to a generalization of  in-in correlators involving non-local operators, where all time evolution takes place after the past singularity and before the future singularity.

\subsection{Studying the Singularity}

Our ability to discover and check our proposal for describing scattering inside the BTZ horizon rests on the existence of the protected set of 
local boundary correlators, which we can prove in a simple way are insensitive to the singularity. However, the ultimate goal is not to merely describe scattering inside the horizon far from the singularity, since such scattering is approximately the same as scattering in a static spacetime. This regime is only useful to vet our proposal, precisely because we know the answers already, dominated by EFT. Rather the goal is to use our non-perturbative CFT proposal to describe scattering close to the singularity where cosmological blueshifts take us out of the EFT domain, and where even perturbative string theory may miss important features.  This interesting kind of sensitivity to the singularity is not outright absent from the protected set of correlators, but it is suppressed by $\sim 1/$blueshift. However, one can study processes with kinematics chosen such that they would not proceed but for such cosmological blueshifts (that is, they would not proceed for $r_S = \infty$), in which case the leading effects are sensitive to the singularity. 

Furthermore, more general (gauge-fixed EFT) bulk correlators are order one sensitive to the singularity and UV physics, but not mathematically divergent. The same is also true for local boundary correlators in the more extended $\text{AdS}_\text{global}/\Gamma$ realization of BTZ, as we will discuss in \cite{btzglobal}.

%suggests that the BTZ case offers
 %the best prospects for an explicit formulation of the interior of the horizon. For example, the Virasoro algebra of $1+1$ CFT is realized as the asymptotic symmetry structure of  AdS$_3$ \cite{brown} and acts on states inside it, such as BTZ.

\subsection{Relation to the literature}
  
Several earlier attacks have been made on more explicitly extending holography into the black hole interior, some specific to BTZ, while others apply also to higher-dimensional eternal black holes. The most direct approach has been to study the thermofield CFT formulation carefully, and to identify those subtle, non-local features that might encode key aspects of the black hole interior  \cite{vijay2}  \cite{marolf} \cite{juanbh} \cite{recentjuan} (see \cite{shenker} for higher-dimensional discussion). 
Our work is certainly in the same spirit, but we claim our non-local CFT operators more sharply and more knowably probe the interior. Another general direction is to try and construct the CFT dual of interior field operators  \cite{kabatbh} \cite{transhorizon} \cite{raju}, in part by using the gravitational EFT equations of motion to evolve exterior field operators in "infaller" time into the interior. This is necessarily restricted to situations in which the bulk metric fluctuates modestly, whereas we propose a non-perturbative formulation.
Yet another general approach is to try to enter the horizon by a variety of analytic continuations of external (Lorentzian or Euclidean) correlators \cite{vakkurikraus} \cite{shenkerkraus} \cite{jared} \cite{bala} \cite{liu}. Our work  has this aspect to it, but it is governed and understood from a physical perspective in which analytic continuation merely provides an efficient means of calculation, rather than a first principle. 
The symmetry-quotient structure of BTZ has led to attempts to construct a "symmetry-quotient" form of a dual CFT \cite{horowitzmarolf}. 
Another BTZ-specific approach is to
take advantage of being able to follow the BTZ geometry beyond the "singularity", where further AdS-like boundary regions exist. One then tries to make sense of CFT on the various boundary regions and how they connect together \cite{martinec} \cite{vakkurikraus}.  
Our work furthers these directions, of making sense of the quotient structure from the CFT perspective, and using it to show how different boundary regions are entangled.  A number of variants of BTZ have also been constructed and studied \cite{sheikh1,sheikh2}.

\subsection{Organization of Paper}

We  start from the symmetry quotient construction of BTZ from AdS$_\text{Poincar\'e}$, and try to make sense of the idea of a "quotient CFT" dual.
In Section~\ref{S:Poincar\'eQuotient}, we review the quotient construction of BTZ geometry from AdS$_\text{Poincar\'e}$ and how this extends the spacetime smoothly past the  singularity, although 
gravitational EFT diagrams ending at the singularity do diverge. 
In Section~\ref{S:BTZboundary}, we identify the boundary regions of the BTZ spacetime, outside the horizon and inside the whiskers.  We point out the central challenges for formulating a dual CFT on the boundary of BTZ, related to the presence of
lightlike and timelike closed curves. 
 In Section~\ref{S:singularity} we explore the BTZ singularity with the simplest examples, before beginning a more general attempt to formulate a CFT dual.  The relevant BTZ correlators, with end points inside and outside the singularity and horizon, are obtained by the method of images applied to AdS$_\text{Poincar\'e}$.
We illustrate how naive divergences encountered as interaction vertices approach the singularity 
in fact cancel to give mathematically well-defined correlators.  In Section~\ref{S:spaceToTime}, in order to massage the CFT on the  BTZ boundary into a non-perturbatively well-defined form, we introduce the transformation switching time and space inside the horizon,  arriving in \eqref{traceFormulaDyson} at our central result, a generalization of the thermofield CFT formulation allowing probes of physics inside the horizon.
 Eq. \eqref{traceFormulaDyson} is manifestly well-defined and manifestly respects the symmetry construction of BTZ.   In Section~\ref{S:thermofieldCFT}, we recast \eqref{traceFormulaDyson} in canonical thermofield form, resulting in \eqref{thermofieldCFT}, with probes inside the horizon appearing as non-local probes of the thermofield-entangled CFTs.  
 Many of our manipulations in sections~\ref{S:spaceToTime} and~\ref{S:thermofieldCFT} 
   are formally based on the CFT path integral. But for concrete confirmation we must turn to the dual AdS diagrammatics.  
 
   In Section~\ref{S:rindlerAdSCFT} we study the Rindler AdS/CFT correspondence ($r_S = \infty$) in detail, and prove the above results in this limit in bulk EFT,  allowing us to probe inside the Rindler horizon by studying specific non-local correlators \emph{outside} the horizon. We  check that our proposal reproduces the AdS$_\text{Poincar\'e}$ correlators everywhere.  In Section~\ref{S:finiteLambda}, we finally check that Eq.  \eqref{traceFormulaDyson} does indeed act as the dual of BTZ by showing that it gives the associated local boundary correlators, including the whisker regions, and that these correlators are finite and dominated by EFT (despite traversing the singularity). This
   follows from the analogous Rindler proof in Section~\ref{S:rindlerAdSCFT} by applying the method of images in EFT. We explain how these local boundary correlators are generally insensitive to the breakdown of EFT near the singularity, allowing us to use EFT to check our CFT 
   proposal is sharply sensitive to scattering inside the horizon just as in the Rindler ($r_S = \infty$) limit. 
  In Section~\ref{S:sensitivityToSingularity},  we demonstrate that bulk correlators are sensitive to the singularity and UV physics there, although still mathematically finite. 
  We also show how to design special boundary correlators  where
  the near-singularity UV physics dominates, so that our CFT proposal is needed to describe them.
      In Section~\ref{S:conclusion}, we comment on our derivations and some aspects of the physical picture that emerges from our work, and  outline future directions.

%%%%%%%%%%%%%%%%%%%%%%%%%%%%%%%%%%%%%%%%%%%%%%%%%%%%%%%%%%%%%%%%%%%%%%%%%%%%%%%%%%%%%%%%%%%%
%%%%%%%%%%%%%%%%%%%%%%%%%%%%%%%%%%%%%%%%%%%%%%%%%%%%%%%%%%%%%%%%%%%%%%%%%%%%%%%%%%%%%%%%%%%%
\section{BTZ as Quotient of AdS$_\text{Poincar\'e}$} \label{S:Poincar\'eQuotient}

In higher-dimensional black holes, the Kruskal extension into the interior ends at a curvature singularity. 
In the BTZ case however,  $uv= 1$ in \eqref{kruskalMetric} does not represent a true curvature singularity and the geometry can be smoothly extended beyond it. Such an extension is most simply given by  a quotient of the Poincar\'e patch of AdS (AdS$_\text{Poincar\'e}$) \cite{carlip} \cite{juanandy}
\cite{danielsson},
\begin{align}
	ds^2 = \frac{dx^+ dx^- - dz^2}{z^2} \qquad (z>0),
\end{align}
where $x^\pm \equiv t \pm x$
 and we identify points related by the discrete rescaling 
\begin{equation}
 	(x^\pm, z) \equiv (e^{r_S} x^\pm, e^{r_S} z). \label{quotient}
\end{equation} 
As straightforwardly checked, the Poincar\'e coordinates are related to the Kruskal coordinates by
\begin{align}
	x^+ = \frac{2e^\sigma v}{1- uv} && x^- = \frac{2e^\sigma u}{1-uv} && z= \frac{1+uv}{1-uv} e^\sigma,
\end{align}
and to the Schwarzschild coordinates by 
\gatherStart
	\begin{aligned}
		x^\pm &=
		\begin{cases}
			\pm\sqrt{1-\frac{1}{r^2} } \, e^{\pm \sigma^\pm},	 &\text{if $r > 1$;} \\
			\phantom{\pm} \sqrt{\frac{1}{r^2} - 1 } \, e^{\pm \sigma^\pm},		&\text{if $r < 1$}
		\end{cases}
		&&
		z = \frac{e^\sigma}{r} \label{schwarzschildCoordinates}
	\end{aligned} \\
	\sigma^\pm \equiv \tau \pm \sigma.
\gatherEnd

The horizon, singularity and asymptotic AdS boundaries now reside at:
\alignStart
	\text{boundary:} \quad & z =0 \\
	\text{horizon:} \quad &x^\pm = 0  \\
	\text{singularity:} \quad &  z^2 - x^+ x^- = 0.
\alignEnd
The true nature of the apparent black hole singularity becomes clearer. While the BTZ black hole spacetime has locally AdS geometry and finite curvature everywhere, the singularity surface consists of closed lightlike curves, given by $x = t \cos \gamma$, $z = t \sin \gamma$, parametrized by $\gamma$.
The region inside this surface consists of closed timelike curves,  which we will call the "whisker" region similarly to \cite{whiskers}.

The presence of such curves does not in itself constitute a geometric singularity\footnote{In the BTZ realization as a quotient of AdS$_\text{global}$ the singularity also includes a breakdown of the spacetime manifold (Hausdorff) structure itself. But the points at which this further complication takes place are pushed off to infinity in our Poincar\'e patch realization of BTZ. This breakdown is relevant to some of the studies in the literature but not to the correlators discussed in this paper. We will more thoroughly clarify this point in \cite{btzglobal}.}, but it does
pose a conceptual challenge for physical interpretation, and on general grounds implies the breakdown of quantum (effective) field theory in the vicinity of the closed lightlike curves \cite{wald}.
See \cite{visser} for a concise review, and  \cite{steif} \cite{ortiz} for computations of stress-tensor divergences at the BTZ singularity.
Similar features have also been studied in string theory (as reviewed in \cite{berkooz}.)
We illustrate the basic problem by looking at an EFT  amplitude for a scalar field in the BTZ background. Following \cite{shenkerkraus} we focus on the scalar propagator from a point on an AdS boundary, $x^\pm$,  external to the black hole to a point inside the horizon and near the "singularity", $(y_{\pm}, z)$. 
Because BTZ is a quotient of AdS$_\text{Poincar\'e}$, we can easily work out this propagator by the method of images applied to the boundary-bulk propagator of AdS$_\text{Poincar\'e}$ \cite{vakkuri} \cite{shenkerkraus}:
\begin{align}
	K_\text{BTZ}(x^\pm, y^\pm, z) &= \sum_{n \in \mathbb Z} \frac{(e^{nr_S} z)^\Delta}{ \left[ e^{2nr_S} z^2 - (x^+ - e^{nr_S} y^+)(x^- - e^{nr_S} y^-)  \right]^\Delta}
	\label{kWittenImaged} 
\end{align}
where $m^2 = \Delta(\Delta -2)$ and we have summed over images of the bulk point. 
Generically, the image sum clearly converges, the summand behaving asymptotically as $\sim e^{-|n| r_S \Delta}$. The exception is the singular surface $z^2 - y^+ y^- = 0$, where the summand becomes $n$-independent for large $n$, and the series diverges. (We omit a discussion of the $i \epsilon$ prescription in the propagator as it does not avoid the divergence as $\epsilon \rightarrow 0$, although it will play an important role later in the paper.)
 This feature is general for correlators with some points ending on the surface 
$z^2 - y^+ y^- = 0$, the perturbative incarnation of divergent backreactions that justifies this surface being called the "singularity".
 A subtler question is whether one can propagate or scatter \emph{through} the singularity within gravitational EFT. If so, one can just avoid probes (correlator endpoints) very near the singularity and trust EFT calculations elsewhere. This question is particularly relevant for correlators ending on the full BTZ boundary (including inside the whiskers) since these would define local operator correlators of a possible CFT dual to BTZ. Before tackling this question, we first study the BTZ boundary itself.

%As shown in \cite{Kraus:2002iv}, this results in the breakdown of the derivative expansion of non-renormalizable EFT as interaction points approach the singularity and the incoming propagators diverge (Fig.\@~\ref{coneOfSilence}). Therefore, the surface 
%$x^+ x^- + z^2 = 0$ \emph{is} a serious singularity, not in terms of classical geometry but in terms of (classical or quantum) gravitational EFT. 
%One of the goals of a holographic dual to BTZ quantum gravity is to provide us with a resolution and deeper understanding of this singularity.

%%%%%%%%%%%%%%%%%%%%%%%%%%%%%%%%%%%%%%%%%%%%%%%%%%%%%%%%%%%%%%%%%%%%%%%%%%%%%%%%%%%%%%%%%%%%
%%%%%%%%%%%%%%%%%%%%%%%%%%%%%%%%%%%%%%%%%%%%%%%%%%%%%%%%%%%%%%%%%%%%%%%%%%%%%%%%%%%%%%%%%%%%
\section{The Extended BTZ Boundary and Challenges for the CFT Dual} \label{S:BTZboundary}

Since BTZ is at least locally AdS-like, it seems very natural to guess that BTZ quantum gravity  has a holographic dual given by a $1+1$ dimensional CFT "living" on the BTZ boundary. Within the Kruskal extension, this boundary, $uv =-1$, consists of the two disjoint solutions with $u>0$, $v<0$ or $u<0$, $v>0$, corresponding to the two asymptotically AdS$_\text{global}$ regions outside the horizon, as in higher-dimensional AdS black holes. This is then consistent with the now-standard thermofield picture of two CFTs living on two copies of the boundary of AdS$_\text{global}$, namely two separate CFTs each living on a spatial circle $\times$ infinite time, but in an entangled state. 
However, this Kruskal boundary corresponds in our AdS$_\text{Poincar\'e}$ coordinates to $z = 0$, $x^+ x^- < 0$, whereas  in the AdS$_\text{Poincar\'e}$ realization the boundary is straightforwardly all of $z=0$. The regions $z=0$, $x^+ x^- > 0$ are missed in the Kruskal extension because they lie inside the singularity, while the Kruskal extension stops there.  The question then arises whether these inside-singularity boundary regions play an important role in the CFT dual of BTZ (the view taken in \cite{martinec} \cite{vakkurikraus}), even for "projecting" the part of BTZ outside the singularity but inside the horizon. We will show  that there are in fact two equivalent formulations of the CFT dual of BTZ, one in which the entire boundary region is needed for the CFT, and a second one in terms of two entangled CFTs on just the disjoint boundary regions outside the horizon. 

\subsection{BTZ Boundary as Disconnected Cylinders} \label{disconnectedBoundary}

We begin by identifying the full boundary region of BTZ, $\d$BTZ, within the AdS$_\text{Poincar\'e}$ realization,   regardless of where this takes us with respect to the singularity. Even the simple identification of $\d$BTZ as $z=0$ is subtle because of the quotienting. Naively this would yield $1+1$ Minkowski spacetime with the identification $x^\pm \equiv e^{r_S} x^\pm$. Such an identification does not make straightforward sense because rescaling is \emph{not} an isometry of Minkowski space.  The subtlety is that the boundary geometry is only determined from the bulk geometry up to a Weyl transformation \cite{wittencft}, which conformally invariant dynamics cannot distinguish. Therefore more precisely,
\begin{align}
	ds^2_{\d\text{BTZ}} &= f(x^\pm)dx^+ dx^-,
\end{align}
where $f$ is a Weyl transform of $1+1$ Minkowski spacetime, with the 
identifications $x^\pm \equiv e^{r_S} x^\pm$ and hence $f$-periodicity $f(x^\pm) = e^{2 r_S} f(e^{r_S} x^\pm)$. 

Two choices of $f$ will prove insightful. The first is 
\begin{align}
	f = \frac{1}{|x^+ x^-|}. \label{weylHorns}
\end{align}
It is useful to break up the $1+1$ Minkowski plane into the four regions, 
\alignStart
	\text{Right ($R$)} \quad&x^+ > 0, x^- < 0 \\
	\text{Future ($F$)}\quad &x^\pm > 0 \\
	\text{Left ($L$)}\quad&x^+ < 0, x^- > 0 \\
	\text{Past ($P$)}\quad &x^\pm < 0. 
\alignEnd

We will refer to $R, L$ as Rindler wedges and $F,P$ as Milne wedges. We can adapt Rindler-like coordinates for each wedge,
\begin{align}
	x^\pm &=
	\begin{cases}
		\pm e^{\pm \sigma^\pm},	&x \in R \\
		\phantom{\pm} e^{\pm \sigma^\pm},	&x \in F \\
		\mp e^{\pm \sigma^\pm},		&x \in L \\
		- e^{\pm \sigma^\pm},	&x \in P,
	\end{cases}
	&& \sigma^\pm \equiv \tau \pm \sigma
\end{align}
so that the quotienting and Weyl transformation take the simple forms
\gatherStart
	\sigma \equiv \sigma + 2\pi r_S \\
	f = e^{-2\sigma}. 
\gatherEnd
We can simply restrict $\sigma$ to the fundamental region $-\pi r_S \leq \sigma \leq \pi r_S$. 
We see that $\d$BTZ is then given by \emph{four} disjoint spacetime cylinders,
\begin{align}
	ds^2_{\d\text{BTZ}} &=
	\begin{cases}
		+d\sigma^+ d\sigma^-,	&\text{R, L} \quad \text{spacelike circle} \times \text{infinite time} \\
		-d\sigma^+d\sigma^-,	&\text{F, P} \quad \text{infinite space} \times \text{timelike circle!}
	\end{cases} 
\end{align}

\begin{figure}
        \centering
        \begin{subfigure}[b]{0.33\textwidth}
                \centering
                \includegraphics[width=\textwidth]{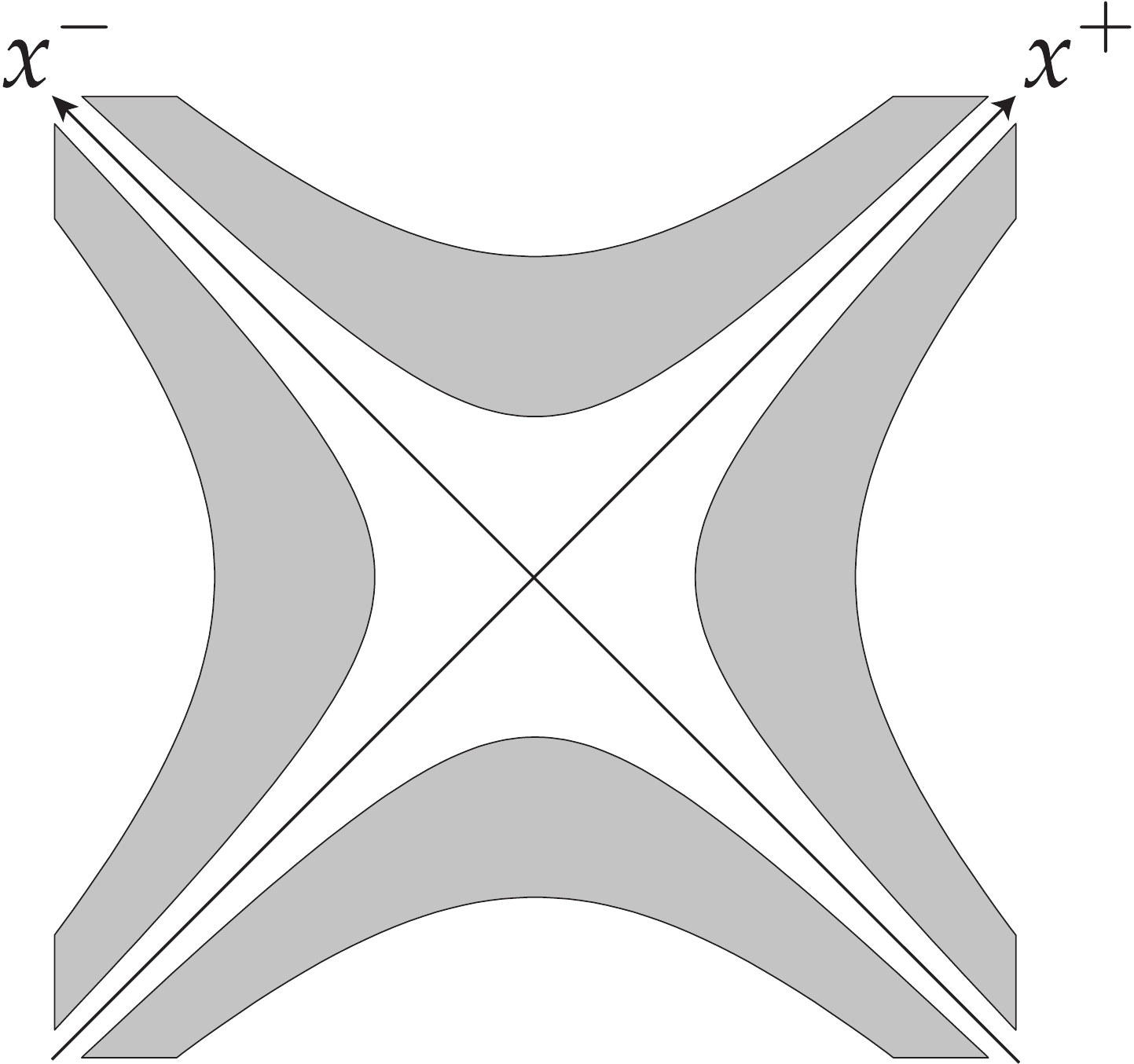}
                \caption{The fundamental region in $\sigma$ mapped to the Minkowski plane by the Weyl transformation \eqref{weylHorns}}
                \label{horns}
        \end{subfigure}
        \qquad
        \begin{subfigure}[b]{0.33\textwidth}
                \centering
                \includegraphics[width=\textwidth]{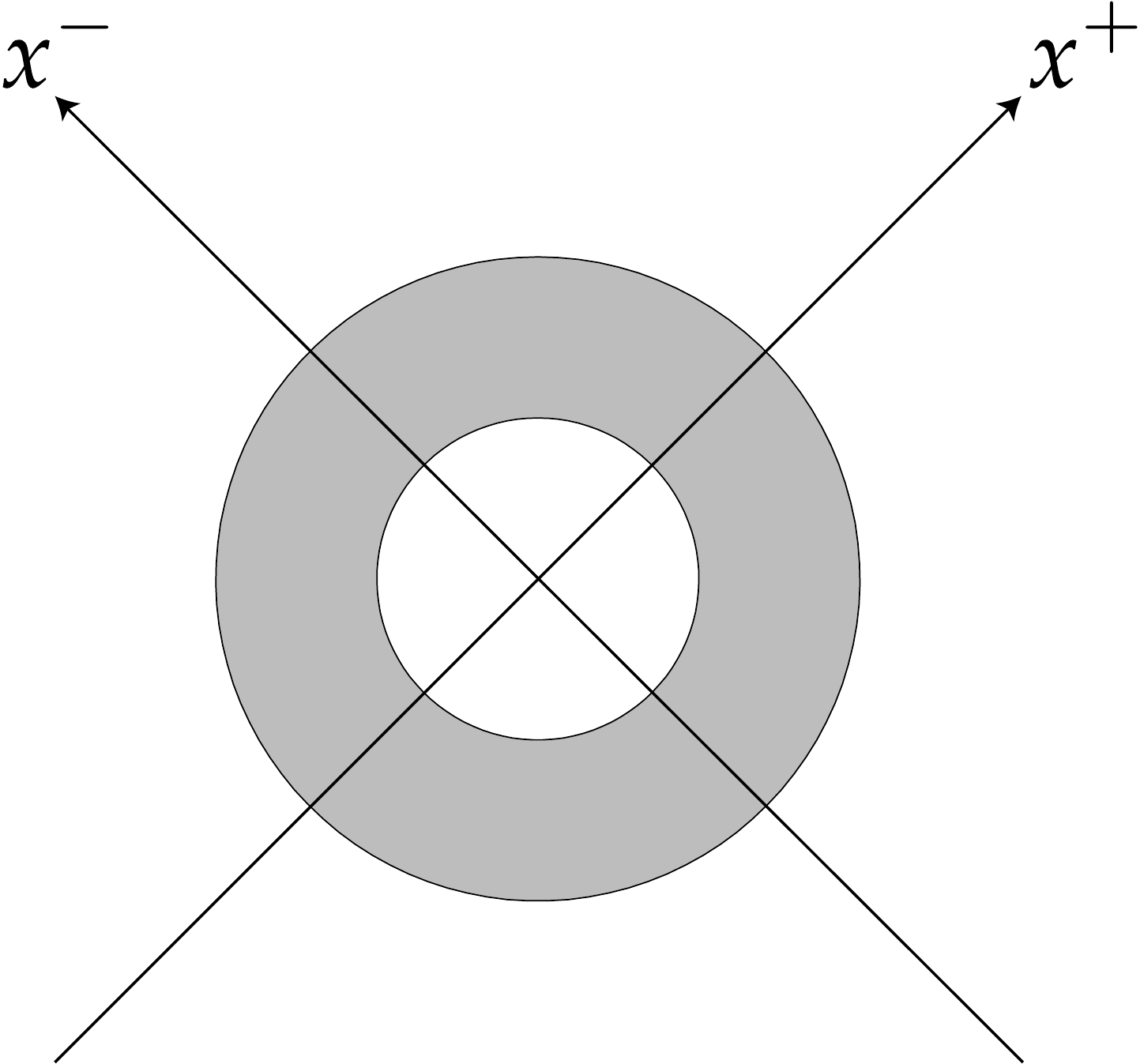}
                \caption{The fundamental region in $\alpha$ mapped to the Minkowski plane by the Weyl transformation \eqref{weylTorus}}
                \label{torus}
        \end{subfigure}
        \caption{Two different choices of fundamental region for the BTZ boundary}\label{fundamentalRegions}
\end{figure}
The cylinders in the Rindler wedges are just the boundaries of the AdS$_\text{global}$ asymptotics outside the horizon described above. But the cylinders in the Milne wedges are the boundary region inside the singularity.\footnote{These four disjoint boundary components are the AdS$_\text{Poincar\'e}$ subset of the larger set of boundary components arising in the further extension of BTZ as a quotient of AdS$_\text{global}$.}	  The Weyl transform maps the cylinders to the four shaded regions of $1+1$ Minkowski spacetime in Fig.\@~\ref{horns}. In this way we can think of the shaded region as a fundamental region for the quotienting procedure on the CFT-side. 
	
We see that the four cylinders present two challenges for hosting a dual CFT. The first is that they remain disjoint and therefore we need some sort of generalization of the thermofield entanglement of two CFTs with which to connect them. For related early work in this direction, see \cite{vakkurikraus}.  The second issue is that the Milne wedge cylinders have circular time.

\subsection{Connected View of $\partial$BTZ}

To guess how to move forward we use a different choice of Weyl transformation, which gives us a different view of $\d$BTZ (the CFT being insensitive to such choices),	
\begin{align}
	f = \frac{1}{(x^+)^2 + (x^-)^2}. \label{weylTorus}
\end{align}
Using "polar" coordinates on the Minkowski plane, 	
\begin{align}
	t = e^\alpha \sin \theta && x = e^\alpha \cos \theta, \label{polarMinkowski}
\end{align}
with the usual identification $\theta \equiv \theta + 2\pi$ and the BTZ quotient identification $\alpha \equiv \alpha + 2\pi r_S$, we find (Fig.\@~\ref{torus}):
\begin{align}
	ds^2_{\d\text{BTZ}} &= \cos 2\theta (d\theta^2 - d\alpha^2) + 2 \sin 2 \theta \, d\theta d \alpha \label{torusMetric} \\
	&= \text{Lorentzian Torus!} \notag
\end{align}

\begin{figure}
	\centering
	\includegraphics[width=0.33\textwidth]{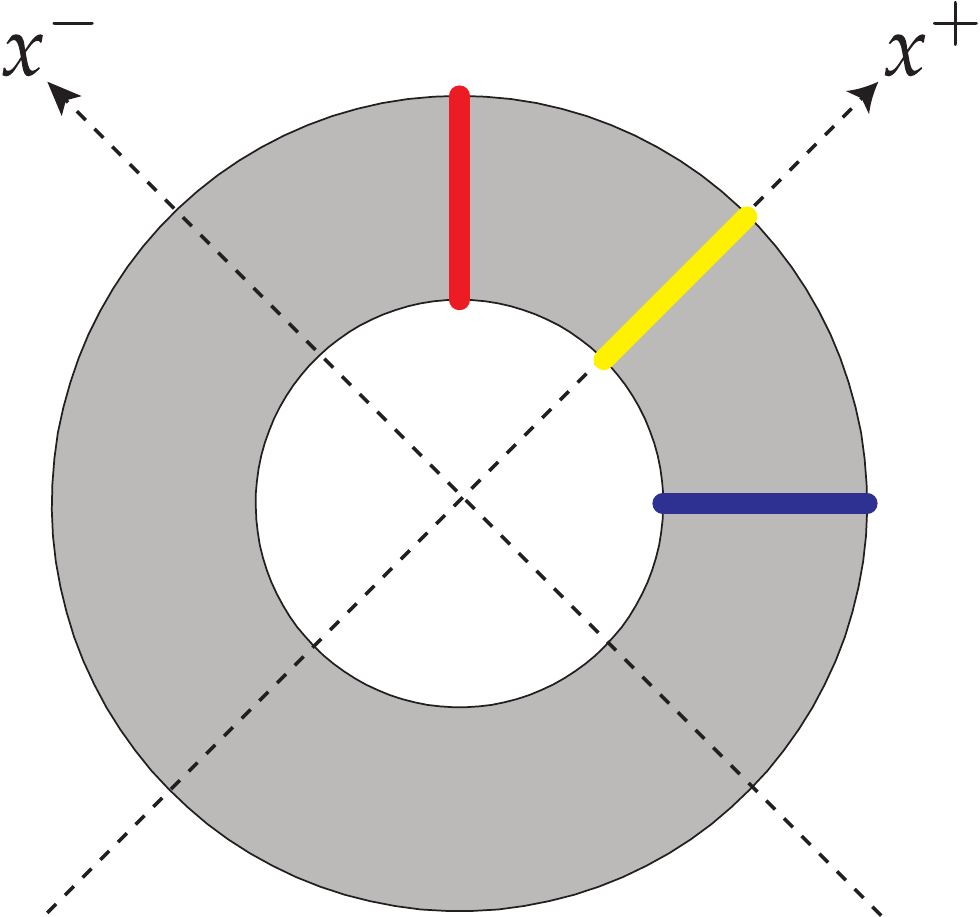}
	\caption{The Lorentzian torus contains closed timelike (red), lightlike (yellow), and spacelike (blue) curves.  The inner and outer edges of the annulus are identified after Weyl transformation \eqref{weylTorus} to make the Lorentzian torus of \eqref{torusMetric}.}
	\label{torusCircles}
\end{figure}

This geometry \cite{misner} is smooth and \emph{connected}, but still contains alarming features (Fig.\@~\ref{torusCircles}). There are still timelike circles in the Milne wedges 
$  \tfrac{\pi}{4} < \theta < \tfrac{3\pi}{4},  \tfrac{-\pi}{4} > \theta > \tfrac{-3\pi}{4}$, oriented in the $\alpha$-direction. (There are still only spacelike closed curves in the Rindler wedges.) The "joints", $\theta = \frac{-3\pi}{4}, \frac{-\pi}{4}, \frac{\pi}{4}, \frac{3\pi}{4}$, while smoothly connecting the geometries of the different wedges, are themselves light-like circles in $\alpha$. So it does not appear that a CFT on this boundary region will allows us to evade the difficult problem of doing field theory at lightlike circles presented by the BTZ singularity \cite{wald} \cite{visser}. 
(For discussion of a very similar $1+1$ context see \cite{misnerpathology}).)
If we try to excise these lightlike circles we are left with the disjointedness of the boundary and the CFT living on it. 

~

Before making any interpretation, we will first try to simply define local operator correlators of the CFT on the full $\partial BTZ$, using the method of images applied to AdS$_\text{Poincar\'e}$ boundary correlators. However, we must check that these are even mathematically well-defined in the face of the BTZ singularity. Therefore we 
 first study simple examples and then general features of how the singularity enters into correlators. We then show that 
 boundary correlators on all of $\partial BTZ$ are mathematically well-defined, although the  singularity does represent a breakdown of gravitational EFT.

%%%%%%%%%%%%%%%%%%%%%%%%%%%%%%%%%%%%%%%%%%%%%%%%%%%%%%%%%%%%%%%%%%%%%%%%%%%%%%%%%%%%%%%%%%%%
%%%%%%%%%%%%%%%%%%%%%%%%%%%%%%%%%%%%%%%%%%%%%%%%%%%%%%%%%%%%%%%%%%%%%%%%%%%%%%%%%%%%%%%%%%%%
\section{Boundary Correlators and the Singularity} \label{S:singularity}

In this section we study the simplest examples which illustrate the implications of the singularity for defining correlators within gravitational EFT, and for identifying them with equivalent CFT correlators. For this purpose we will not need to study these BTZ correlators in a UV-complete framework such as string theory, although we assume such a framework exists. We will compute these BTZ correlators using the method of images.  It is convenient to define
\begin{equation}
	\lambda \equiv e^{r_S},
\end{equation}
so that the quotient identification \eqref{quotient} can be written
\begin{equation}
 	(x^\pm, z) \equiv (\lambda x^\pm, \lambda z).
\end{equation} 
\subsection{Approaching Singularity from Outside}

We start by noting the full $i \epsilon$ structure of the bulk-boundary propagator of AdS$_\text{Poincar\'e}$, 
which is important for what follows here:
\begin{equation}
K_\text{AdS} = \left[ \frac{z}{z^2 - (x^+ - y^+)(x^- - y^-) + i \epsilon(x^0 - y^0)^2} \right]^\Delta.
\end{equation}
This structure for Lorentzian $K$ is most easily derived from the well-known Euclidean $K$ \cite{wittencft} by analytic continuation in time.
The boundary point, $x^\pm, z' = 0$ can be in any of the boundary regions, $L, R, F, P$. The analogous BTZ propagator is given by summing over images of the bulk point, $e^{nr_S} y^{\pm}, e^{nr_S} z$, as in \eqref{kWittenImaged}. 
It is important that the $i \epsilon$ is also thereby imaged. To study the near singularity region it is useful to follow \cite{shenkerkraus} and switch to AdS Schwarzschild coordinates, where the bulk point is at $\sigma, \tau, r$, and the boundary point is given by $\sigma', \tau', r' = \infty$. We can zoom in on the region where the bulk point approaches the singularity, $r \rightarrow 0$, and the image sum divergence for positive large $n$ dominates: 
\alignStart
	K_\text{BTZ} &\underset{r \to0, }{\sim} \sum_\text{$n>0$ large} \left( \frac{1}{e^{\sigma'^+} e^{-\tau} - e^{-\sigma'^-} e^\tau + e^\sigma r \lambda^n + i \epsilon \lambda^n}  \right)^\Delta \\
	&\sim \left( \frac{1}{e^{\sigma'^+} e^{-\tau} - e^{-\sigma'^-} e^\tau } \right)^\Delta \ln\left( e^\sigma r + i \epsilon \right). \label{kWittenDivergence}
\alignEnd
The approximation in the first line is to drop terms even more subdominant for large $n>0$. In the second line we noted that the $\lambda^n$-dependent terms are subdominant for small $r$ for the 
first $\sim \ln(\lambda/(e^{\sigma} r))$ terms in $n>0$,  with the sum rapidly converging for larger $n$. Therefore, the sum is given by the $n$-independent constant multiplied by $\sim \ln (e^{\sigma} r)$, for small $r$. Crucially, the $i \epsilon$ appears inside the logarithm by the first line's analyticity
 in $e^{\sigma} r + i \epsilon$. 
 
 \subsection{Flawed attempt to Scatter through Singularity}
 
 Let us now explore the possibility that a dual CFT resides on $\partial$BTZ, as identified in the previous section, by trying to construct the leading in $1/N_\text{CFT}$ planar contribution to a 3-point local operator correlator in terms of a tree level BTZ diagram: 
 \begin{equation}
	\vev{\tilde \O(x_F) \O(x_{R_1}) \O(x_{R_2})}_\text{CFT $\d$BTZ} \equiv \vev{\tilde \O(x_F) \O(x_{R_1}) \O(x_{R_2})}_\text{tree BTZ EFT}, \label{wittenDiagram}
\end{equation}
where as usual, on the left-hand side the operators are defined operationally as limits of bulk fields, 
\begin{equation}
	\O = \lim_{z\to0} \frac{\phi(x^\pm, z)}{z^\Delta}.
\end{equation}

 We choose two scalar primary operators to be on the $R$ Rindler wedge, and the remaining operator to be in the $F$ Milne wedge, so that the diagram is forced to pass through the singularity and we can test what difficulties it poses. In this section, we will seek to understand if such a correlator is even mathematically well-defined, not yet addressing its physical interpretation, given that one operator lies inside the singularity where there are time-like closed curves. In subsection~\ref{S:arbitrarySources} we will give a physical interpretation of the $F$ endpoint as associated to a conceptually straightforward but \emph{non-local} operator
in a hot (thermofield) CFT.  
  For convenience, we have chosen the $F$ scalar primary to be different from the $R$ operators with different scale dimension, 
 $\tilde \Delta \neq \Delta$, so that there are two dual scalar fields in BTZ. We consider a typical non-renormalizable interaction term in BTZ EFT, 
\begin{equation}
	\L_\text{int} = \sqrt g \, \tilde \phi\,  g_\text{BTZ}^{MN} \, \d_M \phi \d_N \phi. 
\end{equation}

\begin{figure}
	\centering
	\includegraphics[width=0.33\textwidth]{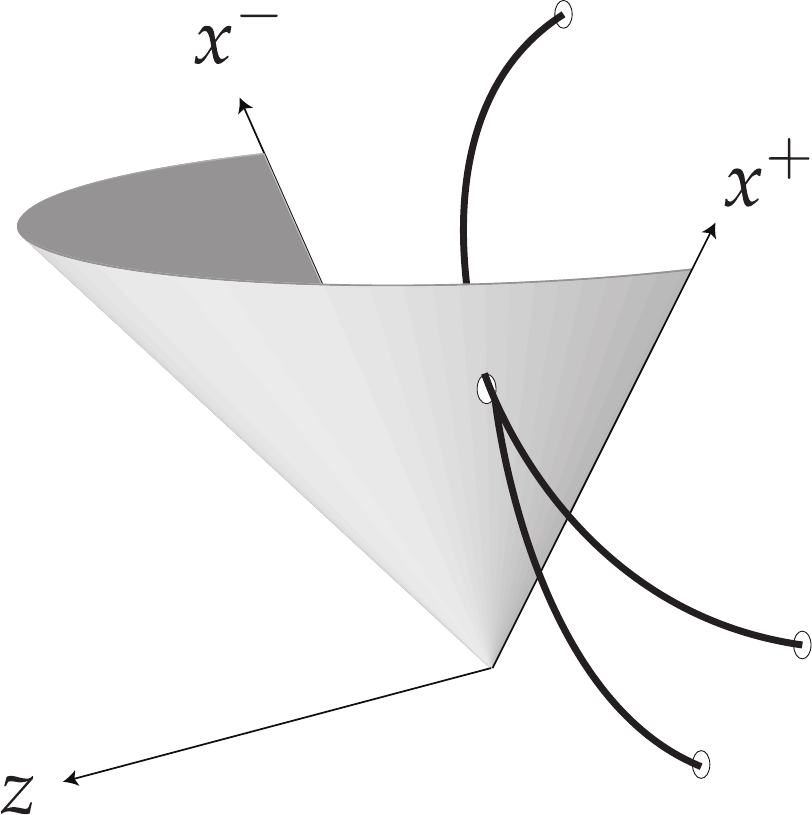}
	\caption{Interaction vertex approaching the singularity as seen in the AdS$_\text{Poincar\'e}$ covering space.  The singularity is the half-cone $x^2 + z^2 = t^2$ with $z>0$.  All lines end on the boundary $z=0$.}
	\label{coneOfSilence}
\end{figure}

It is straightforward to then write the resulting 3-point correlator in terms of $K_\text{BTZ}$ and an integral over a fundamental region of our quotient for the above bulk interaction vertex, and see that it receives divergent contributions as the interaction vertex approaches the singularity (Fig.\@~\ref{coneOfSilence}), 
\alignStart
	\vev{\tilde \O_F \O_{R_1} \O_{R_2}} &= \int_\text{fund.} \neg d^2y dz \sqrt g \, K_\text{BTZ}(x_F,y,z) g^{MN} \d_M K_\text{BTZ}(x_1,y,z) \d_N K_\text{BTZ}(x_2,y,z) \\
	&\underset{r \to0}{\sim}  \int_0^\infty dr \int_{-r_S/2}^{r_S/2} d\sigma \int d\tau \, (\text{function of $\sigma$, $\tau$}) \times r \, \frac{\ln(r + i\epsilon)}{(r+i\epsilon)^2} \xrightarrow[\epsilon \to 0]{} \infty.
\alignEnd
Ref. \cite{shenkerkraus} earlier worked through a very similar calculation.
Naively, this blocks us from defining such correlators. However, this calculation is in error. 

\subsection{Approaching Singularity from Inside}

Schwarzschild coordinates are useful for cleanly separating out the direction of approach to the singularity from the direction which is being imaged, but they are restricted to only the outside of the singularity, $z^2 - y^2 >0$. 
We should also include the asymptotic contributions as the 
interaction vertex approaches the singularity from inside it, $z^2 - y^2 < 0$.
This requires new Schwarzschild-like coordinates for $z^2-y^2<0$:
\gatherStart
	\begin{alignedat}{2}
		y^\pm &= \sqrt{\tfrac{1}{\tilde r^2} + 1} \, e^{\pm \sigma^\pm} &&\qquad (\tilde r > 0) \\
		z &= \frac{e^\sigma}{\tilde r} &&\qquad (\sigma^\pm \, \text{real})
	\end{alignedat} \\
	ds^2 = -(\tilde r^2 +1) d\tau^2 - \frac{ d\tilde r^2}{\tilde r^2 + 1} + \tilde r^2 d\sigma^2.
\gatherEnd
Both sets of Schwarzschild coordinates together cover half of AdS$_\text{Poincar\'e}$, $x^+ >0$.  
(We can obviously cover 
$x^+ <0$  analogously, but will not need to.)

We can now repeat the analysis of near-singularity asymptotics for $K_\text{BTZ}$ but approaching from inside.
For example, considering the boundary point in $K_\text{BTZ}$ to be in the $R$ wedge for concreteness (since the asymptotics will not distinguish other choices), we have
\begin{equation}
\begin{aligned}
	K_\text{BTZ} &\underset{r \to0}{\sim} \sum_\text{$n>0$ large} \left( \frac{1}{e^{\sigma'^+} e^{-\tau} - e^{-\sigma'^-} e^\tau  
		- e^\sigma \tilde r \lambda^n + i \epsilon \lambda^n}  \right)^\Delta \\
	&\sim \left( \frac{1}{e^{\sigma'^+} e^{-\tau} - e^{-\sigma'^-} e^\tau} \right)^\Delta \ln\left( -e^\sigma \tilde r + i \epsilon \right).
\end{aligned}
\end{equation}

\subsection{Proper account of Scattering through Singularity} \label{rotatingZ}

Putting this together with the near-singularity asymptotics from the outside, we obtain the correct behavior,
\alignStart
	\vev{\tilde \O_F \O_{R_1} \O_{R_2}} &= \int_\text{fund.} \neg d^2y dz \sqrt g \, K_\text{BTZ}(x_F,y,z) g^{MN} \d_M K_\text{BTZ}(x_1,y,z) \d_N K_\text{BTZ}(x_2,y,z) \\
	&\underset{r \to0}{\sim}
		  \int_0^\infty dr \int d\sigma d\tau \, r \, \frac{\ln(r + i\epsilon)}{(r+i\epsilon)^2} 
	- \int_0^\infty d\tilde r \int d\sigma d\tau \, \tilde r \, \frac{\ln(- \tilde r + i\epsilon)}{(- \tilde r+i\epsilon)^2} \\
	&\underset{r \equiv - \tilde r}{=}  \int_{-r_0}^{r_0} dr \int d\sigma d\tau \, r \, \frac{\ln(r + i\epsilon)}{(r+i\epsilon)^2},
\alignEnd
where we have defined $r \equiv - \tilde{r}$ for $r < 0$, and $r_0$ is small enough that we trust our asymptotics. 
It is straightforward to see that the $r$ integral converges near the singularity, $r \sim 0$. For example, the integral can clearly be deformed into the upper half complex $r$-plane, completely avoiding $r =0$. In this way, our $3$-point correlator is  mathematically well-defined despite having to traverse the singularity. (As mentioned earlier, we will give a physical interpretation in subsection~\ref{S:arbitrarySources}.) For a similar BTZ correlator, \cite{shenkerkraus} found a similar cancellation of divergence from positive and negative $r$, although the 
 $r<0$ region in this case arose from  the past singularity, so that the cancellation was non-local in spacetime. For us however, the cancellation is local, just from the two sides very close to the future singularity surface.

In fact, there is a technically simpler way, directly in Poincar\'e coordinates, to see the above finiteness. It is not quite as physically transparent  as the local and Lorentzian account above, but it will generalize to other correlators, and it will arise very naturally in our final dual CFT construction. The trick is to note that one can
rotate the interaction $z$ coordinate in the complex plane whenever $y^\pm$ is in the $F$, $P$ wedges,  \emph{before} doing the image sums. In our example, 
\begin{multline}
	\vev{\tilde{\cal O}(x_F) \O(x_{R_1}) \O(x_{R_2})} = \sum_{k,m,n} \int_\text{$F$ fund.} \neg d^2y dz \, \frac{1}{z} 
		\left[ \frac{z \lambda^k}{z^2 \lambda^{2k} - (x_F - \lambda^ky)^2 + i\epsilon} \right]^{\tilde \Delta} \\
		\times \eta^{MN} \d_M \left[ \frac{z \lambda^m}{z^2 \lambda^{2m} - (x_{R_1 }- \lambda^my)^2 + i\epsilon} \right]^\Delta 
		\d_N \left[ \frac{z \lambda^n}{z^2 \lambda^{2n} - (x_{R_2} - \lambda^ny)^2 + i\epsilon} \right]^{\Delta}  \\
		+ \text{other wedges}, 
\end{multline}
where we integrate over a fundamental region in $(y,z)$ for each wedge, and $\eta_{MN}$ is the $2+1$ Minkowski metric.
We can  combine one sum, say $\sum_k$, and $\int_\text{fund.}$ into $\int$ over all AdS$_\text{Poincar\'e}$, so that 
after re-naming the other image indices, $m \rightarrow m+k, n \rightarrow n+k$, 
\begin{multline}
	\vev{\tilde{\cal O}_F \O_1 \O_2}_\text{BTZ} = \sum_{m,n} \int_\text{AdS$_\text{Poincar\'e}$ $F$-wedge} \neg d^2y dz \, \frac{1}{z} 
		\left[ \frac{z}{z^2 - (x_F - y)^2 + i\epsilon} \right]^{\tilde \Delta} \\
	\times \eta^{MN} \d_M \left[ \frac{z \lambda^m}{z^2 \lambda^{2m} - (x_{R_1} - \lambda^my)^2 + i\epsilon} \right]^\Delta 
		\d_N \left[ \frac{z \lambda^n}{z^2 \lambda^{2n} - (x_{R_2} - \lambda^ny)^2 + i\epsilon} \right]^{\Delta} \\
	+ \text{other wedges}. \label{boundaryCorrelator}
\end{multline}
Note the $m$, $n$ summand is analytic in $z$ for $\re z, \im z >0$ so that we can rotate the $z$ contour to the imaginary axis. In general we do this only in the $F$, $P$ wedges, \emph{but not} in the $L$, $R$ wedges.
Here, we just track the $F$ wedge integration region explicitly for the interaction vertex:
\begin{multline}
	\vev{\tilde{\cal O}_F \O_1 \O_2}_\text{BTZ} =  \sum_{m,n} \int_F d^2y \int_0^\infty dz \frac{1}{z} 
		\left[ \frac{i z}{-z^2 - (x_F - y)^2 + i\epsilon} \right]^{\tilde \Delta} \\
		\times \eta^{MN} \d_M \left[ \frac{i z \lambda^m}{ -z^2 \lambda^{2m} - (x_{R_1} - \lambda^my)^2 + i\epsilon} \right]^\Delta
		\d_N \left[ \frac{i z \lambda^n}{ -z^2 \lambda^{2n} - (x_{R_2} - \lambda^ny)^2 + i\epsilon} \right]^{\Delta} \\
		+ \text{other wedges}. \label{boundaryCorrelatorRotated}
\end{multline}
With this rotated $z$, it is straightforward to see that 
 $y^2 + z^2 \neq 0$ since $y^2 >0$ in $F$, so the image sums can now be safely performed and will converge.  
 
 The finiteness of correlators with endpoints at the BTZ boundary generalizes to finiteness of \emph{bulk} correlators with endpoints away from the singular surface. The deeper reason for such finiteness will emerge in Section~\ref{S:finiteLambda}. 
 This does not mean the singularity has disappeared. As we saw straightforwardly in the discussion of \eqref{kWittenImaged}, there are divergences when correlators end \emph{on} the singularity. Furthermore, we will demonstrate in Section~\ref{S:sensitivityToSingularity} one can isolate UV-sensitive effects even at significant distances/times away from the singular surface (but $< R_\text{AdS} \equiv 1$).  
  
 In \cite{btzglobal}, we will give a more detailed account of  $2 \rightarrow 2$ "scattering" through the singularity into the whisker region, in a manner that allows more direct comparison with related studies in the literature, in string theory and EFT (reviewed in \cite{berkooz}). We will reproduce the pathologies in the literature related to the singularity, but also point out how the whisker regions play an important role in resolving them.

 \subsection{Matching to CFT on $\partial$BTZ} \label{matchingToCFT}
 
 In the above sequence of operations, rotating the $z$ contours before doing the sum over two of the images, 
 the method of images applied to AdS$_\text{Poincar\'e}$ correlators does seem to provide us with well-defined boundary correlators in BTZ, which in turn can be interpreted as defining local primary correlators  for a dual CFT "living" on the Lorentzian torus $\partial$BTZ. Indeed if one were to directly define a CFT 3-point correlator on the Lorentzian torus by viewing it as a quotient of $1+1$ Minkowski spacetime, it would be given in planar approximation by the analogous correlator in Minkowski space, but with image sums over two of the positions of the three local operators, with one position kept fixed. One can then use AdS$_\text{Poincar\'e}$ tree diagrams to compute these Minkowski correlators. Eqs.\@~\eqref{boundaryCorrelator} and \eqref{boundaryCorrelatorRotated} can be put in precisely this form by dividing imaged numerators and denominators by $\lambda^{2m}$ or $\lambda^{2n}$. This gives
\begin{multline}
	\vev{\tilde{\cal O}_F \O_1 \O_2}_\text{BTZ} = \sum_{m,n} \int_\text{AdS$_\text{Poincar\'e}$ $F$-wedge} \neg d^2y dz \, \frac{1}{z} 
	\lambda^{m \Delta} \lambda^{n \Delta}
	\left[ \frac{z}{z^2 - (x_F - y)^2 + i\epsilon} \right]^{\tilde \Delta} \\
	\times \eta^{MN} \d_M \left[ \frac{z}{z^2 - (\lambda^m x_{R_1} - y)^2 + i\epsilon} \right]^\Delta
		\d_N \left[ \frac{z}{z^2 - (\lambda^n x_{R_2} - y)^2  + i\epsilon} \right]^{\Delta} \\
		+ \text{other wedges},
\end{multline}
before rotating $z$, and 
\begin{multline}
	\vev{\tilde{\cal O}_F \O_1 \O_2}_\text{BTZ} = \sum_{m,n} \int_F d^2y \int_0^\infty dz \, \frac{1}{z} 
	      \lambda^{m \Delta} \lambda^{n \Delta}
		\left[ \frac{i z}{ -z^2 - (x_F - y)^2 + i\epsilon} \right]^{\tilde \Delta} \\
		\times \eta^{MN} \d_M \left[ \frac{i z}{ -z^2 - (\lambda^m x_{R_1} - y)^2 + i\epsilon} \right]^\Delta
		\d_N \left[ \frac{i z}{ -z^2 - (\lambda^n x_{R_2} - y)^2  + i\epsilon} \right]^{\Delta} \\
		+ \text{other wedges},
\end{multline}
after rotating $z$ into the manifestly summable form. We have also re-defined $m, n \rightarrow -m, -n$ above. 

The summands now have the forms of $1+1$ Minkowski CFT correlators computed by AdS/CFT, at images of the operator positions, $x$. The $x$ themselves are chosen from a fundamental region. Given the universal form of these $3$-point CFT correlators, we have
\begin{equation}
	\vev{\tilde{\cal O}_F \O_1 \O_2}_\text{BTZ} = 
	\sum_{m,n} \frac{\lambda^{m \Delta} \lambda^{n \Delta}}{(\lambda^m x_{R_1} - \lambda^n x_{R_2})^{2\Delta - \tilde{\Delta} }
	(\lambda^m x_{R_1} - x_F)^{\tilde{\Delta}} (\lambda^n x_{R_2} - x_F)^{\tilde{\Delta}}}.
\end{equation}
This has precisely the form of a correlator for a CFT "living" on the boundary of BTZ, computed in the planar limit of a $1/N_{\CFT}$ expansion using the method of images, where Minkowski spacetime is interpreted as the covering space of the BTZ boundary modulo a Weyl transformation
as discussed in subsection~\ref{disconnectedBoundary}. 
The powers of $\lambda^{m \Delta}, \lambda^{n \Delta}$ in the numerator are accounted for 
now as the local primary operator responses to this Weyl-rescaling. In this form, it is straightforward to check that the sums over images converge as long as the CFT local operators do not lie exactly on the lightlike circles of $\partial$BTZ.

 We could proceed to generalize in this direction but instead prefer to first reformulate our whole problem in a way that makes the non-perturbative (in $1/N_\text{CFT}$) structure clear.

%%%%%%%%%%%%%%%%%%%%%%%%%%%%%%%%%%%%%%%%%%%%%%%%%%%%%%%%%%%%%%%%%%%%%%%%%%%%%%%%%%%%%%%%%%%%
%%%%%%%%%%%%%%%%%%%%%%%%%%%%%%%%%%%%%%%%%%%%%%%%%%%%%%%%%%%%%%%%%%%%%%%%%%%%%%%%%%%%%%%%%%%%
\section{Space $\leftrightarrow$ Time Inside the Horizon} \label{S:spaceToTime}

We now describe a strategy for making sense of the boundary regions and their interconnections, and thereby framing the 
CFT dual non-perturbatively.
 The central observation is that in $1+1$ dimensions, and in particular in conformal field theory, there is very little to distinguish  "time" from "space". Indeed we can view the switch $x \leftrightarrow t$, or equivalently $x^\pm \rightarrow \pm x^\pm$, as an "improper" conformal transformation, in that it changes the metric only by an overall factor (the defining feature of conformal transformations), but that factor is~$-1$: 
\begin{align}
	ds^2 = dt^2 - dx^2 \to dx^2 - dt^2 = - (dt^2 - dx^2).
\end{align}
This makes it a plausible symmetry of CFT. 
If we can make the switch $x \leftrightarrow t$ in just the Milne wedges, then the timelike circles would be turned into ordinary spacelike circles, as already the case in the Rindler wedges. 

\subsection{$x \leftrightarrow t$ in Free CFT}

Eventually, we will have to check this proposition for CFTs with good AdS duals, but let us first gain intuition by studying \emph{free} scalar CFT 
and seeing in what concrete sense $x \leftrightarrow t$ is a symmetry of the theory. Let us focus on the path integral for correlators in $1+1$ Minkowski spacetime for the primary operators of the form 
$e^{iq \chi(x^\pm)}$, with scaling dimension $\Delta = q^2/2$, and scalar field $\chi$. If these operators transform as "primaries" with respect to the improper conformal transformation, 
$x^\pm \rightarrow \pm x^\pm$, then we expect that  
\alignStart
	& \int \D\chi e^{i \int dt dx \, (\d_t \chi)^2 - (\d_x \chi)^2} e^{i q_1 \chi(x^\pm_1)} \cdots e^{i q_n \chi(x^\pm_n)} \\
	&= \int \D\hat\chi e^{i \int dt dx \, (\d_x \hat\chi)^2 - (\d_t \hat\chi)^2} (-1)^\frac{\Delta_1}{2}e^{i q_1 \hat\chi(\pm x^\pm_1)} \cdots (-1)^\frac{\Delta_n}{2}e^{i q_n \hat\chi(\pm x^\pm_n)} \\
	&= \int \D\hat\chi e^{-i \int dt dx \, (\d_t \hat\chi)^2 - (\d_x \hat\chi)^2} e^{i q_1 \hat\chi(\pm x^\pm_1)} \cdots e^{i q_n \hat\chi(\pm x^\pm_n)} (-1)^\frac{\Delta_n+ \cdots + \Delta_n}{2}, 
\alignEnd
where $\chi(x^\pm) \equiv \hat \chi(\pm x^\pm)$ and the $q_j$ satisfy charge conservation, $\sum_j q_j = 0$. 
 Formally, the powers of $-1$ are the standard powers of $\partial \hat{x}_-/\partial x^-$ for scalar primaries, although here there is clearly an ambiguity in how to take fractional powers.  
Comparing the first and last lines we arrive at the equivalent statement for the $n$-point Greens function,
\begin{align}
	G_{q_1, \cdots, q_n}(x^\pm_1, \cdots, x^\pm_n) = (-1)^\frac{\Delta_n+ \cdots + \Delta_n}{2} G^*_{-q_1, \cdots, -q_n}(\pm x^\pm_1, \cdots, \pm x^\pm_n).
\end{align}

It is straightforward to check this guess, since $G$ is known explicitly (reviewed in~\cite{joestring}):
\alignStart
	G &= \prod_{i>j} \left[ \frac{1}{(x_i^+ - x_j^+)(x_i^- - x_j^-) + i \epsilon(t_i - t_j)^2} \right]^\frac{q_i q_j}{2} \label{scalarPropagator} \\
	&= \prod_{i>j} e^{\frac{-i\pi q_i q_j}{2}} \left[ \frac{1}{(x_i^+ - x_j^+)(-x_i^- - (-x_j^-)) - i \epsilon(x_i - x_j)^2} \right]^\frac{(-q_i) (-q_j)}{2}  \\
	&= \prod_k e^{\frac{i\pi \Delta_k}{2}} G^*_{-q_1, \cdots, -q_n}(\pm x^\pm_1, \cdots, \pm x^\pm_n).
\alignEnd
The trade of $(x_i-x_j)^2$ for $(t_i - t_j)^2$ in the second line follows because the $i \epsilon$ only matters when the rest of the denominator vanishes, which is where these two expressions coincide. The product of phase factors in the second and third lines coincide by charge conservation, 
\gatherStart
	0 = \left( \sum_i q_i \right)^2 = \sum_{ij} q_i q_j = 2 \sum_{i>j} q_i q_j + \sum_k q_k^2 \\
	\implies \sum_k q_k^2 = \sum_k 2 \Delta_k = - 2 \sum_{i>j} q_i q_j.
\gatherEnd
We see that this resolves the ambiguity of fractional powers of $-1$ in our formal derivations,  
$e^{\frac{i\pi \Delta_k}{2}} \equiv (-1)^\frac{\Delta_k}{2}$.

In \cite{xtot}, we will prove the $x \leftrightarrow t$ property for  time-ordered correlators with  up to four external points for a general $1+1$ CFT, using the conformal bootstrap approach. In Section~\ref{S:rindlerAdSCFT} of this paper, we prove (a refinement of) this property on the dual AdS side within EFT for any number of external points.

\subsection{$x \leftrightarrow t$ in Milne Wedges and Reconnecting to Rindler Wedges}

Having made this plausible case for $x \leftrightarrow t$ symmetry, we will apply the transformation on the $F$ Milne wedge to make it more $R$-Rindlerlike. More generally, we also compound it with $x^\pm \rightarrow - x^\pm$ as required to make all four wedges appear $R$-Rindlerlike, so that we can use $R$-Rindler coordinates in all of them:
\begin{align}
	\pm e^{\pm \sigma^\pm} = e^{\alpha} (\sin \theta \pm \cos \theta) =
	\begin{cases}
		x'^\pm_R = \phantom{\pm} x^\pm_R \\
		x'^\pm_F = \pm x^\pm_F & x \leftrightarrow t\\
		x'^\pm_L = - x^\pm_L & x, t, \leftrightarrow -x, -t\\
		x'^\pm_P = \mp x^\pm_P & x \leftrightarrow -t,
	\end{cases} \label{rindlerCoordinates}
\end{align}
where $|\theta| \leq \frac{\pi}{4}$.  Now let us take seriously that the dual CFT lives on the Lorentzian torus, say as prescribed by a CFT path integral on this spacetime. After the above space/time interchanges, the path integral on each
 wedge \emph{separately} corresponds to a canonical quantum mechanical time evolution (in $\theta$) for the CFT on a spacelike circle (in $\alpha$). Tentatively, the overall connected toroidal structure suggests that the initial and final states of each wedge evolution are to be identified with those of adjacent wedges, and then summed over. (This neglects any concern that the joints themselves are lightlike circles where field theory is expected to be pathological, but we will be more careful about this below.) 
 
It thereby appears that the full torus path integral is the \emph{CFT-trace} of the product of quantum evolution operators for the four separate wedges, akin to the thermal trace structure of finite temperature field theory and its equivalence to the thermal path integral in cyclic Euclidean time. However, we have neglected to take into account that adjacent wedges have differing space/time interchanges, so that the initial/final state of a particular wedge has to be reinterpreted before being "handed off" to an adjacent wedge. For example, consider the $x^- \rightarrow -x^-$ transformation in passing from the $R$ to $F$ wedge. Noting that under the standard CFT symmetry operator
$e^{i\beta(S-K)}$, where $S, K$ are the generators of $x^\pm$ dilatations and boosts respectively, 
\alignStart
	x^- &\to e^{2\beta} x^- \\
	x^+ &\to x^+,
\alignEnd
it is natural to guess that  under $e^{\frac{\pi}{2}(S-K)}$,
\alignStart
	x^- &\to e^{i\pi} x^-  = - x^-\\
	x^+ &\to x^+.
\alignEnd
It has the form of an analytic continuation \cite{vakkurikraus} around the  complex $x^-$ plane (by $\pi$) and therefore makes sense 
 if the $x^\pm$ dependence in states/amplitudes  is analytic enough at the "joints" of the torus.  Let us continue to hope for the best and take $e^{\frac{\pi}{2}(S-K)}$ as the 
requisite operator to reinterpret states at the hand-off between the $R$ and $F$ wedges. Similar operators can be constructed to act at the other "joints" of the torus. 

Although we are inspired by the connectedness of the Lorentzian torus, and could proceed directly in this language, it is more convenient to implement the above insights using our first realization of $\d$BTZ as four disjoint cylindrical spacetimes, with the standard Rindler coordinates, $\sigma^\pm$. The CFT should be indifferent to these different realizations by different Weyl transformations.
We do this for three reasons. Firstly, after the space/time exchanges above, all the four cylindrical spacetimes have the same very simple geometry, $ds^2 = d \sigma^+ d \sigma^-$, with circular $\sigma$ and infinite time. Secondly,  this form of  $R$ and $L$ wedges is just the standard home of the CFTs in the thermofield proposal for the black hole dual, so it will make translation of our results into thermofield language easier. Thirdly, 
the CFT operators involved in the hand-off of states between wedges are simply energy and momentum in Rindler coordinates, 
\begin{align}
	P_\pm^{(\sigma)} \equiv \frac{H^{(\tau)} \mp P^{(\sigma)}}{2} = \frac{K^{(x)} \pm S^{(x)}}{2}, 
\end{align}
where the $K, S$ are to be interpreted here as (consistently) restricted to the $R$ wedge. 

%This approach to connecting the different boundary regions of BTZ is in a similar spirit to that of \cite{vakkurikraus}, but with a more concrete proposal here, and with more conclusive checks.

\begin{figure}
	\centering
	\includegraphics[width=0.33\textwidth]{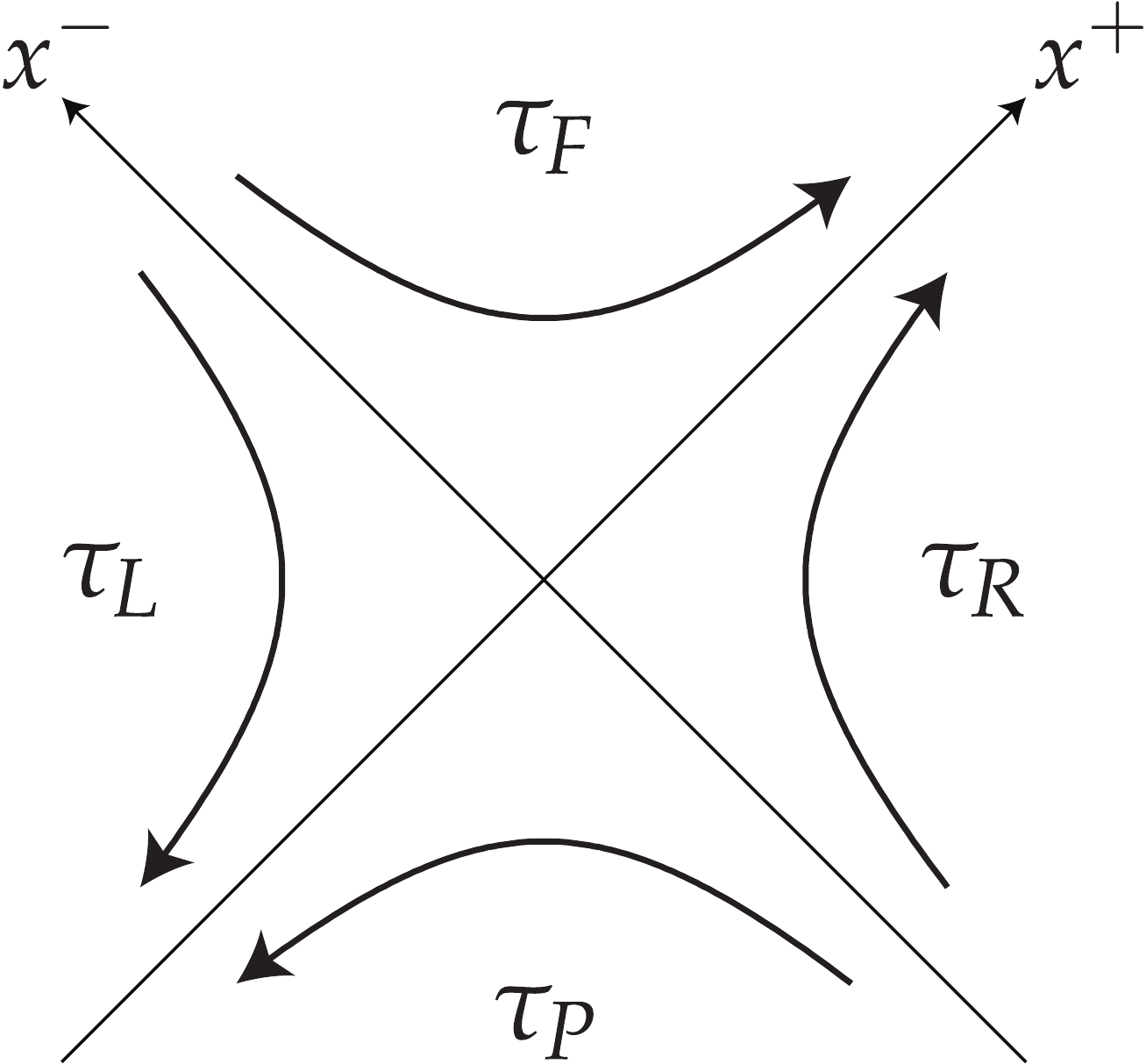}
	\caption{The direction of $\tau$ in each wedge with respect to Minkowski $x^\pm$ as given in \eqref{rindlerCoordinates}.}
	\label{tauDirections}
\end{figure}

\subsection{CFT on $\partial$BTZ as a Trace}

\begin{figure}
	\centering
	\includegraphics[width=0.33\textwidth]{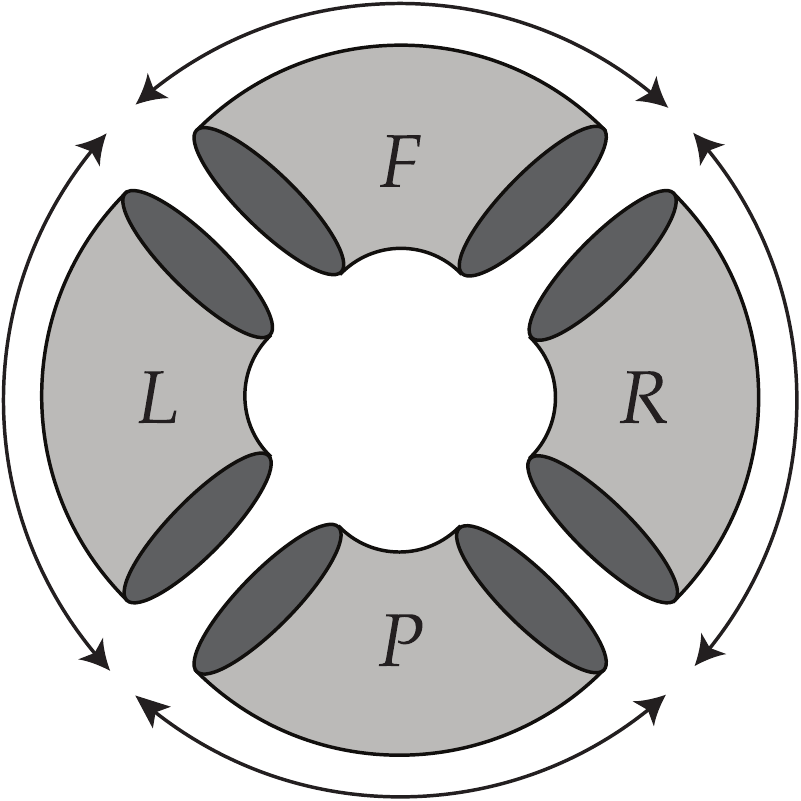}
	\caption{Closed lightlike curves are avoided by a limiting process indicated by the arrows.  Closed timelike curves are avoided by formally transforming $t \leftrightarrow x$ in the $F$ and $P$ wedges.}
	\label{funnyTorus}
\end{figure}

We thereby arrive at a precise and well-defined CFT proposal for interpreting the torus path integral, as the dual of BTZ quantum gravity:
\gatherStart
	Z[J_{L,R,F,P}] = \lim_{\TT\to \infty} \tr \left\{ U_P^\dag e^{-\pi P_-} U_L e^{-\pi P_+} U_F^\dag e^{-\pi P_-} U_R e^{-\pi P_+} \right\} \label{traceFormula} \\
	\text{where} \quad U = \T_\tau e^{-i \int_{-\TT}^{\TT} d\tau \, (H_\text{CFT} - J \O)}.
\gatherEnd
Here, each $U$ is the time evolution operator within each wedge, with possible source terms for CFT operators 
$\cal{O}$ in any wedge. $\T_\tau$ denotes time-ordering with respect to $\tau$ (Fig.\@~\ref{tauDirections}). In order to be careful about the joins between wedges we regulate the evolutions to finite but very large final/initial times $\tau = \pm {\cal T}$, and assume that sources, $J$, vanish outside these times. We then take the limit as ${\cal T} \rightarrow \infty$ in order to obtain the Weyl-equivalent  of the full torus path integral, except that (just) the lightlike circles have been delicately excised (Fig.\@~\ref{funnyTorus}).  
It is straightforward to see that this limit exists once we write the Dyson series expansion for the source terms,
 since all $e^{\pm iH_\text{CFT} \TT}$ factors cancel between the different wedges (using the fact that $H$ and $P$ commute for the CFT on the cylinder):
\gatherStart
	Z = \tr (\T_\tau \hat U_P)^\dag e^{-\pi P_-} (\T_\tau \hat U_L) e^{-\pi P_+} (\T_\tau \hat U_F)^\dag e^{-\pi P_-} (\T_\tau \hat U_R) e^{-\pi P_+} \label{traceFormulaDyson} \\
	\hat U \equiv e^{-i \int_{-\infty}^\infty d\tau d\sigma\, J \O^H}.
\gatherEnd
Here $\O^H$ is the Heisenberg operator (with respect to $H_\text{CFT}$) related to the Schr\"{o}dinger operator $\O$.

Finally, notice that we have chosen specific signs for our exponents in the $e^{- \pi P_{\pm}}$ operators, even though formally rotations by $\pi$ and $-\pi$ are equivalent in the $x^\pm$ complex planes. These choices  ensure convergence of the sum over CFT states implied by the trace, in the face of these exponential weights.
To see this, note that
\begin{equation}
e^{- \pi P_{\pm}} = \sum_n e^{- \frac{\pi}{2} (E^n \mp p^n)}  \ket n \bra n, \label{damping}
\end{equation}
for a complete set of energy-momentum CFT eigenstates. We expect $E^n > |p^n|$ for such states, and this  certainly follows if the CFT is supersymmetric.  Therefore we always have exponential damping of excited states in the trace formula, equation \eqref{traceFormulaDyson}.

We have arrived at a \emph{non-perturbatively} well-defined formulation of a partition functional in terms of a CFT living on a spatial circle $\times$ time. It remains to show that it is a hologram of the extended BTZ black hole if the same CFT on Minkowski space has a low-curvature AdS dual. If so, it necessarily must UV complete the approach to the BTZ singularity and gravitational EFT. 

%%%%%%%%%%%%%%%%%%%%%%%%%%%%%%%%%%%%%%%%%%%%%%%%%%%%%%%%%%%%%%%%%%%%%%%%%%%%%%%%%%%%%%%%%%%%
%%%%%%%%%%%%%%%%%%%%%%%%%%%%%%%%%%%%%%%%%%%%%%%%%%%%%%%%%%%%%%%%%%%%%%%%%%%%%%%%%%%%%%%%%%%%
\section{CFT Dual in Thermofield Form} \label{S:thermofieldCFT}

In this section we translate the trace form of the CFT partition functional motivated above into the thermofield language of two entangled CFTs. This will be useful for proving that its large-$N_\text{CFT}$ diagrammatic expansion reproduces the tree-level (classical) BTZ EFT diagrammatics, and in making contact with the standard framework outside the horizon.

%%%%%%%%%%%%%%%%%%%%%%%%%%%%%%%%%%%%%%%%%%%%%%%%%%%%%%%%%%%%%%%%%%%%%%%%%%%%%%%%%%%%%%%%%%%%
\subsection{Special Case of Purely Rindler Wedge Sources} \label{s:thermofieldCFTPureRindler}

Let us first restrict our sources to local operators in the Rindler wedges, $L$ and $R$, which corresponds to the part of the BTZ boundary lying outside the horizon.  We check that the standard picture \cite{juanbh} emerges straightforwardly from our trace formula.

In this special case, $J_{F,P} = 0$ and \eqref{traceFormulaDyson} becomes
\alignStart
	Z &= \tr e^{-\pi H} (\T \hat U_L) e^{-\pi H} (T \hat U_R) \\ \label{traceToThermofield}
	&= \sum_{n,m} e^{-\pi(E_n + E_m)} \matrixel{n}{\T \hat U_L}{m} \matrixel{m}{\T \hat U_R}{n} \\
	&= \sum_{n,m} e^{-\pi(E_n + E_m)} \matrixel{\bar m}{\T \hat U_L}{\bar n} \matrixel{m}{\T \hat U_R}{n}, 
\alignEnd
where $\ket{\bar \chi} \equiv \CPT \ket \chi$ for arbitrary ket $\ket \chi$.  Note that $\matrixel{\bar m}{\T \hat U_L}{\bar n}$ is still $\T$-ordered with respect to the argument of $\O$, since (taking $x_1^0 \ge x_2^0 \ge \cdots \ge x_n^0$ without loss of generality),
\alignStart
	\matrixel{\psi}{\T\O_1(x_1) \cdots \O_n(x_n)}{\chi} &= \matrixel{\psi}{\O_1(x_1) \cdots \O_n(x_n)}{\chi} \\
	&= \matrixel{\bar \chi}{\CPT \bigl[ \O_1(x_1) \cdots \O_n(x_n) \bigr]^\dag\CPT^{-1}}{\bar \psi}  \\
	&= \matrixel{\bar\chi}{\CPT \O_n^\dag(x_n) \CPT^{-1}  \cdots \CPT \O_1^\dag(x_1) \CPT^{-1}}{\bar \psi}  \\
	&= \matrixel{\bar\chi}{\O_n(-x_n) \cdots \O_1(-x_1)}{\bar \psi} \label{CPT} \\
	&= \matrixel{\bar\chi}{\T\O_n(-x_n) \cdots \O_1(-x_1)}{\bar \psi}.
\alignEnd
The second line is true for any antiunitary operator, while the second-to-last line is how $\O$, which we take as Lorentz scalar (primary) for simplicity, transforms under $\CPT$.  (A more general irreducible representation of the Lorentz group simply receives additional factors of $(-1)$.)

Thus we arrive at standard thermofield form,
\begin{equation}
	Z = \matrixel{\Psi}{T \hat U_L \otimes \T \hat U_R}{\Psi}, \label{thermofieldLR}
\end{equation}
with 
\begin{equation}
	\ket{\Psi} \equiv \sum_n e^{-\pi E_n} \ket{\bar n} \otimes \ket{n} \label{thermofieldState}
\end{equation}
being an entangled state of two otherwise decoupled CFTs on a spatial circle.

%%%%%%%%%%%%%%%%%%%%%%%%%%%%%%%%%%%%%%%%%%%%%%%%%%%%%%%%%%%%%%%%%%%%%%%%%%%%%%%%%%%%%%%%%%%%
\subsection{General Case of Arbitrary Sources} \label{S:arbitrarySources}

Having warmed up as above, let us move to the general case of sources
 $J_{L,R,F,P} \neq 0$, and even possibly non-local sources in these wedges. By inserting complete sets of
 energy-momentum eigenstates again, we can translate our trace formula,
\alignStart
	Z &= \tr e^{-\pi H} e^{\pi P_-} (\T \hat U_P)^\dag e^{-\pi P_-} (\T \hat U_L) e^{-\pi H} e^{\pi P_-} (\T \hat U_F)^\dag e^{-\pi P_-} (\T \hat U_R) \\
	&= \sum_{n,m} e^{-\pi (E_n + E_m)} \matrixel{\bar m}{(\T \hat U_L) e^{-\pi P_-} (\T \hat U_P)^\dag e^{\pi P_-}}{\bar n}
			\matrixel{m}{e^{\pi P_-} (\T \hat U_F)^\dag e^{-\pi P_-} (\T \hat U_R)}{n} \\
	&= \matrixel{\Psi} {(\T \hat U_L) e^{-\pi P_-} (\T \hat U_P)^\dag e^{\pi P_-} \otimes e^{\pi P_-} (\T \hat U_F)^\dag e^{-\pi P_-} (\T \hat U_R)}{\Psi}, \label{thermofieldCFT}
\alignEnd
using $P_\pm \ket{\bar n} = P_\pm \ket{n} = p^n_\pm \ket n$, which follows from $\CPT \O \CPT^{-1}$ rules.  Note that even if the source terms consist of products of local CFT operators, the terms,
\begin{equation}
	e^{-\pi P_-} (\T \hat U_P)^\dag e^{\pi P_-} \equiv \left(\T e^{-i \int \J e^{\pi P_-} \O^H e^{-\pi P_-}} \right)^\dag \label{nonlocalOperators}
\end{equation}
will be products of non-local operators, $e^{\pi P_-} O^H e^{-\pi P_-}$, since $e^{-\pi P_-}$ is non-local.
 Nevertheless, the general partition functional has now been put into thermofield form. 
 
\begin{figure}
	\centering
	\includegraphics[width=0.33\textwidth]{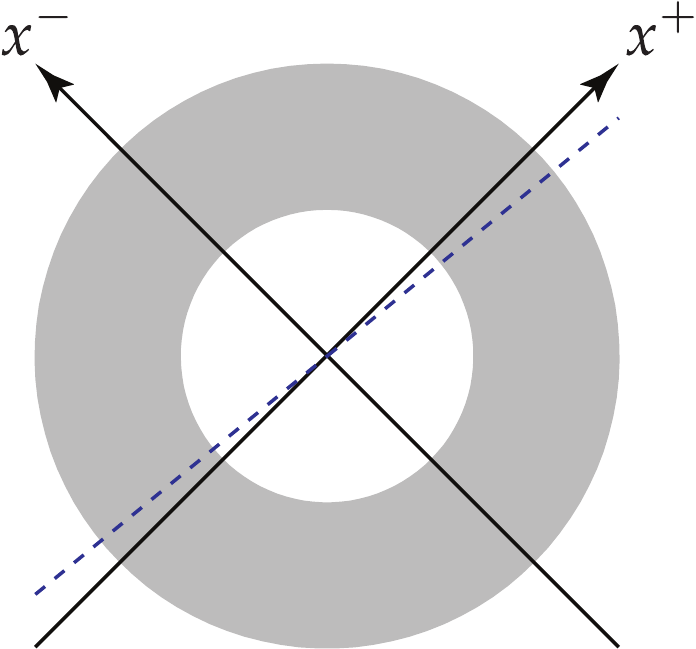}
	\caption{Spacelike hypersurface (blue dashed line) where $\ket N$ and $\ket M$ are located.}
	\label{completeSetOfStates}
\end{figure}

 In order to verify that this CFT formula reproduces BTZ gravitational EFT, it is very useful to separate the 
 Rindler wedges from the $F$ and $P$ wedges by insertions of a complete set of states, $\ket N$,  of $\overline{\text{CFT}} \otimes
 \text{CFT}$ (in contrast to states $\ket n$ of a single CFT). We do this as follows. First we write our thermofield 
 formula in the (obviously equivalent) factorized form,
 \begin{align}
	Z &= \matrixel{\Psi}{ \left[\1 \otimes e^{\pi P_-} (\T \hat U_F)^\dag e^{-\pi P_-} \right] \left[(\T \hat U_L) \otimes (\T \hat U_R)\right] 
	 	\left[ e^{-\pi P_-} (\T \hat U_P)^\dag e^{\pi P_-} \otimes \1 \right]}{\Psi}. \label{thermofieldFactorized}
\end{align}
We insert the resolution of the identity,
\begin{equation}
	\1 = \sum_{N} \left[e^{iH \TT} \otimes e^{-iH \TT} \right]  \ket{N} \bra{N} \left[e^{-iH \TT} \otimes e^{iH \TT} \right],
\end{equation}
between the first pair of square brackets, and again between the second pair of \eqref{thermofieldFactorized}.
Note that the $e^{iH \TT} \otimes e^{-iH \TT} \ket N$ form a complete orthonormal basis of 
$\overline{\text{CFT}} \otimes \text{CFT}$ if the $\ket N$ do because $e^{iH \TT} \otimes e^{-iH \TT} $ is unitary
(and just the boost symmetry operation in Minkowski spacetime language). We thereby get
\alignStart
	Z &= \sum_{N,M} 
	\matrixel{\Psi}{ \left[\1 \otimes e^{\pi P_-} (\T \hat U_F)^\dag e^{-\pi P_-} \right] \left[e^{iH \TT} \otimes e^{-iH \TT} \right] }{N} \\
	&\qquad \matrixel{N}{\left[e^{-iH \TT} \otimes e^{iH \TT} \right] \left[(\T \hat U_L) \otimes (\T \hat U_R)\right] \left[e^{iH \TT} \otimes e^{-iH \TT} \right]}{M}  \\
	&\qquad \matrixel{M}{\left[e^{-iH \TT} \otimes e^{iH \TT} \right] \left[ e^{-\pi P_-} (\T \hat U_P)^\dag e^{\pi P_-} \otimes \1 \right]}{\Psi}. 
\alignEnd
We can view the states $\ket N, \ket M$ as being located on the spacelike hypersurface shown in Fig.\@~\ref{completeSetOfStates}. 

Finally, we massage the exponential weights for our later convenience using $\1 = e^{-\pi P_-} e^{\pi P_-} = e^{\pi P_-}e^{-\pi P_-}$,
\begin{align}
	\begin{split}
		Z &= \sum_{N,M} 
		\matrixel{\Psi}{ \left[e^{-\pi P_-} e^{\pi P_-} \otimes e^{\pi P_-} (\T \hat U_F)^\dag e^{-\pi P_-} \right] \left[e^{iH \TT} \otimes e^{-iH \TT} \right] }{N} \\
		&\qquad \matrixel{N}{\left[e^{-iH \TT} \otimes e^{iH \TT} \right] \left[(\T \hat U_L) \otimes (\T \hat U_R)\right] \left[e^{iH \TT} \otimes e^{-iH \TT} \right]}{M}  \\
		&\qquad \matrixel{M}{\left[e^{-iH \TT} \otimes e^{iH \TT} \right] \left[ e^{-\pi P_-} (\T \hat U_P)^\dag e^{\pi P_-} \otimes e^{\pi P_-}e^{-\pi P_-} \right]}{\Psi}  \\
		&= \sum_{N,M} 
		\matrixel{\Psi}{ \left[\1 \otimes (\T \hat U_F)^\dag \right] \left[e^{\pi P_-} \otimes e^{-\pi P_-} \right] 
			\left[e^{iH \TT} \otimes e^{-iH \TT} \right] }{N} \label{thermofieldCFTSeparated} \\
		&\qquad \matrixel{N}{\left[e^{-iH \TT} \otimes e^{iH \TT} \right] \left[(\T \hat U_L) \otimes (\T \hat U_R)\right] \left[e^{iH \TT} \otimes e^{-iH \TT} \right]}{M}  \\
		&\qquad \matrixel{M}{\left[e^{-iH \TT} \otimes e^{iH \TT} \right]\left[e^{-\pi P_-} \otimes e^{\pi P_-} \right] \left[(\T \hat U_P)^\dag \otimes \1 \right]}{\Psi}, 
	\end{split}
\end{align}
where to get the last equality we have used
\begin{align}
	e^{\pi P_-} \otimes e^{-\pi P_-} \ket \Psi &= \sum_{n} e^{\pi p_-^n} e^{-\pi p_-^n} \ket{\bar n} \otimes \ket n = \ket \Psi.
\end{align}

Note that the above invariance of $\ket \Psi$ can just be thought of as the repeated application of the infinitesimal symmetry invariances of  $\overline{\text{CFT}} \otimes
 \text{CFT}$,
\gatherStart
	\left(\1 \otimes P_\pm - P_\pm \otimes \1 \right) \ket \Psi = 0, \\
	\begin{aligned}
		\1 \otimes H - H \otimes \1 &\equiv \text{Minkowski Boost} \equiv K\\
		\1 \otimes P - P \otimes \1 &\equiv \text{Minkowski Dilatation} \equiv S. \label{HPtoKS}
	\end{aligned}
\gatherEnd
Time and space translations in $\sigma^\pm$-space correspond to boosts and dilatations in $x^\pm$-space. 
The negative sign on the second $P$ is due to the parity operation in $\CPT$.  Even though we are compactifying Minkowski spacetime, we are doing it by quotienting by a discrete dilatation (after a Weyl transformation), which does not break these boost and dilatation symmetries.

In this section and the last, we have made a series of natural guesses to frame the trace formula \eqref{traceFormula}, \eqref{traceFormulaDyson} for the partition functional. In the next sections we use bulk diagrammatics for explicit verification, beginning with the non-quotiented limit, $r_S \rightarrow \infty$.

%%%%%%%%%%%%%%%%%%%%%%%%%%%%%%%%%%%%%%%%%%%%%%%%%%%%%%%%%%%%%%%%%%%%%%%%%%%%%%%%%%%%%%%%%%%%
%%%%%%%%%%%%%%%%%%%%%%%%%%%%%%%%%%%%%%%%%%%%%%%%%%%%%%%%%%%%%%%%%%%%%%%%%%%%%%%%%%%%%%%%%%%%
\section{$r_S = \infty$: Rindler AdS/CFT} \label{S:rindlerAdSCFT}

In the limiting case of $r_S = \infty$, we are no longer quotienting AdS to get BTZ, we simply have AdS. In this case we know that there is a CFT dual on $1+1$ Minkowski spacetime. Nevertheless all our considerations and derivations above apply for any $r_S$, including $r_S = \infty$, and therefore \eqref{thermofieldCFTSeparated} should give us a second, very different looking, dual description. 
It is a non-trivial check of our proposal for these two dual descriptions to agree and holographically "project" quantum gravity and matter on AdS. In this section, we verify this at  EFT tree level.

We begin with AdS EFT, with bulk sources, 
%\footnote{In general, source terms can be thought as small perturbations to the action, in say a 
%path integral.
%Strictly speaking, the explicit spacetime dependence in $\J$ in the source term \eqref{bulkSources} would then violate general coordinate invariance, which is the key "gauge invariance" 
%of quantum gravity, suggesting that such source terms are unphysical. We can however include such non-invariant terms after gauge-fixing coordinate invariance in some way, just as we do in ordinary gauge-theories. In this way, \eqref{bulkSources} allows us to define a generating functional for 
%$\phi$ correlators, but the subset of physical (gauge-invariant) observables is far smaller. We will derive relationships among the full set of correlators inside and outside the Rindler (and later, BTZ) horizon, which then reduces to analogous statements on just  the physical observables. We will adopt this viewpoint even though, for simplicity, we do not explicitly track metric fluctuations in this paper.   Of course, for scalar EFT on a \emph{rigid} AdS (or BTZ) spacetime, no such subtleties arise, and much of our technical development can be read  as relating the interior of the horizon to the exterior at just this level. Of course to ultimately make contact with a dual CFT we must consider that we are in principle doing quantum gravity (fluctuatiing geometry).},
\begin{gather}
	\text{Sources} = \int d^2 x dz \sqrt{g_\text{AdS}} \,  \J(x^\pm, z) \phi(x^\pm, z). \label{bulkSources}
\end{gather}
For simplicity, we consider AdS scalar fields explicitly, but we can clearly generalize our discussion to higher spin fields, including gravitational fluctuations about AdS (as long as we do this in the context of diffeomorphism gauge-fixing, as discussed in the Introduction).
We can break up AdS into four wedges, $F$, $P$, $R$, $L$, just based on $x^\pm$ and extending for all $z$.

\subsection{Special Case of Purely Rindler Wedge Sources}

We will warm up with the special case of only Rindler wedge sources, $\J_{F,P} = 0$.
The usual Rindler construction for any field theory on a spacetime containing a (warped) $1+1$ Minkowski spacetime factor implies \cite{bisowich1, bisowich2} \cite{sewell1, sewell2} \cite{unruhweiss}
\alignStart
	Z &= \matrixel{0}{\T \hat U_L \hat U_R}{0} \\
	&= \matrixel{\Psi}{\T_\tau \hat U_L \T_\tau \hat U_R}{\Psi}.
\alignEnd
On the first line we just have correlators for the Dyson series in the source perturbations in the AdS vacuum, where
\begin{equation}
\hat U = e^{i \int d^2x dz \sqrt {g_\text{AdS}} \, \J \phi}.  
\end{equation}
On the second line, we have replaced the AdS vacuum by its Rindler description (for a general non-conformal field theory), 
\begin{gather}
	\ket \Psi \equiv \sum_k e^{-\pi K_k} \ket{\bar k} \otimes \ket k = \ket{0}_\text{AdS},
\end{gather}
where $K_k$ and $\ket k$ are boost eigenvalues and eigenstates, respectively. Since $\O_L$ and $\O_R$ commute by their spacelike separation, the $\T$-ordering factorizes into separate $\T$-ordering on the $L$ and $R$ operators. We can take the $\T$-ordering on the second line to be with respect to the Rindler time, $\tau$, since the $\tau$-direction is timelike.   We use $\T_\tau$ to denote ordering with respect to $\tau$ and $\T$ for time ordering with respect to Minkowski time, $t$.

We change to coordinates in which the AdS$_\text{Poincar\'e}$ boost symmetry becomes $\tau$-time translation symmetry, and the symmetry of rescaling $x^\pm, z$ becomes a spatial $\sigma$ translation symmetry,
\begin{align}
	x^\pm =
	\begin{cases}
		\mp \sqrt{1-\frac{1}{r^2}} \, e^{\mp \sigma_\mp}, 	&x \in L \\
		\pm \sqrt{1-\frac{1}{r^2}} \, e^{\pm \sigma^\pm}, 	&x \in R
	\end{cases}
	&&
	z =\frac{e^\sigma}{r}, && (r>1). \label{AdSRindler}
\end{align}
The new coordinates cover each Rindler wedge, $L, R$, which now look precisely like the $r_S \rightarrow \infty$ limit of the \emph{exterior} of the BTZ black hole in Schwarzschild coordinates \eqref{schwarzschildCoordinates}, namely the BTZ black \emph{string} (since $\sigma$ is not compact now). But it is also just the Rindler coordinate view of the $R$ (or $L$) wedge of  AdS$_\text{Poincar\'e}$. 
We will refer to the portion of AdS$_\text{Poincar\'e}$ covered by these coordinates for $r>1$ as "AdS$_\text{Rindler}$", and to $r$, $\sigma^\pm$ as "AdS$_\text{Rindler}$ coordinates."
We can write the Dyson series for the Rindler wedge source perturbations as
\alignStart
	\hat U_L &= e^{i \int d^2\sigma \int_1^{\infty} dr \sqrt{g_\text{AdS$_\text{Rindler}$}} \, \J_L \phi} \\
	\hat U_R &= e^{i \int d^2\sigma \int_1^{\infty} dr \sqrt{g_\text{AdS$_\text{Rindler}$}} \, \J_R \phi}.
\alignEnd
Because AdS$_\text{Poincar\'e}$ boosts correspond to $\pm$ AdS$_\text{Rindler}$ time ($\tau$) translations, we now have
\begin{align}
	\ket \Psi = \sum_k e^{-\pi E_\text{AdS$_\text{Rindler}$}^k} \ket{\bar k} \otimes \ket k = \ket{0}_\text{AdS$_\text{Poincar\'e}$} \label{thermofieldStateAdS},
\end{align}
where $E_\text{AdS$_\text{Rindler}$}^k$ and $\ket k$ are eigenvalues and eigenstates of the AdS$_\text{Rindler}$ Hamiltonian.

\subsection{Comparison with Dual CFT}

Now let us invoke standard AdS/CFT duality.  For greater familiarity, first specialize our AdS side further to just boundary sources,
\begin{align}
	\int d^2x dz \sqrt g \, \J \phi \to \int d^2x \, J \lim_{z\to 0} \frac{\phi(x,z)}{z^\Delta} = \int d^2 x \, J \O_\text{primary}.
\end{align}
 Comparing with the dual CFT expression, 
\alignStart
	\ket{0}_\text{AdS} &= \sum_k e^{-\pi E_\text{Rindler}^k} \ket{\bar k} \otimes \ket k \\
	&= \sum_n e^{-\pi E_\text{CFT}^n} \ket{\bar n} \otimes \ket n = \ket{0}_\text{CFT},
\alignEnd
we see that the AdS$_\text{Rindler}$ spacetime must be interpreted as a coarse-grained, classical (planar in large $N_\text{CFT}$) description of an excited stationary CFT state which dominates the thermal sum over states.
In the thermofield gravity description, 
one can think of it as a large excitation of the gravitational field, turning the AdS$_\text{Poincar\'e}$ vacuum
\begin{equation}
	ds^2 = \frac{1}{z^2} \left[ d\tau^2 - d\sigma^2 - dz^2 \right] \qquad ( z > 0 ) \\
\end{equation}
into
\begin{equation}
	ds^2 = \frac{1}{z^2} \left[ d\tau^2 - (1+z^2) d\sigma^2 - \frac{dz^2}{1+z^2} \right],
\end{equation}
where we have rewritten \eqref{AdSschwarzschildMetric} using $z \equiv 1/\sqrt{r^2-1}$, for $r>1$.
We must also sum over metric and other EFT fluctuations away from this dominant state, and these are dual to the CFT deviations from the dominant CFT state. In the gravity description, these deviations from the AdS$_\text{Rindler}$ geometry include Unruh radiation (in the language where we are just seeing AdS$_\text{Poincar\'e}$ from the Rindler observer viewpoint) or Hawking radiation (in the language where we view the dominant geometry as the BTZ black string with horizon). For simplicity, we are making the following approximations on the gravity side of the duality:
\begin{equation}
	\sum_{\substack{\text{quantum} \\ \text{gravity states}} } \approx \sum_{\substack{\text{gravitational} \\ \text{EFT states}} } 
		\approx \sum_{\substack{\text{scalar $\phi$ fluctuations on} \\ \text{fixed AdS$_\text{Rindler}$ metric}  }}.
\end{equation}
That is, in the sum over metrics we are keeping the dominant AdS$_\text{Rindler}$ metric but dropping fluctuations of it. Instead, we are keeping just scalar field fluctuations about this geometry as the simplest illustration of how more general fluctuations will work. 

\subsection{General Case of Arbitrary Sources}

Having interpreted the special status of the AdS$_\text{Rindler}$ metric, let us return to the case of sources in all regions, $ \J_{L,R,F,P} \neq 0$.  We now take the gravity dual of our proposed construction of CFT$_\text{Mink}$ correlators in terms of 
$ \ket{\overline{\text{CFT}_\text{Rindler}}} \otimes \ket{\text{CFT}_\text{Rindler}}$, and show that there is a perfect and non-trivial match. With 
AdS$_\text{Rindler}$ as the dominant CFT state in the thermofield sum, 
the gravity dual of our proposed construction on 
$ \ket{\overline{\text{CFT}_\text{Rindler}}} \otimes \ket{\text{CFT}_\text{Rindler}}$, \eqref{thermofieldCFTSeparated}, is given by the analogous construction on $\ket \Psi = \ket{0}_\text{AdS} \in \ket{\overline{\text{AdS$_\text{Rindler}$}}} \otimes \ket{\text{AdS$_\text{Rindler}$}} $, namely
\alignStart
	Z &= \sum_{N,M}
	\matrixel{\Psi}{ \left[\1 \otimes (\T_\tau \hat U_F)^\dag \right] \left[e^{\pi P_-} \otimes e^{-\pi P_-} \right] \left[e^{iH \TT} \otimes e^{-iH \TT} \right] }{N} \label{thermofieldExpandedAdS} \\
	&\qquad\times \matrixel{N}{\left[e^{-iH \TT} \otimes e^{iH \TT} \right] \left[(\T_\tau \hat U_L) \otimes (\T_\tau \hat U_R)\right] \left[e^{iH \TT} \otimes e^{-iH \TT} \right]}{M}  \\
	&\qquad\times \matrixel{M}{\left[e^{-iH \TT} \otimes e^{iH \TT} \right]\left[e^{-\pi P_-} \otimes e^{\pi P_-} \right] \left[(\T_\tau \hat U_P)^\dag \otimes \1 \right]}{\Psi}. 
\alignEnd
Here, $\ket \Psi$ is given by \eqref{thermofieldStateAdS} and all operators relate to excitations on AdS$_\text{Rindler}$. 
$P_\pm$ are conjugate to  $\sigma^\pm$ as before. $H$ refers to the EFT Hamiltonian in the AdS$_\text{Rindler}$ background, $H_\text{AdS$_\text{Rindler}$}$. 

\begin{figure}
	\centering
	\includegraphics[width=0.4\textwidth]{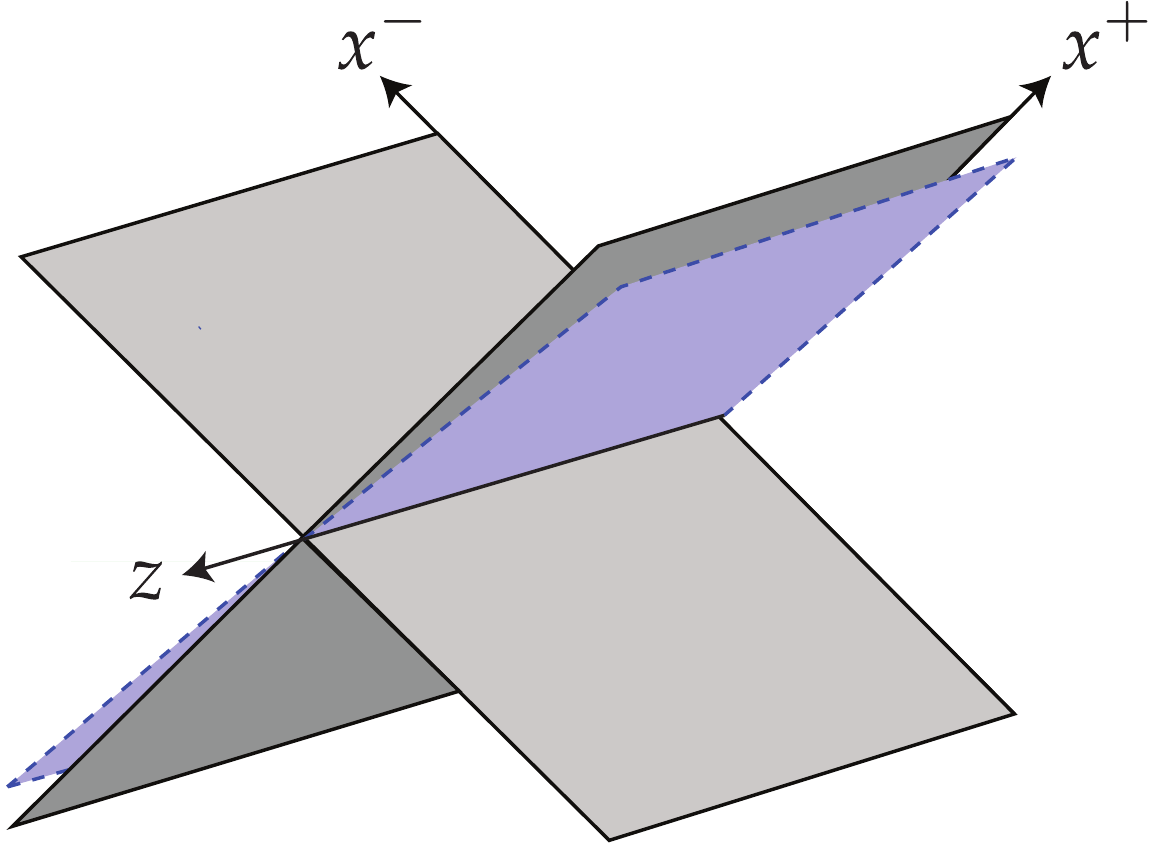}
	\caption{Bulk spacelike hypersurface (blue plane) where $\ket N$ and $\ket M$ are located.}
	\label{completeSetOfStates3D}
\end{figure}

The $\ket N$, $\ket M$ are excitations of  $\ket \Psi$. Converting to AdS$_\text{Poincar\'e}$ coordinates, the time evolution specified localizes these excitations to the spacelike hypersurface 
illustrated in Fig.\@~\ref{completeSetOfStates3D}. We will refer to this as the "$\TT$-hypersurface". We can think of  $\ket N$, $\ket M$ as being given by (multiple) scalar field operators on the $\TT$-hypersurface acting on $\ket \Psi$. 

We now massage and reinterpret the various matrix elements in \eqref{thermofieldExpandedAdS} in AdS$_\text{Poincar\'e}$ language. Let us first focus on the $F$ matrix element
\begin{multline}
	\matrixel{\Psi}{ \left[\1 \otimes (\T_\tau \hat U_F)^\dag \right] \left[e^{\pi P_-} \otimes e^{-\pi P_-} \right] \left[e^{iH \TT} \otimes e^{-iH \TT} \right] }{N} \\
	\equiv \matrixel{N}{\left[e^{-iH \TT} \otimes e^{iH \TT} \right] \left[e^{\pi P_-} \otimes e^{-\pi P_-} \right] \left[\1 \otimes (\T_\tau \hat U_F) \right] }{\Psi}^* \label{futureMatrixElement}.
\end{multline}
Note that 
\alignStart
	\1 \otimes (\T_\tau \hat U_F) &= \1 \otimes \T_\tau e^{i \int d\tau d\sigma \, \sqrt{g_\text{AdS$_\text{Rindler}$}} \, \J_F(\sigma^\pm, \tau) \phi(\sigma^\pm, \tau)} \\
	&= \T e^{i \int d^2x dz \sqrt{g_\text{AdS$_\text{Poincar\'e}$}} \, \J_F( x'^\pm_F, Z) \phi(  x'^\pm_F, z) } \\
	&= \T e^{i \int d^2x' dz \sqrt{g_\text{AdS$_\text{Poincar\'e}$}'} \, \J_F( x'^\pm_F, Z) \phi(  x'^\pm_F, z) } \\ \label{tauToTOrdering}
	&= \T \hat U_F.
\alignEnd
In particular we have replaced Rindler time-ordering by Poincar\'e time-ordering because these agree up to operators with spacelike separations as usual. In the last two lines we have switched to Poincar\'e notation rather than the tensor product Rindler$^2$ notation of the first line.
Also note that 
 $\J_F( x'^\pm, z) \neq 0$ only for both $x^\pm > 0$, which is equivalent to $x'^+ >0$ and $x'^- < 0$.  (See $x'$ definition in \eqref{rindlerCoordinates}).
  So we have sources for $\phi$ in the $R$-wedge. 
 Of course this better not be the full answer since this source term should be for correlators of points in the $F$-wedge, and indeed we must still take into account (and translate to Poincar\'e language) the non-local operation
\begin{align}
	\left[ e^{\pi P_-} \otimes e^{-\pi P_-} \right] \equiv e^{\frac{\pi}{2}(S-K)} \label{nonLocalTranslator}
\end{align}
Using \eqref{HPtoKS}, \eqref{tauToTOrdering}, and \eqref{nonLocalTranslator}, \eqref{futureMatrixElement} becomes
\alignStart
	\matrixel{N}{\left[e^{-iH \TT} \otimes e^{iH \TT} \right] \left[e^{\pi P_-} \otimes e^{-\pi P_-} \right] \left[\1 \otimes (\T_\tau \hat U_F) \right] }{\Psi}^*  \label{futureMatrixElementMink}
	= \matrixel{N}{e^{-i K \TT} e^{\frac{\pi}{2}(S-K)} \T \hat U_F  }{\Psi}^*
\alignEnd

\subsection{Diagrammatic Analysis of Thermofield Formulation} \label{S:thermofieldDiagrams}

To get oriented let us first neglect  the  operation $e^{\frac{\pi}{2}(S-K)}$ in our matrix element \eqref{futureMatrixElementMink} altogether. Then \eqref{futureMatrixElementMink} has the following very general AdS$_\text{Poincar\'e}$ diagrammatic form:
\begin{align}
	\matrixel{N}{e^{-i K \TT} \T \hat U_F  }{\Psi}^* \to \int \Bigl[ \prod \text{AdS-Propagators} \times (-i \, \text{couplings}) \Bigr]^*, \label{schematicFuture}
\end{align}
the complex conjugate of AdS diagrams, with external lines ending on the ${\cal T}$-hypersurface, corresponding to connecting to the $\ket N$ states, or to $\phi$ in the $R$-wedge where $\J_F(x',z) \neq 0$. The integral(s) indicated are over internal interaction vertices.

Now consider the effect of $e^{\frac{\pi}{2}(S-K)}$ acting on $\ket N$. We can break this up in the form $\left[ e^{\frac{\pi}{2k}(S-K)} \right]^k$ for some large number $k$. Then as long as the above AdS-diagrams (AdS propagators) are analytic enough in their $x^-, z$ dependence for the locations of scalar particles in the $\ket N$ state (where external lines attach), the action of each $e^{\frac{\pi}{2k}(S-K)}$ is to just analytically continue the diagram, $x^- \to e^{\frac{i\pi}{k}} x^-$ and $z \to e^{\frac{i\pi}{2k}} z$. One can repeat such small analytic continuations many times 
to analytically continue $x^- \to e^{i \beta} x^-$ and $z \to  e^{\frac{i\beta}{2}} z$ as long as the diagram (propagator) remains analytic along the neighborhood of the path traced out thereby in the complex $x^-, z$ planes, ultimately arriving after $k$ iterations to $x^- \to - x^-$ and $z \to i z$.

This is indeed the case, as we now check. The bulk AdS$_\text{Poincar\'e}$ propagator has the form \cite{ooguri} \cite{burgess},
\begin{align}
	G_\text{AdS}(x^\pm,z; y^\pm, z') = \xi^\Delta F(\xi^2), \label{hypergeometric}
\end{align}
where $F(\xi^2)  \equiv F(\frac{\Delta}{2},  \frac{\Delta}{2} + \frac{1}{2}; \Delta; \xi^2)$ is a hypergeometric function which is analytic in the complex $\xi^2$-plane with a cut along $(1, \infty)$,  and where
\begin{equation}
	\xi \equiv \frac{2 z z'}{z^2 + z'^2 - (x^+ - y^+) (x^- - y^-) + i \epsilon(x^0 - y^0)^2}. \label{xi}
\end{equation}
Because of the branch cuts in $F$ and $\xi^{\Delta}$ we must be very careful in any analytic continutations we perform. Our first step will be to simply rotate \emph{all} $z$ coordinates in the diagrams of \eqref{schematicFuture},  
\begin{equation}
	z \rightarrow e^{\frac{i\beta}{2}} z, \quad \text{from $\beta = 0$ to  $\beta = \pi - \epsilon$}. \label{zRotation}
\end{equation}
It is straightforward to check that $\xi$ never passes through a branch cut of $G_\text{AdS}$ in such a rotation. 

Let us interpret this move. 
If $z$ corresponds to a point on the ${\cal T}$-hypersurface, then this rotation is just part of the action of $e^{\frac{\pi}{2} (S-K)}$ acting on $|N \rangle$, as discussed above. The $e^{\frac{\pi}{2} (S-K)}$ is also supposed to rotate the associated $x^-$, but since we are taking 
${\cal T}$ very large, and well to the future/past of our sources, $x^- \approx 0$ along this hypersurface. Therefore the action of 
$e^{\frac{\pi}{2} (S-K)}$ on it is trivial. 
If instead, $z$ corresponds to an interaction vertex, then this move corresponds to a (passive) contour rotation of the integral over the interaction vertex location. 
The only other possibility is that $z$ corresponds to a source point, where ${\cal J} \neq 0$. For a source localized to the AdS boundary, 
 necessarily $z = 0$, which is insensitive to the rotation.
For a bulk source which is analytic enough in $z$, the above move would again correspond to (passively) rotating the contour of the $z$-integral over the source region. We will discuss subtleties of boundary and bulk source terms further in subsections \ref{S:boundaryCorrelators} and \ref{S:generalCorrelators}, respectively, 
but proceed with allowing rotation of  source points for these broad reasons. 

After completing the above rotation of all $z$ coordinates in the diagrams of \eqref{schematicFuture}, at $\beta = \pi - \epsilon$ it is straightforward to check that we end up with
\alignStart
	\xi( x^\pm, e^{\frac{i(\pi - \epsilon)}{2}} z; y^\pm, e^{\frac{i(\pi - \epsilon)}{2}} z' ) &=  \frac{2 z z'}{z^2 + z'^2 + (x^+ - y^+) (x^- - y^-) - i \epsilon(x^1-y^1)^2 }\\
	&= \xi^*(\pm x^\pm, z; \pm y^\pm, z').
\alignEnd
From this, and the fact that the hypergeometric function in terms of which $G$ is given satisfies $F({\xi^*}^2) = F^*(\xi^2)$, we obtain the 
simple but non-trivial identity,
\begin{equation}
	G\Bigl(x^\pm, e^{\frac{i(\pi - \epsilon)}{2}}  z;  y^\pm, e^{\frac{i(\pi - \epsilon)}{2}} z' \Bigr)  = G^*(\pm x^\pm,z; \pm y^\pm, z').
\end{equation}
The contour rotations of interaction vertices for real to (nearly) imaginary $z$ results in the change of integration measure, 
\begin{equation}
	\int d^2 x \int_0^\infty \frac{dz}{z^3} \cdots  \to i\int d^2 x \int_0^\infty \frac{dz}{i^3 z^3} \cdots,
\end{equation}
resulting in the replacement in diagrams
\alignStart
		(-i \, \text{couplings}) &\to (+i \, \text{couplings}).   \label{analyticContinuation} \\
\alignEnd
We see that both propagators and interactions are thereby complex-conjugated, and the sign of every $x^-$ is flipped in the diagrams 
corresponding to \eqref{schematicFuture}. This all happened as  a consequence of a single active move, namely to act with $e^{\frac{\pi}{2} (S-K)}$ on 
the points ending on the ${\cal T}$-hypersurface. The complex conjugation simply undoes the conjugation already appearing in \eqref{schematicFuture}. 
For interaction vertices $x^- \rightarrow - x^-$ is clearly irrelevant since it is integrated, and for points ending on the ${\cal T}$-hypersurface we are insensitive to $x^- \rightarrow - x^-$ because $x^- \approx 0$ there. Therefore, $x^- \rightarrow - x^-$ is only significant 
for source points. This now corrects the naive "wrong", that we started with $F$-wedge sources for $\phi$ in the $R$-wedge, as noted below \eqref{tauToTOrdering}. The action of $e^{\frac{\pi}{2} (S-K)}$ has performed this "correction". 

We are now poised to recover all AdS$_\text{Poincar\'e}$ correlators from our thermofield formula, but must carefully consider boundary versus 
bulk source options.

\subsection{Testing Boundary Localized Correlators (in all regions)} \label{S:boundaryCorrelators}

Let us first study the familiar case of sources localized to $\d\text{AdS}$. Since
\begin{equation}
	\O = \lim_{z\to0} \frac{\phi(x^\pm, z)}{z^\Delta},
\end{equation}
we are not integrating $z$. Therefore rotating such $z$ as we prescribe in the previous subsection will not be a passive move, but will result in an extra factor of $1/i^\Delta$ from the above limit. This is easily corrected by multiplying the correlator from the trace formula by $i^\Delta$ for 
each external boundary point in the $F$ ($P$) region.  Then, for boundary sources, the diagrammatic analysis of the previous subsection proves that \eqref{futureMatrixElementMink} is
\alignStart
	\matrixel{N}{e^{-i K \TT} e^{\frac{\pi}{2}(S-K)} \T \hat U_F  }{\Psi}^* 
	= \matrixel{\Psi}{\left( \left.\T \hat U_F \right|_{x^- \to - x^-} \right)e^{iK \TT} }{N}. 
\alignEnd
As discussed below \eqref{analyticContinuation}, the $x^- \to - x^-$ applies to all source points in $\hat U_F$, correcting the naive "wrong" of starting with $F$-wedge sources for $\phi$ in the $R$-wedge. A completely analogous analysis can be made for the $P$ wedge. Eq.\@~\eqref{thermofieldExpandedAdS} thereby takes the form,
\alignStart
	Z_\text{Rindler Thermofield}  
	&= \sum_{N,M} \matrixel{\Psi}{\left( \left.\T \hat U_F \right|_{x^-\to - x^-} \right) e^{iK\TT} }{N} \\
	&\qquad \times \matrixel{N}{e^{-iK\TT} \T \hat U_L \hat U_R e^{iK\TT}}{M} \\
	&\qquad \times \matrixel{M}{e^{-iK\TT} \left( \left.\T \hat U_P \right|_{x^-\to - x^-} \right)}{\Psi} \label{nonCompactDuality} \\
	&= \matrixel{\Psi}{\left( \left.\T \hat U_F \right|_{x^-\to - x^-} \right) \T \hat U_L \hat U_R \left( \left.\T \hat U_P \right|_{x^-\to - x^-} \right)}{\Psi} \\
	&= \left. \matrixel{\Psi}{\T \{\hat U_F \hat U_L \hat U_R \hat U_P\}}{\Psi} \right|_{\{x_F^-, x_P^-\} \to \{-x_F^-, -x_P^-\}} \\
	&= Z_\text{Poincar\'{e}},
\alignEnd
where we used the orthonormality of $\ket N$ to get to the second equality, and the fact that all future and past operators lie to the future and past of the $L$, $R$ wedges respectively, and that $L$ wedge operators commute with those of the $R$ wedge, to get to the third equality. Again, the $\{x_F^-, x_P^-\} \to \{-x_F^-, -x_P^-\}$ applies only to source points in $\hat U_F$ and $\hat U_P$, correcting the naive "wrong."  We have thereby demonstrated that our trace formula and its thermofield equivalent correctly reproduce arbitrary (local) CFT correlators in Minkowski space as captured by the dual AdS EFT.

The $i^\Delta$ factors needed to achieve the above agreement may seem unusual, but they
are just what one should expect of a 
conformal transformation law of a scalar primary $\O$, given our  \emph{improper} conformal transformation,
\begin{equation}
	\O' = \left( \frac{dx^+}{dx'^+} \right)^\frac{\Delta}{2} \left( \frac{dx^-}{dx'^-} \right)^\frac{\Delta}{2} \O,
\end{equation}
$x' = t$, $t' = x$ or $x'^+ = x^+$, $x'^- = -x^-$.
Equivalently, in the $F$ wedge, local operators $\O$ in the trace formula are reinterpreted as
\begin{equation}
	e^{\frac{\pi}{2}(K-S)} \O(x^\pm) e^{\frac{\pi}{2}(S-K)} = e^{-\frac{i\pi\Delta}{2} }\O(\pm x^\pm),
\end{equation}
inside $T$-ordered matrix elements.
Therefore there is a perfect match between our trace formula and the Minkowski/Poincar\'e formulation once these 
 transformation factors are included.
 
In detail,  we see that the $F, P$ source terms in the trace formula must have extra 
 $i^{\Delta}$ factors in order to yield a desired set of AdS$_\text{Poincar\'e}$ source terms. 
 If we think of source terms as perturbations of the CFT Hamiltonian, then hermiticity of such 
 perturbations implies that ${\cal J}$ is real for hermitian $\O$. Clearly, to get such sources 
 for the AdS$_\text{Poincar\'e}$ correlators, we must start with \emph{complex} sources ($i^\Delta \times$ real) 
 in  the trace formula, corresponding to \emph{non-hermitian} CFT perturbations there. This appears to be 
 an essential part of our construction following from the improper nature of the conformal transformation
 switching $x$ and $t$.   We will see a generalization of this feature for bulk sources.

\subsection{Testing General Bulk Correlators} \label{S:generalCorrelators}

Finally, consider \emph{bulk} source terms in $F$. As mentioned in subsection~\ref{S:thermofieldDiagrams}, this case is easiest if we have a bulk source which is analytic in 
$z$. Suppose our goal is to end up with a bulk correlator with a $F$ region source
\begin{equation}
	\int \frac{d^2x \, dy}{z^3} \J \phi = \int _0^\infty \frac{dz}{z} e^{-\frac{1}{a^2} \left( \log z -\log \bar z \right)^2} \phi(t = \bar t, x = \bar x, z)
\end{equation}
where
\begin{equation}
	\J(t,x,z) = \delta(t-\bar t) \delta(x - \bar x) z^2 e^{-\frac{1}{a^2} \left( \log z -\log \bar z \right)^2}
\end{equation}
This is a nice Gaussian function of proper distance in the $z$ direction, with size set by $a$, which can be as small as desired. 
Note that this source term is analytic in $z$ throughout the set of rotated values in \eqref{zRotation}, and falls off rapidly as $|z| \rightarrow 0, \infty$. 
To obtain such a source for our AdS$_\text{Poincar\'e}$ correlator, we have seen in subsection~\ref{S:thermofieldDiagrams} that we must begin in the trace formula with a
source which analytically continues to the target source above, as $z \rightarrow i z$. That is, in the trace formula we must begin with 
\begin{equation}
	\int \frac{d^2x \, dy}{z^3} \J \phi = \int _0^\infty \frac{dz}{z} e^{-\frac{1}{a^2} \left( \log z - \log \bar z - i\frac{\pi}{2} \right)^2} \phi(t = \bar x, x = \bar t, z)
\end{equation}
where
\begin{equation}
	\J(t,x,z) = \delta(t-\bar x) \delta(x - \bar t) z^2 e^{-\frac{1}{a^2} \left( \log z -\log \bar z - i\frac{\pi}{2} \right)^2}
\end{equation}
As discussed earlier, the trading of $\bar{x}$ and $\bar{t}$  will be fixed by the action of $e^{\frac{\pi}{2} (S-K)}$. The analytic $z$ integrand clearly becomes the target source integrand upon performing the $ z \rightarrow i z$ move of \eqref{zRotation}.

Again, what is unusual about such a source term for a real bulk field $\phi$ is that it is not real, and therefore corresponds to a non-hermitian perturbation of a (diffeomorphism gauge-fixed) bulk Hamiltonian. Of course, one can break up such complex sources into real and imaginary parts, so that we reproduce our target AdS$_\text{Poincar\'e}$ correlators/sources by taking straightforward complex linear combintations of 
the corresponding trace formula correlators. With this slightly non-trivial matching of source terms, the results of subsection~\ref{S:thermofieldDiagrams} again translate into the trace formula reproducing the AdS$_\text{Poincar\'e}$ correlators (integrated against the target sources).

The non-trivial matching of sources is to be expected once we take into account that the trace formula  reinterprets $x \leftrightarrow t$ in the CFT in the $F$ region (and similarly for $P$), the result of $e^{\frac{\pi}{2}(K-S)} \O(x^\pm) e^{\frac{\pi}{2}(S-K)}$ for any operator ${\cal O}$ whether local or non-local. When a bulk field operator (in some diffeomorphism gauge-fixed formulation of quantum gravity), 
$\phi(x^\pm, z)$,  corresponds to \emph{some} kind of non-local CFT operator by AdS/CFT duality, it should be reinterpreted in the trace formula as
\begin{equation}
	e^{\frac{\pi}{2}(K-S)} \phi(x^\pm, z) e^{\frac{\pi}{2}(S-K)} = \phi(\pm x^\pm, iz),
\end{equation}
if it lies in $F$,  inside $T$-ordered matrix elements. Of course the bulk field for imaginary $z$ on the right-hand side is not \emph{a priori} well-defined, so this equation should be thought of as a short-hand for our main result:  for analytic sources the  AdS$_\text{Poincar\'e}$ sources match the trace formula sources 
via continuation $z \rightarrow -iz$ for $F$/$P$ regions.

%%%%%%%%%%%%%%%%%%%%%%%%%%%%%%%%%%%%%%%%%%%%%%%%%%%%%%%%%%%%%%%%%%%%%%%%%%%%%%%%%%%%%%%%%%%%
%%%%%%%%%%%%%%%%%%%%%%%%%%%%%%%%%%%%%%%%%%%%%%%%%%%%%%%%%%%%%%%%%%%%%%%%%%%%%%%%%%%%%%%%%%%%
\section{Finite $r_S$: BTZ/CFT} \label{S:finiteLambda}

\subsection{Finiteness of BTZ EFT correlators}

We consider bulk or boundary correlators of the BTZ black hole, with sources anywhere in the \emph{extended spacetime} (including inside the horizon, or even beyond the singularity in the whiskers), as long as bulk sources are analytic in $z$ in the manner discussed in subsection~\ref{S:generalCorrelators}. In the gravitational EFT these BTZ correlators are obtained by the method of images applied to 
AdS$_\text{Poincar\'e}$, in particular the (scalar) propagator in BTZ has the form,
\alignStart
	G_{BTZ}(x^{\pm}, z; y^{\pm}, z') &= \sum_{n= - \infty}^{\infty} G_{AdS}(\lambda^n x^{\pm}, \lambda^n z;  y^{\pm}, z') \label{BTZPropagator} \\
	&=  \sum_{n= - \infty}^{\infty} \xi_n^\Delta F(\xi_n^2),
\alignEnd
where, as in Section~\ref{S:singularity}, we define for convenience
\begin{equation}
	\lambda \equiv e^{r_S}.
\end{equation}
The second line of \eqref{BTZPropagator} follows from \eqref{hypergeometric} and \eqref{xi}, where
\begin{equation}
\xi_n \equiv \frac{2\lambda^n z z'}{\lambda^{2n} z^2 + z'^2 - (\lambda^n x^+ - y^+)(\lambda^n x^- - y^-) + i \epsilon(\lambda^n x^0 - y^0)^2 }.
\end{equation}

A central question is the mathematical finiteness of such EFT correlators, given that the associated Feynman diagrams generally traverse the singularity. This can be understood by looking at the large image-number contributions in the above sum, where
\begin{equation}
\xi_n \xrightarrow[n\to\infty]{} \frac{2z z'}{\lambda^{n} (z^2 - x^+ x^- + i \epsilon (x^0)^2) + {\cal O}(1) }
\end{equation}
implies that for generic points the summand $\propto \lambda^{-n \Delta} F(0)$ for large $n$ and hence the sum converges rapidly. 
However, at the singular surface, $z^2 - x^+ x^- = 0$, \emph{if} we neglect the $i \epsilon$,  we see that $\xi_n$ and hence the summand, become $n$-independent for large $n$, and the sum diverges.  This is the diagrammatic root of the singularity. Once we take into account the 
$i \epsilon$ term we see that we always get a convergent sum again, but for diagrams to remain finite after the ultimate $\epsilon \rightarrow 0$ requires major cancellations before that limit is taken. We studied the simplest examples of this situation and such cancellations in Section~\ref{S:singularity}, but in general correlators the requisite cancellations are not immediately apparent. Nevertheless they do take place, as we now show in a simple and general way. 

Let us again perform the complex rotation of all $z$ coordinates as we did in \eqref{zRotation}, but now stopping at an intermediate value of $\beta = \pi/2$, 
\begin{equation}
	z \rightarrow \frac{1 + i}{\sqrt 2} z, ~  ~ z > 0. \label{finiteBeta}
\end{equation}
As discussed in Section~\ref{S:rindlerAdSCFT}, this  simply represents a passive deformation of $z$-integration contours in the complex plane for interaction vertices and bulk endpoints 
(with analytic sources as in subsection~\ref{S:generalCorrelators}), and multiplication by some phases for boundary endpoints. Therefore this "move" does not affect the finiteness of the correlator. But now we see that for \emph{all} points in BTZ, we have
\begin{equation}
	\xi_n \xrightarrow[n\to\infty]{} \frac{2 i z z'}{\lambda^{n} (i z^2 - x^+ x^-) + {\cal O}(1) }, \label{finiteBetaXi}
\end{equation}
so that the propagator summand $\propto \lambda^{-n \Delta} F(0)$ always for large $n$, the sum converges, and the correlator is indeed finite (even as $\epsilon \to 0$).
It is crucial to note that this finiteness required integrating over all $z > 0$ in the first place, so that inside the horizon we are integrating both inside and outside the singularity. Therefore finiteness required inclusion of the whisker regions. 

\begin{figure}
	\centering
	\begin{subfigure}[b]{0.3\textwidth}
		\centering
		\includegraphics[width=\textwidth]{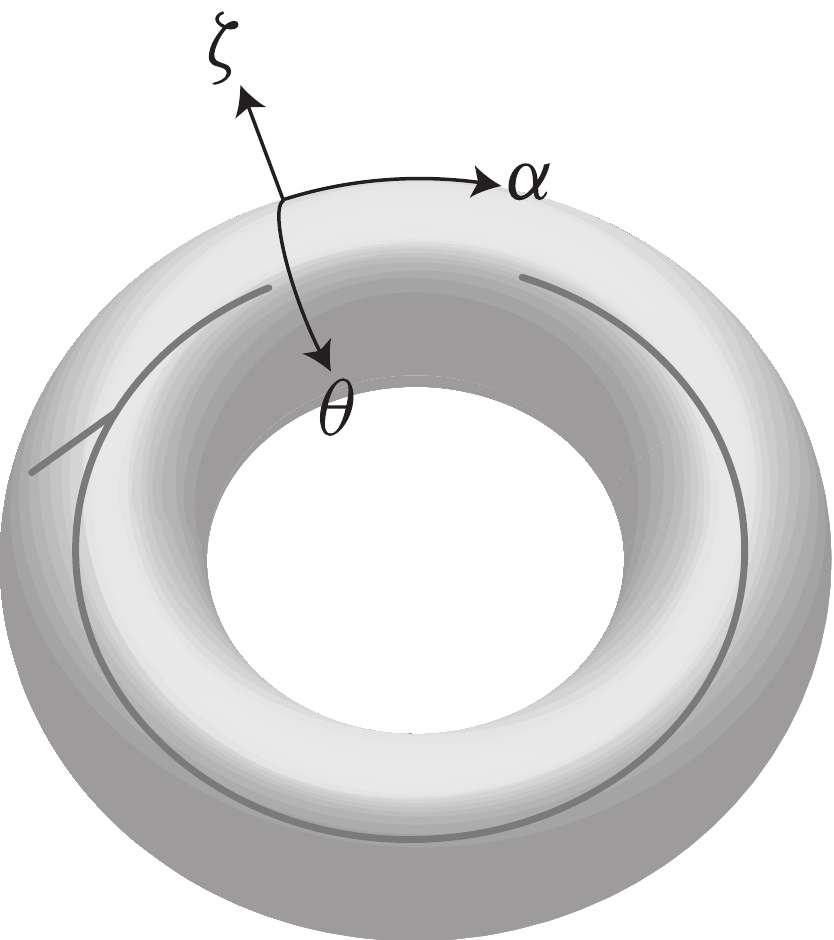}
		\caption{}
		\label{solidTorus}
	\end{subfigure}
	\begin{subfigure}[b]{0.15\textwidth}
		\centering
		\includegraphics[width=\textwidth]{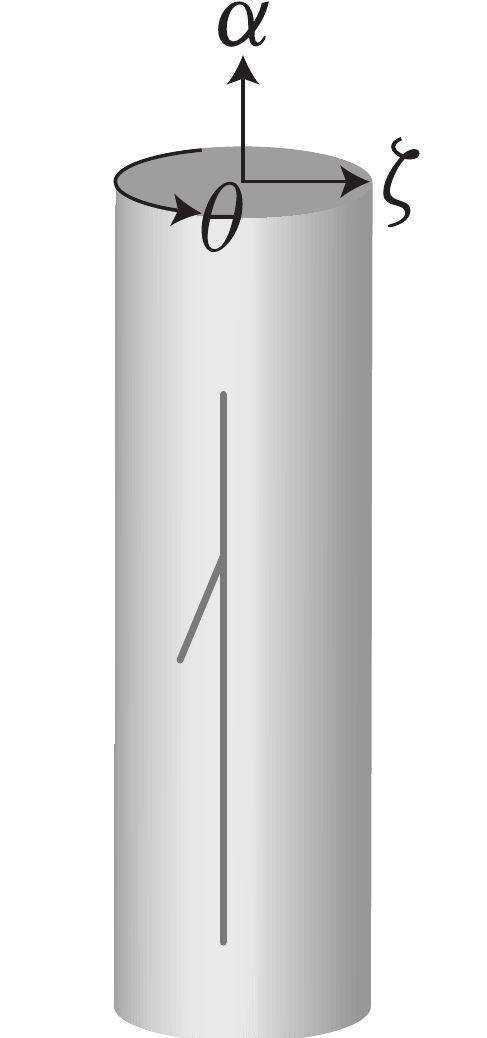}
		\caption{}	
		\label{solidCylinder}
	\end{subfigure}
	\begin{subfigure}[b]{0.3\textwidth}
		\centering
		\includegraphics[width=\textwidth]{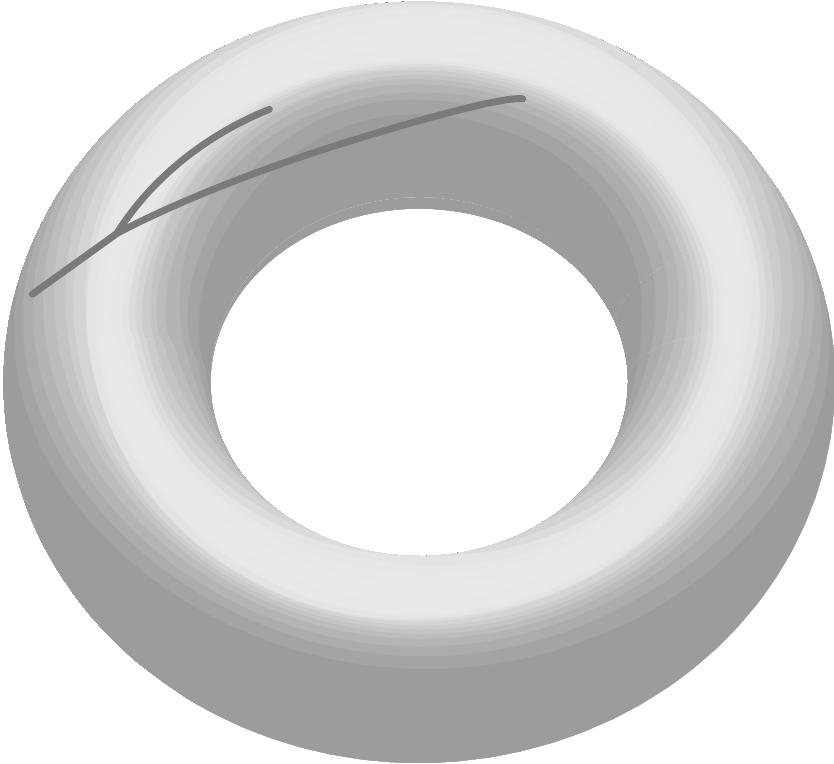}
		\caption{}
		\label{solidTorus2}
	\end{subfigure}
	\begin{subfigure}[b]{0.15\textwidth}
		\centering
		\includegraphics[width=\textwidth]{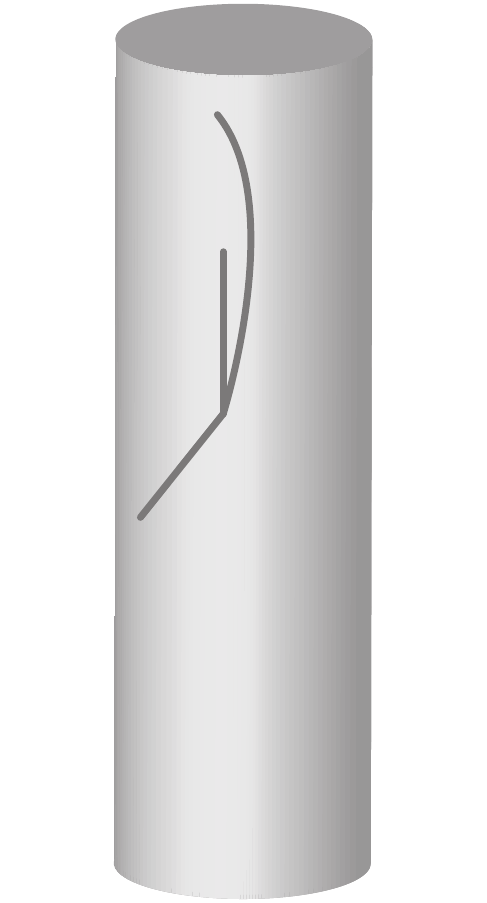}
		\caption{}	
		\label{solidCylinder2}
	\end{subfigure}
	\caption{Relationship between bulk tree level BTZ diagrams and the corresponding diagrams on the AdS$_\text{Poincar\'e}$ covering space.  The dark gray lines are to be interpreted as propagators inside the gray solids (although they may end on the surface).}
	\label{AdSDiagrams}
\end{figure}

The relationship between BTZ and the covering spacetime AdS$_\text{Poincar\'e}$ diagrammatics is most straightforwardly seen in the (leading) tree-level diagrams of EFT, as illustrated in 
Fig.\@~\ref{AdSDiagrams}.  We draw the BTZ spacetime as filling in the Lorentzian torus, to topologically make a \emph{solid} torus with the Lorentzian torus surface as its boundary. In order to view BTZ like this we have switched the roles of the two circles of the Lorentzian torus with respect to Fig.\@~\ref{fundamentalRegions}. Specifically, we generalize \eqref{polarMinkowski} to the bulk, 
\begin{align}
	t = e^{\alpha} \sin \zeta \sin \theta && x = e^{\alpha} \sin \zeta \cos \theta && z= e^{\alpha} \cos \zeta  &&  \left( 0 \leq \zeta \leq \tfrac{\pi}{2} \right).
\end{align}
We compare diagrams in the solid torus with diagrams in AdS$_\text{Poincar\'e}$, which we view in the above coordinates as a solid Lorentzian cylinder by first removing the origin. 
Its boundary, the surface of that cylinder, is interpreted as $1+1$ Minkowski spacetime with the origin removed in $\alpha, \theta$ coordinate space. In this representation, the solid torus is simply the quotient of the solid cylinder by a discrete $\alpha$ translation, periodizing the direction along the cylinder's length.  Figs.\@~\ref{solidCylinder} and \ref{solidCylinder2} show tree diagrams on AdS$_\text{Poincar\'e}$ (as the solid cylinder) where the endpoints of both diagrams are (examples of) images of  the same set of endpoints for a  BTZ (solid torus) correlator. 
 Wrapping the AdS diagrams onto BTZ in Figs.\@~\ref{solidTorus} and \ref{solidTorus2}, the two AdS$_\text{Poincar\'e}$ diagrams appear as different contributions to the same BTZ correlator, but with different image terms for one of the propagators. In this way, by adding up all connected tree
 AdS$_\text{Poincar\'e}$ diagrams with end points being images of the desired BTZ correlator, we get
 the tree-level BTZ diagram, where every BTZ propagator is a sum over AdS image propagators. 
 	
\subsection{Local boundary correlators:  EFT dominance and scattering behind the horizon}

While EFT correlators are finite in BTZ, as described above, this in itself does not prove that these finite correlators dominate the true correlators, which may also include the contributions of heavy states of the UV complete quantum gravity. It is also not immediately obvious that these BTZ correlators sharply probe scattering processes inside the horizon in the same way that AdS$_\text{Poincar\'e}$ correlators probe scattering behind the Rindler horizon. 
However, both these properties are indeed true of the "protected" set of local boundary correlators of BTZ realized as a quotient of AdS$_\text{Poincar\'e}$. (Bulk correlators contain extra UV sensitivity, as do the more general boundary correlators in the BTZ realization as a quotient of 
AdS$_\text{global}$ \cite{btzglobal}.) 

We first demonstrate that for $\lambda \equiv e^{r_S} \gg 1$, local boundary EFT correlators are dominated  by  $n=0$ in the sum over images in each propagator, \eqref{BTZPropagator}. This follows after rotating $z$ by $\beta = \pi/2$ in the complex plane, \eqref{finiteBeta}, so that 
 for large $\lambda$ and $x^{\pm},z, y^{\pm}, z' \sim {\cal O}(1)$ in propagators, all other terms are $\sim {\cal O}(\lambda^{-|n| \Delta})  $.
  The scaling,  $x^{\pm},z, y^{\pm}, z' \sim {\cal O}(1)$ in $\lambda$ follows,  even though these arguments are being integrated, 
 if the boundary endpoints $x_i^{\pm}$ (which determine the region of convergence of the integrals) are chosen ${\cal O}(1)$.  
That is, after rotation of $z$, it is as if there were no singularity, just a very large but compact $\sigma$ direction, and the resulting diagrams are \emph{dominated by the equivalent un-imaged diagrams in the covering AdS$_\text{Poincar\'e}$ spacetime}. In particular, since these un-imaged diagrams describe scattering behind the Rindler horizon, the BTZ correlators must describe scattering behind the quotient of the Rindler horizon, namely the black hole horizon. 

Because we are limiting ourselves to the Poincar\'e patch of AdS and its quotient, we are restricted in how "sharp"  scattering processes can be when initiated and detected from the boundary. The reason is that we have to send and receive scattering waves from the boundary at $z = 0$,
naively suggesting a violation of $z$-momentum conservation. Indeed, $z$-translation invariance is broken by warping but this does not allow us to scatter waves with $z$-wavelengths much smaller than the AdS radius of curvature using boundary correlators. On the other hand, there is no similar obstruction to how small the $x$-wavelength can be. Wavepackets with $z$-wavelengths of order $R_{AdS}$ and much smaller 
$x$-wavelengths can be aimed so that scattering definitely only takes place inside the horizon, and predominantly away from the singularity. 
They thereby give us access to reasonably sharp probes of inside-horizon scattering, but obviously not the most general scattering processes. 
In short, the sharpness of BTZ boundary correlators is the same as for AdS$_{Poincare}$ boundary correlators.

Since we are dominated by the un-imaged AdS$_\text{Poincar\'e}$ correlators, with $ {\cal O}(e^{-|n| r_S \Delta}) $ corrections to ensure BTZ compactness (in $\sigma$), it follows that EFT dominates the boundary correlators as it does in  AdS$_\text{Poincar\'e}$. Even if we included a very heavy particle into the Feynman rules, it can be integrated out in the leading $n=0$ contribution as in AdS, inducing only contact effective interactions among the light EFT states. We will see that this UV-insensitivity is not the case for the subleading $ {\cal O}(e^{-|n| r_S \Delta}) $ effects in Section~\ref{S:sensitivityToSingularity}, and that the effects of large cosmological blueshifts near the singularity are indeed present. 

It may appear that  {\it bulk} correlators are similarly protected by the above reasoning, but it is important to understand why this is not the case. 
 The subtlety is that the above analysis required first performing the complex rotation of \eqref{finiteBeta}. As we have seen, this  only changes Witten diagram contributions to boundary correlators by complex phase factors, so that 
estimates for the magnitudes of different contributions apply straightforwardly to the original correlator before rotation. However, this is not the case for bulk correlators, where bulk sources have to be analytically continued to accomplish \eqref{finiteBeta}, as discussed in subsection~\ref{S:generalCorrelators}. In general, such analytic continuations will completely change the magnitudes of different contributions. Therefore estimates performed after \eqref{finiteBetaXi} do not apply to the original BTZ correlators before \eqref{finiteBetaXi}.  Indeed we will give an example of bulk correlator UV sensitivity in Section~\ref{S:sensitivityToSingularity}. 

If the $\sigma$ circle were always very large there would be no surprise that the correlators approximate those of non-compact $\sigma$, namely AdS$_\text{Poincar\'e}$.   But it is at first surprising here that the  $n \neq 0$ corrections are small even for Witten diagrams passing through the singularity, where the physical size of the $\sigma$ circle is going to zero, as seen in the Schwarzchild metric of \eqref{AdSschwarzschildMetric}. We will see the deeper reason for this in subsection~\ref{S:ininFormalism}.

\subsection{Method of Images applied to Rindler AdS/CFT}

We now use the method of images to go to the finite $r_S$ (compact $\sigma$ direction) analog of \eqref{thermofieldExpandedAdS}, relating local EFT correlators anywhere in BTZ to (non-local) EFT correlators in two copies of the outside-horizon BTZ with thermofield entanglement:
\begin{multline}
	Z_\text{BTZ} = {}_\text{HH}\matrixel{\Psi}{ \left[\1 \otimes (\T_\tau \hat U_F)^\dag \right] \left[e^{\pi P_-} \otimes e^{-\pi P_-} \right]  \label{thermofieldBTZ} \\
	\times \left[(\T_\tau \hat U_L) \otimes (\T_\tau \hat U_R)\right]  \left[e^{-\pi P_-} \otimes e^{\pi P_-} \right] 
		\left[(\T_\tau \hat U_P)^\dag \otimes \1 \right]}{\Psi}_\text{HH}.
\end{multline}
The left-hand side is the generating functional of the bulk or boundary correlators of the BTZ black hole discussed above, 
with any bulk sources being analytic in $z$.   The right-hand side is written in terms of the thermofield state formed by two copies of the \emph{outside-horizon} ($r>1$) portion of  the Schwarzschild view of the BTZ black hole. (Of course these two copies can then be thought of as the outside-horizon portions of a single extended BTZ black hole.) This outside-horizon geometry is just the quotient of the AdS$_\text{Rindler}$ wedge of AdS$_\text{Poincar\'e}$. 
The time and space translation generators on the right-hand side are with respect to the $\tau, \sigma$ directions of the Schwarzschild coordinates for the BTZ black hole, and the fields in all source terms on the right-hand side live only outside the horizon. 

The derivation of \eqref{thermofieldBTZ} from \eqref{thermofieldExpandedAdS} is more transparent when the right-hand side is written in trace form,
\begin{equation}
	Z_\text{BTZ} = \tr_\text{outside} (\T_\tau \hat U_L) e^{-\pi P_+}  (\T_\tau \hat U_F)^{\dagger} e^{-\pi P_-}  (\T_\tau \hat U_R)  e^{-\pi P_+}  (\T_\tau \hat U_P)^{\dagger} e^{-\pi P_-}, \label{traceFormulaBTZ}
\end{equation}	
where the trace is over the Hilbert space on one copy of the outside-horizon region. This equation is just the quotient of the analogous non-compact statement, where the left-hand side is the generating functional for AdS$_\text{Poincar\'e}$ correlators and the right-hand side is a trace over the Hilbert space on AdS$_\text{Rindler}$. On both sides, the compact result follows by imaging the relevant type of propagator and keeping coordinates within a fundamental region. 	As pointed in the discussion below \eqref{damping}, the exponential weights in \eqref{traceFormulaBTZ} are a net suppression of high energy excitations of the Schwarzchild spacetime (outside the horizon), and therefore the right-hand sides of \eqref{thermofieldBTZ} and \eqref{traceFormulaBTZ} are mathematically well-defined, matching the good behavior we have found for the left-hand side. 

As discussed below \eqref{thermofieldCFT}, one can think of local correlators ending inside the horizon (including whiskers) on the left-hand side of \eqref{thermofieldBTZ} as being equal to correlators outside the horizon for non-local operators of the form $e^{\pi P_-} {\cal O}_\text{local} e^{-\pi P_-}$ on the right-hand side.
So far we have established that local boundary correlators in BTZ are EFT-dominated and finite, but we still have not given a physical interpretation of such correlators when they end in the whiskers, problematic due to the time-like closed curves. However, for local boundary correlators, the right-hand side of \eqref{thermofieldBTZ} gives such a simple interpretation. Defining states, 
\alignStart
	\ket{\Psi_P}_\text{outside} &\equiv \left[e^{-\pi P_-} \otimes e^{\pi P_-} \right] 
		\left[(\T_\tau \hat U_P)^\dag \otimes \1 \right]\ket{\Psi}_\text{outside}  \\
	\ket{\Psi_F}_\text{outside} &\equiv \left[e^{-\pi P_-} \otimes e^{\pi P_-} \right] 
		\left[(\T_\tau \hat U_P)^\dag \otimes \1 \right]\ket{\Psi}_\text{outside},
\alignEnd		 		
Eq.\@~\eqref{thermofieldBTZ} can be re-written
\alignStart
	Z_\text{BTZ} &= \matrixel{\Psi_F}{(\T_\tau \hat U_L) \otimes (\T_\tau \hat U_R)}{\Psi_L}
\alignEnd
That is, the correlators including possible endpoints in the whisker boundaries are equal to correlators with endpoints only on the boundaries outside the horizon, but with the thermofield state being replaced by the modified $| \Psi_{P,F} \rangle$ states. Given that we have established 
that such correlators are dominated by the non-compact AdS$_\text{Poincar\'e}$ EFT correlators (image terms being subdominant), we can readily interpret these new states. In non-compact correlators, endpoints in the $F$ (say) boundary just act to "detect" the results of earlier scattering inside the Rindler horizon, or evolving backwards, they set up "out" states, $|\Psi_F \rangle$ which sharply probe the results of the scattering process. The same must therefore be true after quotienting to BTZ, where the $F$ boundary is the whisker boundary. In summary, the whisker regions can be thought of as an auxiliary spacetime in which the local boundary correlator endpoints encode non-local operators that sculpt the thermofield state into a variety of "in" and "out" states that probe the results of scattering inside the horizon, very much as do $F$/$P$ boundary correlators in AdS$_\text{Poincar\'e}$. Furthermore, the local boundary correlators of BTZ are diffeomorphism invariants of quantum gravity.

\subsection{Connecting to CFT Dual on $\partial$BTZ}

\begin{figure}
        \centering
        \begin{subfigure}[b]{0.3\textwidth}
                \centering
                \includegraphics[width=\textwidth]{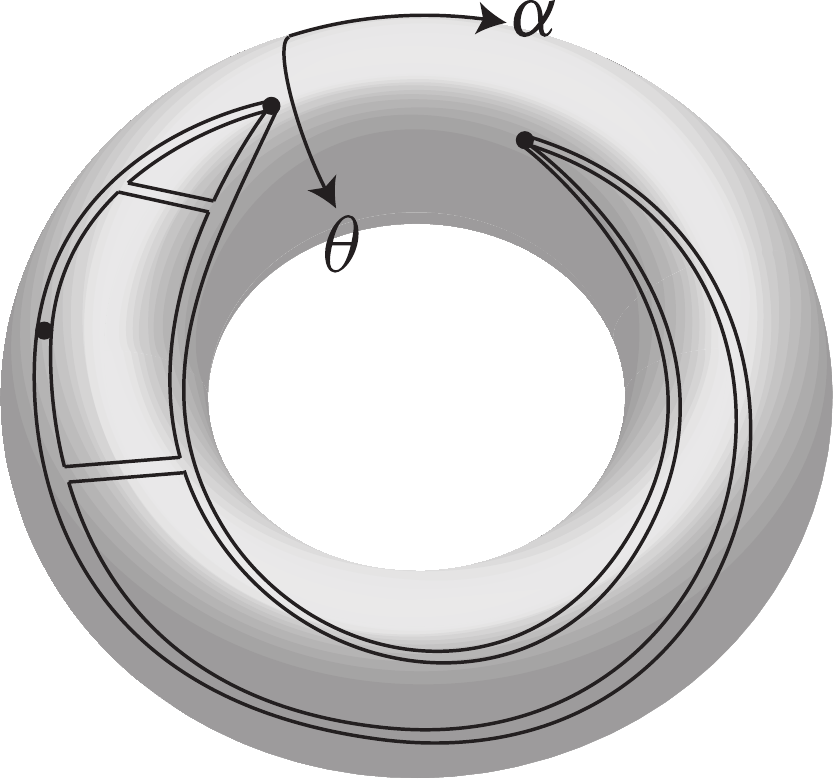}
                \caption{}
                \label{hollowTorus}
        \end{subfigure}
        \qquad
        \begin{subfigure}[b]{0.15\textwidth}
                \centering
                \includegraphics[width=\textwidth]{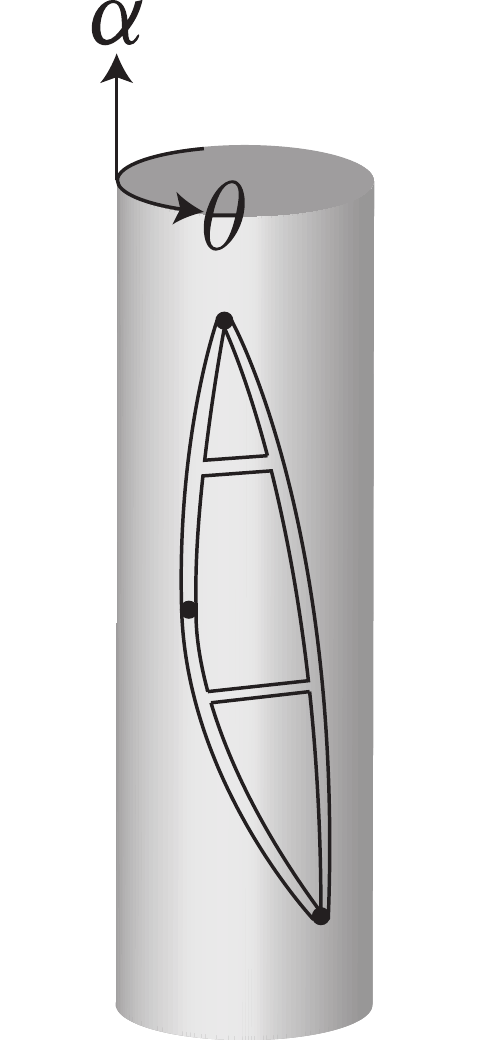}
                \caption{}
                \label{hollowCylinder}
        \end{subfigure}
        \caption{Relationship between planar CFT diagrams in double-line notation (reviewed in~\cite{juantasi}) on the Lorentzian torus and its covering space, the Lorentzian cylinder.  These CFT "gluon" lines are to be interpreted as propagating on the boundary surfaces of the gray solids in Fig.\@~\ref{AdSDiagrams}.  The black dots represent local CFT operators.}
        \label{CFTDiagrams}
\end{figure}

It remains to connect \eqref{thermofieldBTZ} to the CFT  thermofield form of \eqref{thermofieldCFT}, \eqref{thermofieldCFTSeparated}, or equivalently the CFT on the Lorentzian torus $\equiv \partial BTZ$.
The diagrammatic expansion in the bulk theory is dual to a large-$N_{\CFT}$ expansion in a CFT gauge theory.
At infinite $r_S$, tree diagrams such as Fig.\@~\ref{solidCylinder} capture the same physics as the planar diagrams of Fig.\@~\ref{hollowCylinder} in the dual CFT, by standard AdS/CFT duality. Just as Fig.\@~\ref{solidCylinder} maps to contributions to BTZ correlators for specific fixed images in Fig.\@~\ref{solidTorus}, Fig.\@~\ref{hollowCylinder} maps to planar diagrams of the CFT on the Lorentzian torus in Fig.\@~\ref{hollowTorus} , as discussed in more detail for the example of subsection~\ref{matchingToCFT}. Equivalently, we have seen that we can use the right-hand side of \eqref{thermofieldBTZ} for  BTZ tree amplitudes, and these are then identified with  the careful construction of \eqref{traceFormula}, \eqref{traceFormulaDyson} for the CFT on the Lorentzian torus at planar order. That is, the method of images straightforwardly identifies the bulk tree amplitudes to planar CFT amplitudes, either directly on ($\partial$)BTZ or in equivalent thermofield form. The value in the CFT construction of
 \eqref{traceFormula}, \eqref{traceFormulaDyson}  however is that it
includes a UV-complete and non-perturbative (in $1/N_{\CFT}$) description of the approach to the singularity, even when bulk EFT eventually completely breaks down.

%At exactly the same leading order in $N_\text{CFT}$ as the planar CFT diagrams on the Lorentzian torus, Fig.\@~\ref{hollowTorus},  that "descend"  %from planar CFT diagrams on the cylinder of Fig.\@~\ref{hollowCylinder} ($1+1$ Minkowski space with the origin removed),

\begin{figure}
        \centering
        \begin{subfigure}[b]{0.3\textwidth}
                \centering
                \includegraphics[width=\textwidth]{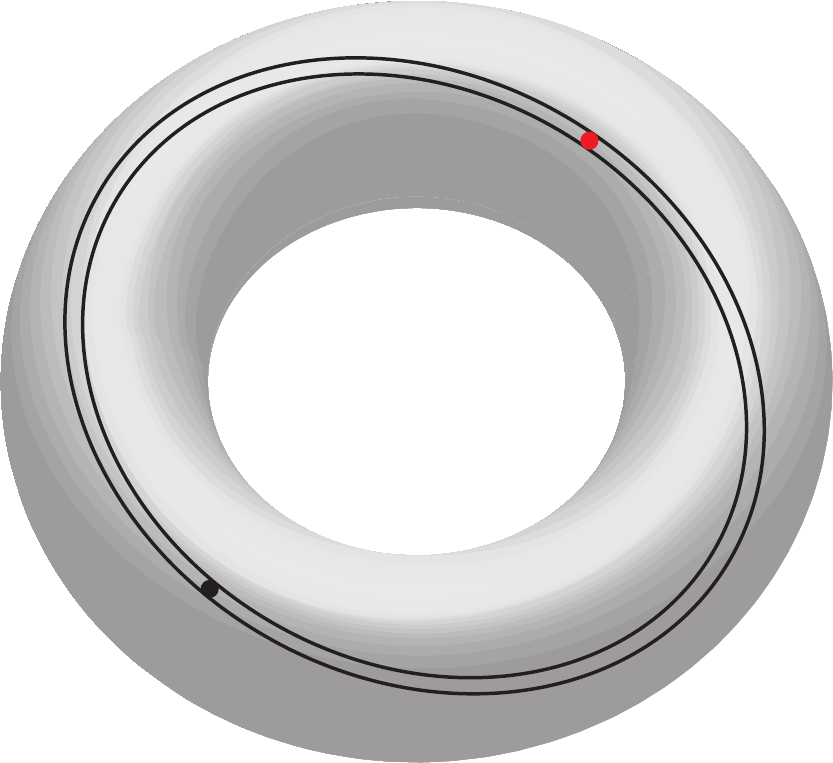}
                \caption{}
                \label{wrapper}
        \end{subfigure}
        \qquad
        \begin{subfigure}[b]{0.15\textwidth}
                \centering
                \includegraphics[width=\textwidth]{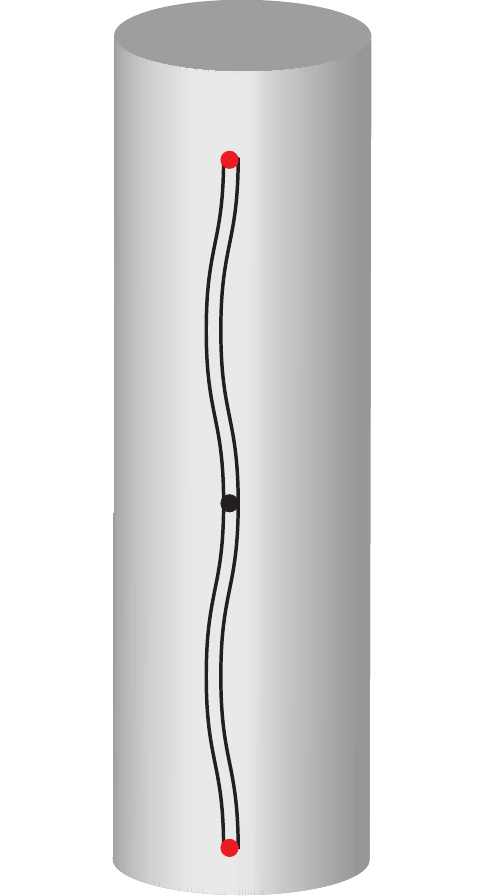}
                \caption{}
                \label{wrapperImages}
        \end{subfigure}
        \caption{Naively, there are diagrams at leading order in $N_\text{CFT}$, such as (a), but which unwrap to diagrams in Minkowski space, such as (b), which violate gauge invariance (for example, gauge non-singlets are created by different images of the same operator).  Such contributions vanish by gauge invariance.}
        \label{wrapperDiagrams}
\end{figure}

Naively, at the same planar order in the CFT
there are also diagrams which "wrap" around the torus in the $\alpha$ direction, such as Fig.\@~\ref{wrapper}, which do not descend from cylinder (Minkowski) planar diagrams
by the method of images, and yet are of the same order in $N_{\CFT}$. These diagrams necessarily
break up a minimal color singlet combination of "gluons" and send some of them to an operator and the remainder to its image (Fig.\@~\ref{wrapperImages}). But in the full gauge-invariant path integral on the torus such diagrams are constrained to vanish. This is a familiar fact if we think of the $\alpha$ direction as "time" (now that we are acclimatized to choosing the "time" direction for our convenience): the non-abelian Gauss Law constraint  says that only gauge-invariant states propagating around the $\alpha$ direction are physical, whereas any part of of a minimal color singlet cannot be gauge-invariant. Closely analogous to this,  in equilibrium thermal gauge theory it is the Gauss Law constraint that enforces that
only gauge invariant states can circle around compact imaginary time (equivalently, the thermal trace is only over gauge-invariant states).

At nonplanar order in the CFT, there are subleading diagrams that can be identified with the loop-level bulk diagrams
that unitarize the tree-level contributions. But there are also new CFT contributions not of this form, namely creation and destruction of finite-energy Wilson-loop states winding around the compact $\sigma$ direction. These have no analog in the non-compact case.  In BTZ, these are dual to quantum gravity states, generically finite-energy "strings", that wind around the bulk $\sigma$ circle, but which have no analog in non-compact
AdS$_\text{Poincar\'e}$. The effects of such extended objects cannot be captured by the simple diagrammatic method of images we have followed for BTZ.  If  the extended objects have tension then the winding states will ordinarily be extremely heavy for large $r_S$, and thereby give exponentially suppressed virtual contributions to correlators between well-separated source points.  But approaching the singularity, the physical $\sigma$-circumference approaches zero, as seen in the Schwarzchild metric, \eqref{AdSschwarzschildMetric}, and the winding states can become light. They are then part of the normally-UV physics which becomes important near the singularity. See \cite{eva} for discussion within string theory. We expect this physics to be contained in our CFT proposal for the non-perturbative BTZ dual, but not part of the EFT checks we have performed in the regime where we argued EFT should dominate.

Finally, beyond any order in $1/N_{\CFT}$, the CFT correlators will have effects, not matching bulk EFT or even perturbative string theory. They may well play an important role near the singularity.

\subsection{Deeper reason for insensitivity to singularity} \label{S:ininFormalism}

Our diagrammatic derivations have non-trivially confirmed our formal CFT expectations for local correlators set forth in \eqref{thermofieldCFT}.
But this does not explain \emph{why} EFT is well-behaved despite the singularity, why technically there was a way to deform the contour 
for interaction vertex integrals so as to avoid the perturbative face of the singularity in image sums, and for boundary correlators why these image sums converge so rapidly.
 We might also worry that EFT misses important UV physics near the singularity, such as heavy particles or the winding states mentioned above. In general, we therefore want to understand whether to trust EFT at all for boundary correlators, especially when some endpoints are in  the whiskers. 

 \begin{figure}
	\centering
	\includegraphics[width=0.4\textwidth]{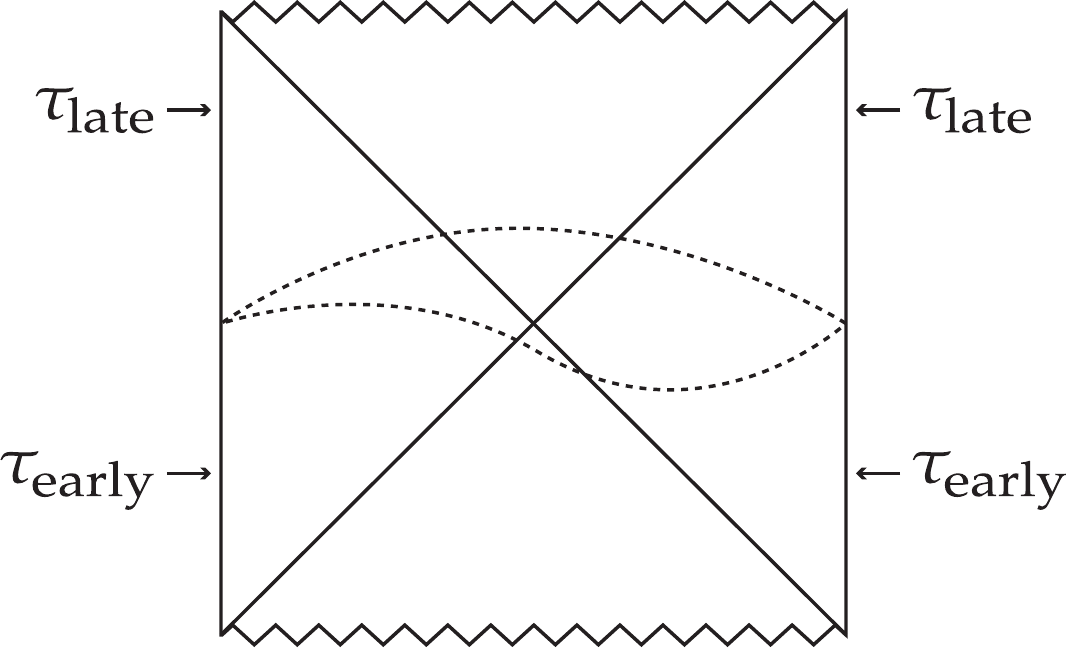}
	\caption{The dashed curves represent two spacelike hypersurfaces that are related by bulk diffeomorphisms.}
	\label{spacelikeSurfaces}
\end{figure}

We begin with non-compact AdS$_\text{Poincar\'e}$ where we understand the diagrammatic expansion. We have formally motivated and then diagrammatically derived the Rindler AdS/CFT result of \eqref{thermofieldExpandedAdS} and more compactly, \eqref{thermofieldCFT}. The typical local boundary correlator of AdS$_\text{Poincar\'e}$ is thereby re-expressed on the right-hand sides using \eqref{nonlocalOperators},
\begin{multline}
	\matrixel{\Psi} { \{ \1 \otimes \overline{T}_{\tau} [  (e^{\pi P_-} {\cal O}^F_1 e^{- \pi P_-}) .... (e^{\pi P_-} {\cal O}^F_{n_F} e^{- \pi P_-}) ] \} \\
		\times \{ T_{\tau} [  {\cal O}^L_1 ....  {\cal O}^L_{n_L}] \otimes T_{\tau} [  {\cal O}^R_1 ....  {\cal O}^R_{n_R}]  \} 
		\{ \1 \otimes \overline{T}_{\tau} [  (e^{\pi P_-} {\cal O}^P_1 e^{- \pi P_-}) .... (e^{\pi P_-} {\cal O}^F_{n_P} e^{- \pi P_-}) ] \} } {\Psi}, 
\end{multline}
where the ${\cal O}$ are local Heisenberg operators of the CFT  or local boundary operators of AdS.  The operators of the form 
$e^{\pi P_-} {\cal O} e^{- \pi P_-} $ are then non-local, but only in the spatial sense. Time evolution, implicit in the Heisenberg operators, ranges between the earliest and latest times that appear in any of the ${\cal O}$ operators above, $\tau_\text{early}, \tau_\text{late}$, say. It is important to note that this time evolution does not go all the way from $\tau = - \infty$ to  $\tau = + \infty$. This is no surprise because we have a (generalized) in-in formalism \cite{schwinger1, schwinger2} (see \cite{weinberg} for a modern discussion and review) in the thermofield form for correlators. We represent this situation in the Penrose diagram of Fig.\@~\ref{spacelikeSurfaces}, where the symmetry $\sigma$ direction is omitted, but is now non-compact $ - \infty < \sigma < \infty $. The spacelike hypersurfaces are pinned on the boundary by the boundary time evolution, but their form in the bulk is otherwise arbitrary
 by diffeomorphism invariance. What is not immediately obvious from the figure, but straightforwardly verified by the AdS$_\text{Poincar\'e}$ metric, 
 is that all such hypersurfaces pinned to the boundary outside the (Rindler) horizon cannot go beyond the jagged lines at any point. This is the only significance of the jagged lines in Fig.\@~\ref{spacelikeSurfaces} since there is of course no singularity in AdS.  
 We have depicted the simplest choice of such hypersurfaces.  
 
 The central point when we move to the compact BTZ case for such correlators, as in \eqref{thermofieldBTZ}, is that Fig.\@~\ref{spacelikeSurfaces} still holds, but now with the omitted $\sigma$ direction of course being compact, and the jagged lines depicting the location of the singularity.  What we see is that in deriving \eqref{thermofieldBTZ} from \eqref{thermofieldExpandedAdS} we are only trusting the diagrammatic expansion and the method of images to compactify $\sigma$ \emph{away from the singularity}. As long as $\tau_\text{early/late}$ are not too early or late, the physical circumference of the $\sigma$ circle can be taken to be large throughout the time evolution and 
 it is not surprising if our correlators are dominated by the non-compact limit and insensitive to the UV physics of the singularity.
In particular, winding states will be very massive throughout this evolution.

\section{Sensing Near-Singularity Physics} \label{S:sensitivityToSingularity}

A good part of this paper has been concerned with the validity and use of bulk EFT and the diagrammatic expansion in order to capture scattering processes behind the BTZ horizon. This has allowed us to test our proposed non-perturbative CFT formulation under 
conditions where we already know what to expect. However, the real importance of such a CFT formulation is that it allows
us to study processes close to the singularity,  where large cosmological blue-shifts make the physics very UV sensitive and EFT breaks down. 
In this section, we want to demonstrate that the UV-sensitive physics near the singularity is certainly present in the BTZ quantum gravity and that whisker correlators and their CFT duals can detect this. To do this, we will show under what circumstances we become sensitive to heavy states beyond EFT, and yet without such sensitivity invalidating our derivations. 

 We know that we see divergences if correlator endpoints are right on the singularity, as simply illustrated just by \eqref{kWittenImaged}. 
 But  EFT should come with some effective cutoff length, below which we do not ask questions. If we simply move correlator endpoints more than the cutoff length away from the singularity they are finite and the cosmological blueshifts are more modest. But mathematical finiteness is not necessarily the same as insensitivity to heavy states. 
We begin by demonstrating that even at distances/times  of order $R_\text{AdS}$ away from the singularity, correlators are sensitive to the UV heavy states outside BTZ EFT. 
To do this we move the point in the $F$ wedge of our Section~\ref{S:singularity} example correlator
from the boundary  to the interior 
%(as we did in subsection~\ref{S:classicalApproximation}), 
and consider
\begin{multline}
	\vev{ \tilde \phi_F(x_F, z') \O(x_{R_1}) \O(x_{R_2})}_\text{tree BTZ} \\
	= \int_\text{fund.} d^2 y dz \sqrt g \, \tilde G_\text{BTZ}(x_F, z'; y,z) g^{MN} \d_M K(x_{R_1}; y,z) \d_N K(x_{R_2}; y, z). \label{bulkCorrelator}
\end{multline}
We assume from now on that $\tilde{\Delta} \gg 1$ and corresponds (via $\tilde m^2 = \tilde \Delta ( \tilde \Delta - 2)$)  to some heavy particle of BTZ quantum gravity that is more massive than the cutoff of BTZ EFT ("string excitations"). We are going to show that we are sensitive to such states at order $R_\text{AdS}$ separations from the singularity.

Choose $x_F$, $z'$ to have timelike geodesic to some points on the singularity, with proper times to these points 
 $ <  R_\text{AdS} (\equiv 1)$, but much larger than the cutoff length.  For example,
 $x_F^\pm = \pm (z' - \delta)$, $\delta < 1$ and near-singularity points $(y^\pm \sim \pm z', z \sim z') $
 are related in this way. 
    For such small separations, the bulk propagator can be approximated by its $2+1$ local Minkowski equivalent, 
\begin{equation}
	\tilde G_\text{AdS} \approx \frac{z'e^{i \frac{\Delta}{z'} \sqrt{(x_F - y)^2 - (z - z')^2 - i\epsilon (x_F^0 -y^0)^2 }} }{\sqrt{(x_F - y)^2 - (z - z')^2 - i\epsilon (x_F^0 -y^0)^2}},
\end{equation}
where $z'$ is the approximately constant redshift of  the inertial $2+1$ Minkowski patch.    $\tilde G_\text{BTZ}$ is of course obtained by images of $(y,z)$ from $\tilde G_\text{AdS}$.  Combining this sum with integration of interaction points over the fundamental region to get integration over all AdS$_\text{Poincar\'e}$, similarly to \eqref{boundaryCorrelator}, 
\begin{equation}
	\vev{\tilde \phi \O_1 \O_2} \sim \int_\text{AdS} d^2y dz \sqrt g \, \frac{z'e^{i \frac{\Delta}{z'} \sqrt{(x_F - y)^2 - (z - z')^2 - i\epsilon (x_F^0 -y^0)^2}} }
		{\sqrt{(x_F - y)^2 - (z - z')^2 - i\epsilon (x_F^0 - y^0)^2 }} g^{MN} \d_M K_1 \d_N K_2 + \cdots .
\end{equation}
The  ellipsis corresponds to integration over interaction points outside $R_\text{AdS}$ of $(x^\pm_F, z)$ and interaction points spacelike separated from $(x^\pm_F, z)$.  For either of these, the Minkowski-dominance approximation breaks down, but precisely so as to suppress these contributions for  
very large $\tilde \Delta$. We are therefore correctly focused on the small timelike separation region.

We again zoom in on the contribution from $(y,z)$ near the singularity and switch to Schwarzschild coordinates:
\alignStart
	\vev{\tilde \phi \O_1 \O_2} &\underset{r \to 0}{\sim} \int_{-r_0}^{r_0} \frac{dr d\sigma d \tau \, r}{(r+i\epsilon)^2} 
		\frac{e^{i \tilde \Delta \sqrt{2 - 2 rr' \cosh(\sigma - \sigma') + (r^2 + r'^2-2) \cosh(\tau - \tau')}} }{\sqrt{2 - 2 rr' \cosh(\sigma - \sigma') + (r^2 + r'^2-2) \cosh(\tau - \tau')}} \\
	&\sim \int_{-r_0}^{r_0} \frac{dr d\sigma d \tau \, r}{(r+i\epsilon)^2}  
		\frac{e^{i \tilde \Delta \sqrt{(r' - r)^2 - (\tau - \tau')^2}}}{\sqrt{(r'-r)^2 - (\tau - \tau')^2} },
\alignEnd
with $(\sigma - \sigma')$, $(\tau - \tau')$, $r$, $r'$ all small, but only $r \to 0$. As $r \to 0$, timelike separation to $(x_F, z)$ requires $r'^2 > (\tau - \tau')^2$, so 
\begin{equation}
	\vev{\tilde \phi \O_1 \O_2}_\text{BTZ} \sim \int_{-r_0}^{r_0} \frac{dr d\sigma d\tau}{r + i\epsilon} e^{\pm i \tilde \Delta(r-r')}.
\end{equation}
The behavior at $\pm r_0$ is smooth so we are basically computing the Fourier transform of $\frac{1}{r+i\epsilon}$.  As $\epsilon \rightarrow 0$, we have unsuppressed Fourier components and there is no suppression for large $\tilde{\Delta}$. In other words the particles sent in from the $R$ wedge are able to produce cutoff scale heavy particles that can propagate far away from singularity. Therefore these heavy states cannot simply be integrated out  by  $r'$, even though $l_\text{cutoff} \sim \frac{1}{\tilde \Delta} \ll r' < 1$. So EFT cannot be trusted at $r'$.

We can contrast this situation with with the analogous AdS$_\text{Poincar\'e}$ correlators (not BTZ):
\begin{equation}
	\vev{ \tilde \phi(x_F, z') \O_{R_1} \O_{R_2} }_\text{AdS} = \int d^2 y dz \sqrt g \, \tilde G_\text{AdS} g_\text{AdS}^{MN} \d_M K_1^\text{AdS} \d_N K_2^\text{AdS}.
\end{equation}
Without any infinite image sums, the $K$'s are smoothly varying on $R_\text{AdS} \equiv 1$ length scales, 
except on light cones from $x_{1,2}$. We  take $(x_F^\pm, z')$ to be away from these lightcones.  For $\tilde \Delta \gg 1$ not to be the exponent of a suppression, $(y,z)$ must be timelike separated with separation $< R_\text{AdS}$.  Therefore in looking for unsuppressed contributions, $(y,z)$ can also be taken away from $x_{1,2}$ lightcones.  But then $\tilde G$ rapidly oscillates on $\frac{1}{\tilde \Delta}$ lengths, so its integral with the smooth $\d K \d K$ is highly suppressed. 
 This is the standard reason for why 
 we can integrate out heavy $\tilde \phi$ in long-wavelength processes.
  In AdS$_\text{Poincar\'e}$ we do not see the kind of breakdown of gravitational EFT that we see in BTZ. 
  
 It is important to note that the BTZ sensitivity to heavy particles, just illustrated, takes place in a correlator with one \emph{bulk} endpoint. 
 Thus it is not in contradiction with our general observation that the purely local boundary correlators are dominated by EFT. But it is the 
 local boundary correlators that are most straightforwardly matched non-perturbatively to CFT correlators and ideally we want to use just these to detect the UV physics near the singularity. Fortunately, while we have shown that EFT dominates local boundary correlators, and even more strongly that the non-compact limit (AdS$_\text{Poincar\'e}$) dominates, this does not preclude the UV physics from residing in the small corrections to these leading approximations. The key then is to look at boundary correlators that vanish at the leading approximation, so that the small UV-sensitive effects dominate.  
 
\begin{figure}
        \centering
        \begin{subfigure}[b]{0.3\textwidth}
                \centering
                \includegraphics[width=\textwidth]{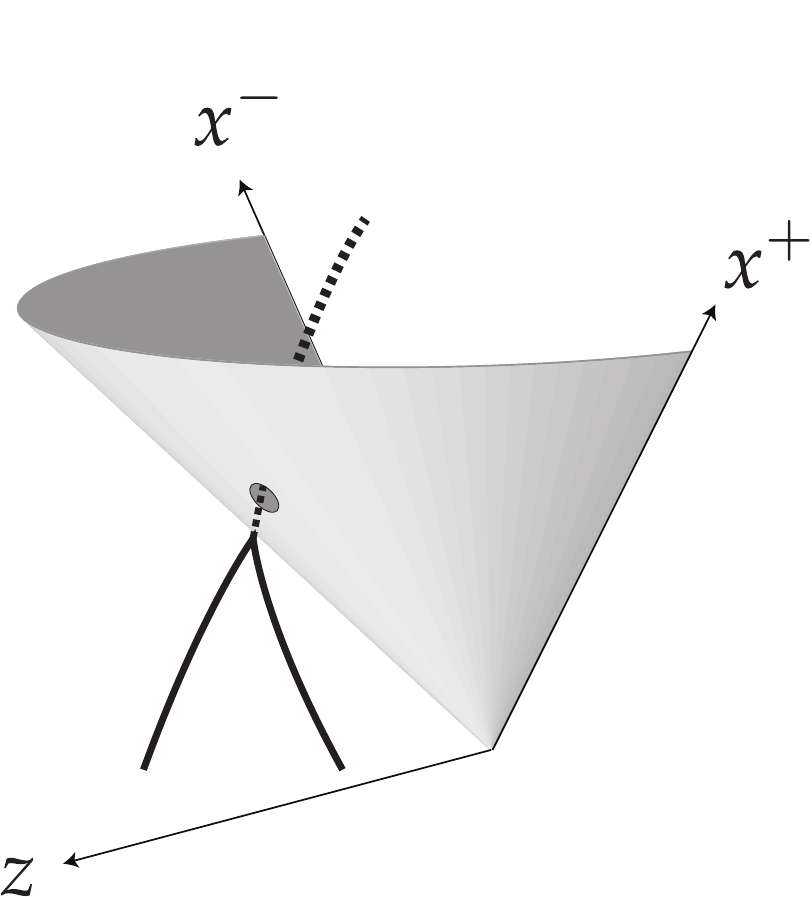}
                \caption{}
                \label{bulkSensitivity}
        \end{subfigure}
        \qquad
        \begin{subfigure}[b]{0.3\textwidth}
                \centering
                \includegraphics[width=\textwidth]{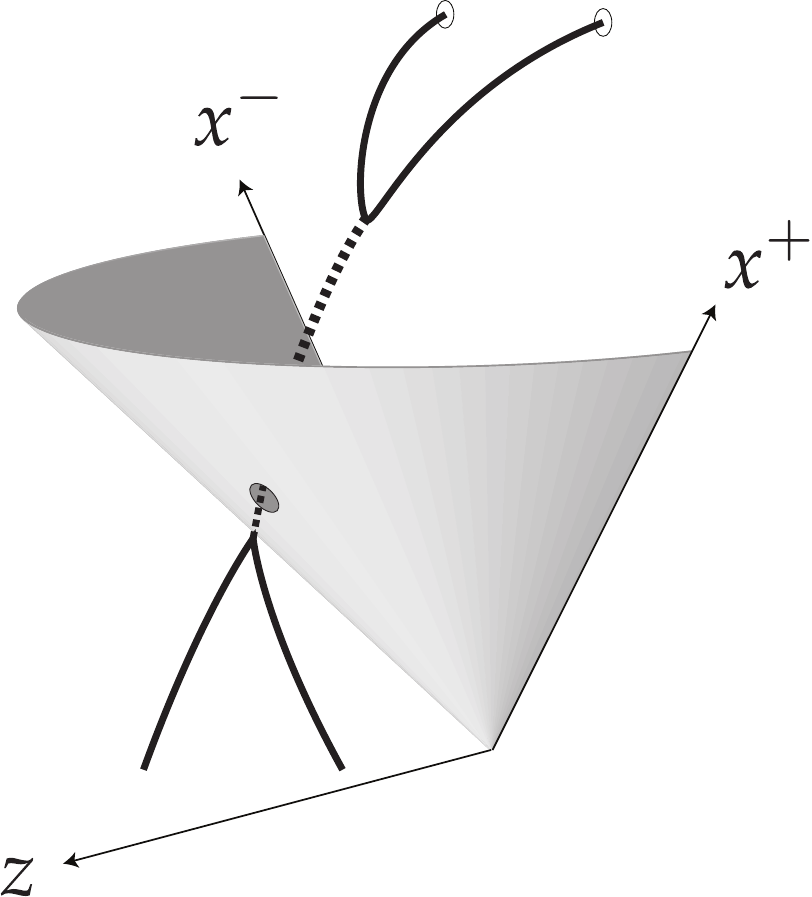}
                \caption{}
                \label{boundarySensitivity}
        \end{subfigure}
        \caption{Sensitivity to the singularity. The cone marks the location of the singularity and the dashed line represents a heavy particle. The lower black lines represent two incoming particles that are initially subthreshold. The heavy particle can be produced due to blueshifting as the singularity is approached. In (b), the heavy particle subsequently decays and its decay products are received at the boundary.}\label{strategy}
\end{figure}
 
 In the process we have considered that creates a heavy particle near the singularity using the large cosmological blueshift there, 
  and propagates it into the whisker, the obvious way to get a purely local boundary correlator is to attach two light particle lines to the 
  bulk point in Fig.\@~\ref{bulkSensitivity} and then connect these to the whisker boundary, as in Fig.\@~\ref{boundarySensitivity}. Using the ability to choose the boundary sources for the four boundary points, we can insist that the incoming beams are softer than the threshold for heavy particle production \emph{unless} one takes into account the cosmological blueshifts, that is unless one looks at large image numbers in the propagators and not just $n=0$ in the notation of \eqref{BTZPropagator}.  Similarly, we can choose sources for the boundary whisker points to be "looking" for hard particles coming from a point away from the singularity. In this way, we have chosen the boundary correlator to vanish at the usual leading approximation of the non-compact limit, but clearly the full correlator  captures the heavy particle production near the singularity and its distant propagation. In fact it might seem that this UV sensitive BTZ boundary correlator is order one, in violation of our general result. But it is easy to see the source of suppression: if it were not for the warp factor we would expect $z$-momentum to be conserved, in which case the heavy particle produced from light particle beams originating at $z=0$ would not "decay" into light particle beams which return to $z=0$. Instead we would expect the final light particle beams to escape to large $z$ and not contribute to this purely boundary correlator. The warp factor can indeed violate $z$-momentum conservation, but it is a very mild effect for hard incoming beams. This is the source of suppression of the boundary correlator that is in keeping with our general result. 
 Therefore, the four-point correlator depicted in Fig.\@~\ref{boundarySensitivity} is  small, but the UV-sensitivity dominates this small correlator. Using \eqref{thermofieldCFT} and \eqref{thermofieldBTZ}, we can write this correlator as a non-perturbatively well-defined CFT thermofield correlator. Of course there might be other UV physics which is harder to model, which would be picked up in similar fashion by our non-perturbative formulation. This is the central payoff of our work.

\section{Comments and Conclusions} \label{S:conclusion}

We have made a precise proposal for the non-perturbative CFT dual of quantum gravity and matter on a BTZ black hole, in terms of $1+1$ Minkowski CFTs with weakly-coupled, low-curvature AdS$_\text{Poincar\'e}$ duals, and provided several non-trivial checks.
It extends the now-standard duality by making sense of a CFT  "living" on the full BTZ boundary realized as a quotient of AdS$_\text{Poincar\'e}$, which includes "whisker regions" beyond the singularity containing timelike closed curves.  We did this by observing that there are well-defined non-local generalizations,
$e^{- \pi P_{\pm}}$,  of the familiar Boltzmann weight, $e^{- \beta H}$, which effectively switch the roles of space and time inside the horizon, and turn the timelike circles into familiar spacelike circles.  We then gave an equivalent thermofield construction of our CFT dual in which non-local correlators in the entangled CFTs are responsible for capturing the results of scattering inside the horizon, giving a concrete realization of complementarity.

We chose to realize BTZ as a quotient of AdS$_\text{Poincar\'e}$, rather than of AdS$_\text{global}$, based on its greater technical simplicity,
and because the set of local boundary correlators in this smaller spacetime are "protected", in the sense of being dominated by gravitational effective field theory even when the contributing Witten diagrams traverse the singularity.
 This construction gave us the minimal extension of BTZ beyond the singularity to make contact with boundary components within and to explore the role they play, even in just ensuring the mathematical finiteness of bulk amplitudes. But both AdS$_\text{Poincar\'e}$, and the portion of the extended BTZ spacetime it covers, are
geodesically incomplete. Our CFT proposal "projects" this geodesically incomplete portion of BTZ in an analogous manner to the way in which  CFT on Minkowski spacetime "projects" quantum gravity on geodesically incomplete  AdS$_\text{Poincar\'e}$.
Our CFT  dual of BTZ lives on the Lorentzian torus, which is also incomplete because of geodesics that can "escape" by passing close to the lightlike circles. But in our careful construction we are cutting out thin wedges around the lightlike circles so this does not arise. Alternatively phrased, in our final construction we only use CFT on spacetime "pieces" of the cylindrical form circle $\times$ time.  We will address the maximally extended BTZ spacetime arising from the quotient of AdS$_\text{global}$ in future work.

While analytic continuation played a role in this paper, we believe it was a matter of calculational  efficiency, rather than as a conceptual tool. For example, in subsection~\ref{rotatingZ} studying scattering through the singularity, we arrived at the same conclusion by direct computation of BTZ diagrams and by  rotating the interaction integral contour of the  $z$-coordinate. In Section~\ref{S:rindlerAdSCFT}, we used analytic continuations as the simplest way of computing the non-local consequence of the $e^{- \pi P_{\pm}}$ "generalized Boltzmann weights". In principle one could directly do the integral over such weights without any continuations but it would be technically much harder. We have checked that the direct computation in \emph{free} CFT  gives the same result as analytic continuation.

We believe our approach should be closely generalizable to quotients of higher-dimensional AdS spacetimes \cite{higherD} \cite{brill} \cite{higherD2}.
These yield interesting black objects with horizons and singularities. Of course it would be a greater technical feat to obtain the dual of higher-dimensional black holes or higher-dimensional cosmologies, without the advantage of a quotient construction from AdS, and with even worse (looking) singularities. It remains of great interest to understand the dual of evaporating black holes. We hope that the "Ising model" of black holes, BTZ, shares enough in common with other systems with horizons and singularities to provide hints on how to proceed.

In the paper, we have viewed the whisker regions, in particular their boundary, as an auxiliary spacetime grafted onto the physical spacetime which is useful in defining states on the physical region, much as Euclidean spacetime grafts are useful in defining Hartle-Hawking states on physical spacetime. However, since the whiskers do have Lorentzian signature, it is intriguing to also see if they can be accorded any more direct physical reality. Once the whisker boundaries are added to the usual boundary regions outside the horizon, we saw that we arrive at a Lorentzian torus.
 Because of the existence of circular time in the whisker boundaries, the CFT path integral does not have a canonical quantum mechanical interpretation, in that we cannot simply specify any initial state in a Hilbert space and let it evolve. Instead the path integral gives us an entire quantum spacetime which we can ask questions of, in the form of correlations of Hermitian observables. In this sense, it has the form of a kind of wavefunction of the Universe.

Alternatively, we can think of our results as simply demonstrating that the extended black hole is a \emph{robust emergent phenomenon} within
 a (single) "hot" CFT.  For instance, we saw in subsection~\ref{s:thermofieldCFTPureRindler} that with sources restricted to being \emph{outside} the horizon, in either exterior region $L$ or $R$, our trace formula reduces to \eqref{traceToThermofield},  which is equivalent to the standard thermofield description, \eqref{thermofieldLR}, \eqref{thermofieldState}.
Local sources in  $L$  can be thought of as specific non-local sources in $R$, so that there is a single CFT in a thermal heat bath,
\gatherStart
        Z[J_{L,R}] =  \tr \left\{ e^{- \beta H} \left[ e^{ \pi H} U_L e^{-  \pi H} \right] U_R  \right\} \\
        \text{where $\beta \equiv 2 \pi$.}
\gatherEnd
This is just a re-writing of the thermofield description as a thermal trace in a single CFT, rather than pure quantum mechanical evolution in two copies of the CFT. To describe observables in $L$, we see we have to take standard observables and "smear" them between $e^{ \pi H}$ and $e^{-\pi H}$. In other words, local $L$ observables are secretly just non-local observables in $R$. In this view there is only the $R$ CFT in a heat bath, and the $L$  is an "emergent" description to track certain non-local correlators. This is related to the discussion of the emergence of "doubling" of CFTs in subsection 5.1 of \cite{raju}. Now, the results of our paper, in particular the last line of \eqref{thermofieldCFT},   has shown  that the $U_{F, P}$ probes of
the inside-horizon $F, P$ regions can be thought of as "emerging" from non-local probes in the outside-horizon $R$ and $L$ regions, arising
from "smearing" standard $R, L$ observables between $e^{\pi P_-}$ and $e^{-\pi P_-}$. Putting all these observations together, we can think of probes anywhere in the extended black hole spacetime as emerging from non-local correlators in a single CFT with thermal density-matrix: Eq.\@~\eqref{traceFormula} can be re-expressed as
\gatherStart
        Z[J_{L,R,F,P}] =  \tr \left\{ e^{- \beta H} \left[ e^{ \pi H} ( e^{\pi P_-} U_P^\dag e^{-\pi P_-}) U_L e^{- \pi H} \right] ( e^{\pi P_-} U_F^\dag e^{-\pi P_-}) U_R  \right\}  \\
        \text{where  $\beta \equiv 2 \pi$.}
\gatherEnd
Non-local correlators in the thermal density matrix "project" the extended black hole, including the singularity. This follows from our results.
In this way, there is a modest "landscape" of regimes of the gravitational dual, connected by horizons.
Possibly other non-local operators, not of the forms above, may project other parts of the "landscape" of the quantum gravity dual.

\section*{Acknowledgments}
We are tremendously indebted to Ted Jacobson for generously sharing his 
 insights, advice and criticism, and  extensive knowledge of the literature. We are also grateful to Dieter Brill, Juan Maldacena, Stephen Shenker and Eva Silverstein  for enlightening discussions and suggestions.  We thank Nima Arkani-Hamed for alerting us to \cite{raju}. This research was supported by the Maryland Center for Fundamental Physics, as well as by NSF grant PHY-0968854 and by NSF grant PHY-1315155.

\bibliographystyle{utphys}
\bibliography{BHreferences}

\end{document}